# The Science Case for Multi-Object Spectroscopy on the European ELT

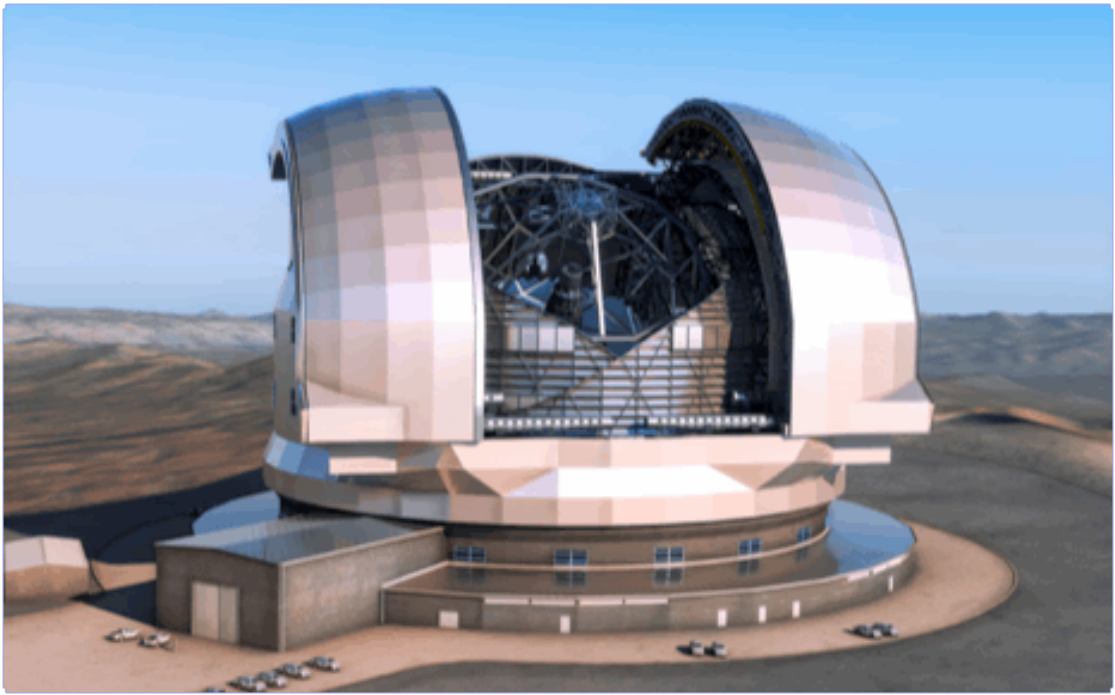

## Contributors

Jose Afonso, Omar Almaini, Philippe Amram, Hervé Aussel, Beatriz Barbuy, Alistair Basden, Nate Bastian, Giuseppina Battaglia, Beth Biller, Piercarlo Bonifacio, Nicholas Bouché, Andy Bunker, Elisabetta Caffau, Stephane Charlot, Michele Cirasuolo, Yann Clenet, Francoise Combes, Chris Conselice, Thierry Contini, Jean-Gabriel Cuby, Gavin Dalton, Ben Davies, Alex de Koter, Karen Disseau, Jim Dunlop, Benoît Epinat, Chris Evans, Fabrizio Fiore, Sofia Feltzing, Annette Ferguson, Hector Flores, Adriano Fontana, Thierry Fusco, Dimitri Gadotti, Anna Gallazzi, Jesus Gallego, Emanuele Giallongo, Thiago Gonçalves, Damien Gratadour, Eike Guenther, Francois Hammer, Vanessa Hill, Marc Huertas-Company, Roridgo Ibata, Lex Kaper, Andreas Korn, Søren Larsen, Olivier Le Fèvre, Bertrand Lemasle, Claudia Maraston, Simona Mei, Yannick Mellier, Simon Morris, Göran Östlin, Thibaut Paumard, Roser Pello, Laura Pentericci, Celine Peroux, Patrick Petitjean, Mathieu Puech, Myriam Rodrigues, Lucía Rodríguez-Muñoz, Daniel Rouan, Hugues Sana, Daniel Schaerer, Eduardo Telles, Scott Trager, Laurence Tresse, Niraj Welikala, Stefano Zibetti, Bodo Ziegler

# EXECUTIVE SUMMARY

This White Paper presents the scientific motivations for a multi-object spectrograph (MOS) on the European Extremely Large Telescope (E-ELT). The MOS case draws on all fields of contemporary astronomy, from extra-solar planets, to the study of the halo of the Milky Way and its satellites, and from resolved stellar populations in nearby galaxies out to observations of the earliest 'first-light' structures in the partially-reionised Universe. These cases are used to identify the top-level requirements on an 'ELT MOS'.

The material presented here results from thorough discussions within the community over the past four years, building on the past competitive studies to agree a common strategy toward realising a MOS capability on the E-ELT. The cases have been distilled to a set of common requirements which will be used to define the MOSAIC instrument, entailing two observational modes ('high multiplex' and 'high definition'). When combined with the unprecedented sensitivity of the E-ELT, MOSAIC will be the world's leading MOS facility. In analysing the requirements we also identify a high-multiplex MOS for the longer-term plans for the E-ELT, with an even greater multiplex ($\geq 1000$ targets) to enable studies of large-scale structures in the high-redshift Universe.

Following the green light for the construction of the E-ELT the MOS community, structured through the MOSAIC consortium, is eager to realise a MOS on the E-ELT as soon as possible. We argue that several of the most compelling cases for ELT science, in highly competitive areas of modern astronomy, demand such a capability. For example, MOS observations in the early stages of E-ELT operations will be essential for follow-up of sources identified by the *James Webb Space Telescope (JWST)*, providing a unique method to investigate the reionisation of the Universe and the origin of the first structures. In particular, multi-object adaptive optics and accurate sky subtraction with fibres have both recently been demonstrated on sky, making fast-track development of MOSAIC feasible.

# INTRODUCTION

In parallel to the Phase B design of the European Extremely Large Telescope (E-ELT; e.g. Gilmozzi & Spyromilio, 2008), nine Phase A instrument studies were undertaken (see Ramsay et al. 2010). These spanned a vast range of parameter space, in part to evaluate the relative merits of different capabilities toward the scientific cases advanced for the E-ELT, while also exploring the instrument requirements and technology readiness of likely components. Three of the Phase A studies were of multi-object spectrographs (MOS): EAGLE (Cuby et al. 2010), OPTIMOS-EVE (Navarro et al. 2010), and OPTIMOS-DIORAMAS (Le Fèvre et al. 2010). Each explored different parameter space in terms of the image quality provided by adaptive optics (AO), number of targets, spectral coverage, spectral resolution, and imaging capability.

The ESO instrument roadmap identified two first-light instruments for the E-ELT (Ramsay et al. 2014): a near-IR imager and a (red-)optical/near-IR integral field unit (IFU) spectrograph, with requirements similar to the Phase A studies for MICADO (R. Davies et al. 2010) and HARMONI (Thatte et al. 2010), respectively. These will exploit the limits of image quality of the E-ELT via high-performance AO, but will necessarily be limited in their spatial extent on the sky: i.e. a ~1 arcmin field-of-view imager, and a monolithic (i.e. single-target) IFU. HARMONI will be well suited to spectroscopy of individual high-$z$ galaxies, and stars in very dense regions (e.g. inner parts of spirals, cluster complexes), but the larger samples needed to explore galaxy evolution – at high and low redshifts – will require MOS observations.

Ahead of the expected Call for Proposals from ESO for a MOS instrument for the E-ELT, in the following sections we give an overview of illustrative science cases for such a MOS, followed by consideration of the instrument requirements to enable the proposed observations.

The E-ELT has been designed to have excellent image quality over a 10 arcmin diameter field-of-view, but the infrastructure required for AO guide stars and other sensors limits it to an effective field of ~7 arcmin. In this document we therefore assume a patrol field with an equivalent diameter of 7 arcmin, offering the attractive prospect of a ~40 arcmin$^2$ field on a 40m-class telescope for spectroscopy.

We note that the E-ELT has been designed with AO integrated into the telescope, with a large adaptive mirror ('M4') and a fast tip-tilt mirror ('M5') to correct for the turbulence in the lower layers of the atmosphere via ground-layer adaptive optics (GLAO). This is envisaged as the basic level of image quality for E-ELT observations; higher-performance AO will be delivered by dedicated modules, or within the instruments themselves. The level of image quality (i.e. AO correction) required for each case is discussed in the following sections; these are divided into one of two types: 'high definition', where high-performance AO is required for tens of objects, and 'high multiplex', where GLAO/seeing-limited performance is sufficient for 100s of objects.

# 'FIRST LIGHT' – SPECTROSCOPY OF THE MOST DISTANT GALAXIES

## 1.1. Probing the epoch of reionisation

Some 380,000 years after the Big Bang, the temperature of the Universe was low enough that the hydrogen-dominated intergalactic medium (IGM) which pervades space became neutral. The IGM today is fully ionised, heated by the integrated ultraviolet (UV) emission from galaxies and active galactic nuclei (AGN). However, how and when the IGM turned from neutral to fully ionised is a matter of great debate.

Observations of the highest-redshift quasars indicate the transition to a fully-ionised IGM occurred no later than ~1 Gyr after the Big Bang, by $z$~6 (e.g. Fan et al. 2006). Quasar spectra can be used to measure the evolution of the Ly-$\alpha$ and Ly-$\beta$ effective optical depth with redshift, and at $z$~6-6.5 they show a complete 'Gunn-Peterson trough', which is sensitive to very small amounts of neutral hydrogen (neutral fraction $x_{HI} < 10^{-3}$), thus constraining the end of reionisation. Another constraint comes from measurements of the electron-scattering optical depth from the polarisation of the cosmic-microwave background, which has demonstrated the presence of ionising sources in the early Universe at $z$~10-15 (Spergel et al. 2007). The results from *Planck* (Planck Collaboration – Ade et al. 2014) indicate that the IGM may have been half-ionised by $z$~11. The reionisation history between these limits remains a mystery and is impossible to trace with present-day instruments. Whether the reionisation occurred slowly over this period, or if there were sporadic periods of ionisation from more than one generation of ionizing sources, is completely unknown.

The prime sources of reionisation have remained elusive so far, most likely due to their faintness (e.g. Bunker et al. 2004, 2010; Choudhury & Ferrara 2007; Robertson et al. 2010; Bouwens et al. 2012). The quest for these sources, which produced the UV radiation field that reionised the IGM, is therefore intimately related to the search for the first, most distant galaxies (Lorenzoni et al. 2011, 2013; Ellis et al. 2013; McLure et al. 2013; Robertson et al. 2013) and questions such as the role of AGN, see Giallongo et al. (2012) and Science Case (SC) 4 in this White Paper.

Ultra-deep ELT-MOS observations of faint continuum-selected sources (from visible and near-IR imaging) will provide the necessary observational constraints to robustly determine the Ly-$\alpha$ properties and hence the ionisation state of the IGM from redshifts of 5 to 13 (Sect. 1.2). Such a programme will also determine the properties of these first, bright galaxies including their ISM, outflows and stellar populations (Sect. 1.3). This will ideally complement the high-quality spectral energy distributions (SEDs) which will be obtained from near-to-mid IR imaging with the *JWST* (but which will be more limited in its spectroscopic capabilities, particularly for continuum observations, compared to the E-ELT).

## 1.2. Looking for the low-luminosity sources responsible for the reionisation

The evolution of the luminosity function of Ly-$\alpha$ emitters, LF(Ly-$\alpha$), as a function of redshift, can be used to derive constraints on the volume-weighted neutral-hydrogen fraction in the IGM (e.g. Malhotra & Rhoads 2004; Dijkstra et al. 2007). The LF(Ly-$\alpha$) method is sensitive to a larger neutral-hydrogen fraction than the Gunn-Peterson test and is therefore one of the most promising methods to constrain the reionisation history of the Universe. However, current estimates are subject to considerable uncertainties and, most importantly, samples are limited to below z~7 (Clément et al. 2012). Among the main uncertainties are the very small samples of Ly-$\alpha$ emitters, the small fraction of spectroscopic confirmations, and sparse knowledge of source properties such as their star-formation rates (SFRs), SF histories, ages, outflow properties, etc.

An alternative approach is to target continuum-selected Lyman-break Galaxies (LBGs) and measure the equivalent width of their Ly-$\alpha$ emission, the fraction of them showing emission, and other related quantities. Such measurements, together with other methods (e.g. Hayes et al. 2011), indicate that beyond z~7 we are starting to probe an epoch at which the IGM is increasingly neutral. For instance, attempts to detect Ly-$\alpha$ at higher redshifts, using many tens/hundreds of hours on 8-10m class telescopes, have been less successful (e.g., Pentericci et al. 2011; Schenker et al. 2012; Caruana et al. 2012, 2014), and the current spectroscopic redshift record is 'only' z = 7.231 (Ono et al. 2012). There is no indication that the physical properties of these galaxies change quickly at these redshifts, so it is plausible to infer that the reduced visibility is due to an increased amount of neutral hydrogen in their vicinity. Depending on the model, the neutral hydrogen required to match the observations ranges from 20 to 60% (e.g., Dijkstra, Mesinger & Wyithe 2011; Jensen et al. 2013).

The current observations, combined with disappointing returns from narrow-band Ly-$\alpha$ searches at z>7 (see review by Dunlop 2012), show that significant improvements are required to probe the increasingly neutral Universe at z>7. Ultra-deep *Y*- to *H*-band E-ELT spectroscopy of large samples of faint continuum-selected sources is required to determine their Ly-$\alpha$ properties and, hence, the ionisation state of the IGM over the range z~7-15. These objects will be known from visible/near-IR imaging with facilities such as the VLT, *HST*/WFC3, and *JWST*. As illustrated by Fig. 1, with an ELT-MOS we aim to:

- Measure Ly-$\alpha$ line fluxes up to ~40 times deeper than current samples at z~7 (blank fields), ~40 times deeper than current searches for Ly-$\alpha$ emission from z~7-8.5 (lensed galaxies), and >100 times deeper than the current searches at z~8-10 (Stark et al. 2007);

- Extend Ly-$\alpha$ searches up to z~13, and undertake spectroscopy of large numbers of galaxies down to $m_{AB}$~30, which are presumably the dominant population of reionisation sources, unachievable with other planned E-ELT instruments (and well matched to the expected source densities).

- Reach an effective Ly-$\alpha$ transmission ($f_{eff}$) of 10% for the faintest objects (and lower values for the brighter objects). Even for sources behind lensing clusters, transmission factors of 1 (10)% will be reachable for objects with $m_{AB}$ = 30 (32), benefiting from gravitational magnification by a factor of ten.

Taken together this unique information will allow us to accurately determine the hydrogen ionisation fraction in the IGM and its evolution with redshift, thus measuring the cosmic reionisation history and determining important properties of the first galaxies in the Universe.

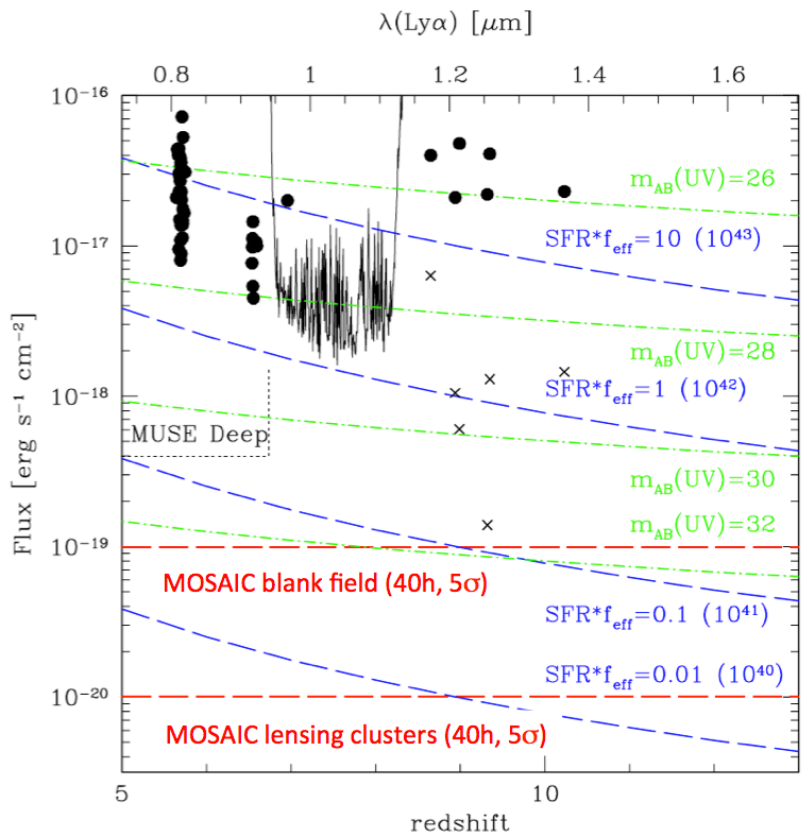

Figure 1: Ly-α sensitivities vs. redshift for an ELT-MOS cf. the expected fluxes and the deepest Ly-α observations available at z>5. *Blue dashed lines:* Expected fluxes for a range of SFR×$f_{eff}$, ranging from 10 to 0.01 $M_\odot$ yr$^{-1}$ (corresponding to L(Ly-α) from $10^{43}$ to $10^{40}$ erg s$^{-1}$) and taking into account the effective Ly-α transmissions ($f_{eff}$). *Green dash-dotted lines:* Expected Ly-α flux for star-forming galaxies, with their rest-frame UV continuum magnitudes ($f_{eff}$ = 1). *Filled circles:* Observations from Shimasaku et al. (2006), Kashikawa et al. (2006), and Ota et al. (2008) at z = 5.7, 6.5, and 7, respectively. C*rosses:* Intrinsic fluxes for lensed candidates from Stark et al. (2007). *Black spectrum:* Flux limit from deep Keck-NIRSPEC spectroscopy (Richard et al. 2008). An ELT-MOS will provide a sensitivity gain of a factor of ≥40, potentially extending up to z~13. For galaxies lensed by clusters, we will be able to measure IGM transmissions down to even fainter physical limits.

From the simulations of Dayal & Ferrara (2012), typical Ly-α transmissions are expected to be of the order of ~20% at z~7 and between 5-40% at z~8 for galaxies with $m_{UV}$ = −18 to −22 (i.e. $m_{AB}$ = 29 to 25). These models, accounting for clustering effects, predict a higher Ly-α transmission for brighter (more massive) sources. They are compatible with the current Ly-α fraction measured spectroscopically in LBGs at redshifts z~6-7, obtained from the latest VLT and Keck campaigns (e.g. Pentericci et al. 2011; Schenker et al. 2012). Measuring the Ly-α fluxes and equivalent widths of a large sample of galaxies (e.g. more than 50 galaxies with $m_{AB}$ = 28 down to fluxes of 2x10$^{-18}$ erg s$^{-1}$cm$^{-2}$) allows one to already distinguish different reionisation scenarios – e.g. patchy versus smooth reionisation – as shown by Treu et al. (2012). Fluxes of 10-20 times fainter will be reached by an ELT-MOS, meaning that more stringent tests of different reionisation models will be possible with the E-ELT. The corresponding source density, e.g. 2-5 arcmin$^{-2}$ mag$^{-1}$ at $m_{AB}$ = 29 for LBGs at z~8 (see Fig. 2) translates to 80-200 sources in the E-ELT patrol field. An ELT-MOS will ideally complement *JWST*-NIRSpec observations of the brightest objects (at moderate spectral resolution).

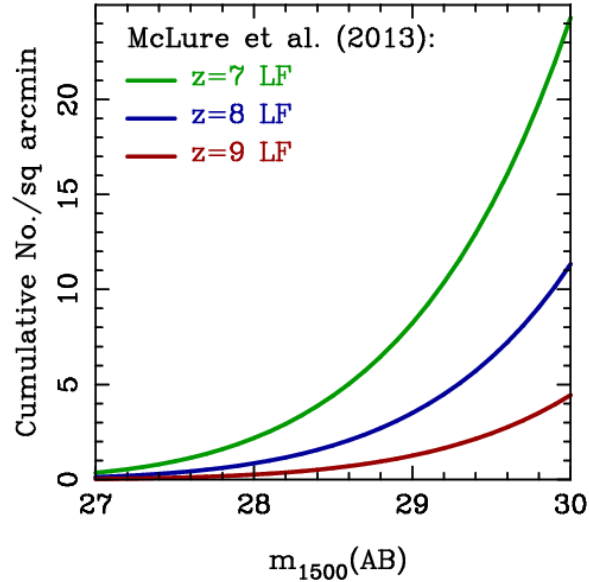

**Figure 2:** Integrated surface density of high-redshift star-forming galaxies per unit redshift interval at z = 7, 8, and 9, as a function of observed near-IR (rest-frame UV) magnitude, as derived from the galaxy luminosity functions from the UDF12 and CANDELS WFC3/IR *HST* programs (McLure et al. 2013).

For objects where Ly-α will remain elusive to detect (even with the E-ELT) other emission lines are also expected from these distant galaxies. Indeed, there are encouraging signs that any demise in observable Ly-α emission may be compensated by increasingly bright, high-ionisation UV lines, such as CIII] at $\lambda_{rest}$ = 1909Å. Support for this comes from observations of extreme [OIII] emission in some low-luminosity, (likely) low-metallicity, star-forming galaxies at z~2 (van der Wel et al. 2011), as well as indications from broad-band photometry that such strong high-ionisation emission may become increasingly prevalent at high-*z* (e.g. Labbé et al. 2013; Stark et al. 2013; de Barros et al. 2014) due to higher specific star-formation rates and/or to harder UV emission from massive stars with significantly subsolar metallicities.

The CIII] line is expected to have a rest-frame $EW_{CIII]}$ ~ 0.15 x $EW_{Ly-\alpha}$. Thus, at *z*~7.5, CIII] line fluxes of ~3x10$^{-18}$ and 3x10$^{-19}$ erg s$^{-1}$cm$^{-2}$ are expected from galaxies with $H_{AB}$ = 27.5 and 30.0 mag, respectively. If CIII] transpires to be a key emission-line for redshift determination at z>7, then the importance of following it out to z~10 (where there are now several LBG candidates – Zheng et al. 2012; Coe et al. 2013; Ellis et al. 2013) argues for near-IR spectroscopy in the *K*-band. Lastly, we note that the goal of detecting HeII ($\lambda_{rest}$ = 1640Å) remains vital in establishing when high-*z* galaxies start to be dominated by light from massive, extremely metal-poor ('Pop III') stars.

## 1.3. Probing the physical properties of the 'first-light' galaxies

For the brightest sub-population of these first galaxies, an ELT-MOS observing programme will also determine the physical properties of their ISM, stellar populations and outflows (if present). In addition to detecting emission-lines in such distant galaxies, an ELT-MOS will have a sufficiently large patrol field to follow-up the near-IR imaging surveys that will be used to identify LBGs within the reionisation epoch.

Observations extending into the *H*-band will enable us to potentially detect the Lyman-break ($\lambda_{rest}$ = 912Å) out to z~17, and rest-frame UV absorption lines at $\lambda_{rest}$ = $\lambda\lambda$1200-2000Å out to z~10. The latter range includes prominent features from Ly-$\alpha$ $\lambda$1215, NV $\lambda$1240, SiII $\lambda$1260, OI $\lambda$1303, SiIV $\lambda\lambda$1393,1402, CIV $\lambda$1549, HeII $\lambda$1640, and CIII] $\lambda$1908. These may permit redshift determinations if Ly-$\alpha$ is completely absorbed, while also providing important information on the ISM, outflows, and metal enrichment.

Low-resolution spectroscopy can be sufficient to derive redshifts, but moderate resolution (*R*~5,000) is required to study the detailed physics of galaxies, as derived from the line strengths and profiles of the ISM lines (such as SiIV, OI and CIV) and stellar absorption lines (NV, CIV, and SiIV, with some ISM contribution in these). Furthermore, the identification of significant velocity offsets between UV ISM lines, Ly-$\alpha$, and photospheric lines in galaxies at z~3 led to the discovery that a large fraction of star-forming galaxies at z~3 are driving strong winds – 'superwinds' – a feedback process thought to be the dominant mechanism which expels baryons from galaxies at these early times (e.g., Wilman et al. 2005). The likely shallower potential wells of typical galaxies at z>7 suggests that outflows from high-z galaxies could have cleared material from the galaxies and their surroundings, allowing ionising photons to escape, as well as enriching the IGM with metals. This will shed light on the problem of how the IGM was reionised and enriched.

Rest-frame UV spectroscopy of this type was pioneered in the 1990s using the Keck telescope to observe galaxies at z~3 (Shapley et al. 2003) and progressively extended up to z~4-5 using Keck and the VLT (Vanzella et al. 2005, 2009; Douglas et al. 2010; Jones et al. 2012). This will likely continue up to z~6 over the coming decade with new instruments such as VLT-KMOS and Keck-MOSFIRE but continuum and absorption-line studies in the UV at z>7 will remain out of reach for the current generation of 8-m class telescopes and of the JWST. In other words, this rest-frame UV continuum work can only be done efficiently with a wide-field, multi-object capability on an ELT.

Large samples over sizeable volumes are essential to avoid cosmic variance, and to test the large-scale structure of galaxies and the inhomogeneity of the IGM as the Universe is reionised. Thus we need:

- A multi-object capability to mitigate the relatively long integration times necessary to detect continuum and absorption lines in these objects (down to $H_{AB}$~28);
- An IFU capability to resolve their clumpy internal structures (e.g., Elmegreen et al. 2007);
- Good image quality, because the z>7 sources are compact (with half-light radii <0.15 arcsec, Grazian et al. 2012; Ono et al. 2013).

Interestingly, the exploitation of strong lensing by massive galaxy clusters to push instruments to the faintest physical limits is now well established (e.g., Smail et al. 1997; Richard et al. 2006, 2008; Stark et al. 2007). Gravitational lensing can provide access to significantly fainter objects (~2-3 mag below the normal detection limits, although their number per field will be relatively small). For example, with a Ly-$\alpha$ sensitivity of ~$10^{-19}$ erg s$^{-1}$ cm$^{-2}$, an ELT-MOS will be able to determine an effective Ly-$\alpha$ transmission down to ~0.1 out to z~9 for objects with $m_{AB}$(UV) = 32, and magnification by ~2 mag. In addition to the magnification, the image of the background galaxy is stretched, so it is possible to study the spatially-resolved ISM physics in detail, which is impossible to do without lensing (e.g., Swinbank et al. 2004; Combes et al. 2012).

The ultimate objective is spectroscopy of >1000 galaxies at the largest redshifts, to enable robust description of their luminosity function. A further goal is to understand the co-evolution of galaxies and AGN in the formation and subsequent growth of the first galaxies; in particular, the link of galaxy and black hole growth in the first Gyr of the Universe, when the first stars were formed and when the earliest galaxies were assembled. The AGN luminosity function beyond the extreme bright end probed by the SDSS (e.g. Fan et al. 2002) is unknown, and an ELT-MOS will enable measurements of the space density and evolution of AGN at z>6, to compare the growth of stellar mass and black holes, to investigate the metallicity evolution of AGN-hosting galaxies (e.g. Nagao et al. 2006) and to investigate the spectral properties of galaxies hosting AGN and those without (e.g. Heckman & Kauffmann, 2006).

Finally, spatially-resolved spectroscopy of the brightest sub-sample of the first galaxies will enable us to probe the geometry and sizes of the outflows and infer the mass-outflow rates (e.g., Förster-Schreiber et al. 2014). Although the high-z galaxies are small, their faintness means that diffraction-limited spectroscopy is not required; the basic requirement is to reach sufficient angular resolution to concentrate the light into 50-100 mas scales with ~10 angular resolution elements across the targets. Such sampling will allow us to probe the spatially-resolved stellar populations within each galaxy, and between multiple components within the same system. Many of the LBGs detected at z=5–6 show multiple components or companions of 1-2 arcsec scales in projection (e.g., Conselice & Arnold 2009). If this trend continues to higher-z (as expected in a hierarchical mass-assembly model, and also based on results on lensed objects, e.g., Swinbank et al. 2009), then spatially resolving the velocity offsets and stellar populations within multiple components/merging systems provides an important way of constraining the dynamical masses and stellar mass-to-light ratios of these galaxies (e.g., Förster-Schreiber et al. 2009).

## 1.4. Finding & following-up the targets

Current and future near-IR facilities (Keck-MOSFIRE, VLT-KMOS, VLT-MOONS) will be used in the coming decade to explore targets beyond z~7. However, if the observed decrease in Ly-$\alpha$ emission is confirmed, such work will become increasingly hard with 8-10m class telescopes. Moreover, the latest *HST*/WFC3 observations are already revealing sources that are clearly beyond the capabilities of current ground-based facilities. Thus, most of the current and future candidates – probably in the hundreds – are likely to remain of interest when the E-ELT arrives. Ahead of this, work will be required to extend the existing ultradeep imaging to a larger number of fields, e.g., data taken for the *HST* Frontiers Fields (near gravitational lenses).

We forsee at least three future sources of high-z candidates for ELT spectroscopy (Table 1), namely the *JWST* and *Euclid* missions, and MICADO ('ELT-CAM'). Other new projects will also likely emerge such as, for example, the *Wide-field Imaging Surveyor for High-redshift* (*WISH*) concept under study for JAXA, which aims to reach about 1.5 mag deeper than the *Euclid* deep survey in the near-IR.

### *JWST*-NIRCam

Near-IR imaging from the *JWST* should deliver 10σ sensitivities of 11.2 nJy in 10,000s (over a 2.2 × 4.4 arcmin field), i.e. $m_{AB}$ = 28.7; this translates to a 5σ sensitivity of $m_{AB}$ = 29 in one hour. The expected number of sources in the notional ELT-MOS patrol field (40 arcmin$^2$) is summarised in Table 1. Imaging from the *JWST* will clearly provide more than enough targets for MOS observations of high-z galaxies, but there are two strategic aspects to keep in mind:

- There is always an associated risk with a major space mission, e.g. if the *JWST* were to suffer problems, or if the telescope/instrument did not deliver the expected sensitivity.

- The synergy between *JWST* and the TMT is an important aspect of the US Decadal Plan, so the European community must also be ready to secure follow-up of *JWST* targets.

### *Euclid*

The wide-area survey of *Euclid* will observe ~15,000 deg$^2$, reaching (5σ) point-source sensitivities of 24.5 mag with the VIS instrument (a very broad filter in the visible) and 24.0 mag in the *Y*, *J*, and *H* bands with the NISP instrument. A deep survey reaching two magnitudes fainter is planned over an area of 40 deg$^2$; assuming a survey depth of $J_{AB}$ and $H_{AB}$ = 26.0. The number of sources per ELT-MOS field for different redshifts are summarised in Table 1, from which it is clear that *Euclid* alone does not provide large samples of high-z galaxies for ELT follow-up; only a survey at the depth envisaged for the *WISH* mission starts to provide meaningful samples of distant targets.

### MICADO

The Phase A design of MICADO had a relatively small field-of-view (~0.8 arcmin$^2$) with a fine pixel scale of 3 milliarcseconds. Nonetheless, its imaging sensitivities are impressive, potentially reaching $H_{AB}$ = 29.9 in one hour (5σ). This results in a large enough number of high-*z* sources to provide targets for the potential ELT-MOS patrol field (see Table 1). With these figures, MICADO will not be dramatically slower than *JWST* (reaching an equivalent depth in one fifth of the time, but with a factor of 12.4 smaller in area).

An additional factor is that *JWST*-NIRCam can observe in two bands simultaneously, boosting its survey speed. Thus, in 1hr, *JWST* will be able to survey a 9.68 arcmin$^2$ field in two bands, down to $H_{AB}$ = 29. In contrast, MICADO will survey a 4.32 arcmin$^2$ in a single band, and will take 4.5 hrs to observe the same area as *JWST* in two bands. That said, given the cost of *JWST* observing time and its nominal 5 year mission lifetime, MICADO observations should provide a competitive and attractive alternative.

In summary, MICADO observations should provide sufficiently rich samples of high-z targets which will then require spectroscopic follow-up for characterisation – the ELT-MOS will be well suited to this task. A wide-field imager as part of an ELT-MOS could be an interesting capability (depending on the image quality achievable), but it is not essential in the early years of the observatory. Both *JWST* and MICADO will require large amounts of observing time to survey large areas of the sky to properly sample the bright end of the luminosity function (e.g. a ~1 deg$^2$ field *a la* COSMOS). This leaves an interesting future niche between *Euclid* (not deep enough) and the *JWST*/ELTs, which a *WISH*-like mission could profitably fill. A prerequisite for such a mission/instrument is that it would need to be cooled, in order to survey longwards of the *H*-band and confidently detect the very high-z tail of the galaxy population.

Table 1: Number of detected sources per 40 arcmin$^2$ ELT-MOS field, as a function of redshift, for the *Euclid* deep survey, the potential *WISH* survey, and for 1 hr exposures with *JWST*-NIRCam and MICADO. Quoted depths are 5σ point-source sensitivities; *JWST* and MICADO observations are for 1 hr integrations. Note the expected *Euclid* densities in the *J*- and *H*-bands are identical except for the last bin (as the Ly-break enters the *J*-band at *z*=8). These numbers were derived from the UV luminosity functions of Bouwens et al. (2007; at z=3.8, 5, 5.9) and McLure et al. (2013; at z=7, 8, 9).

|  |  | $5 < z \leq 6$ | $6 < z \leq 7$ | $7 < z \leq 8$ | $8 < z \leq 9$ |
|---|---|---|---|---|---|
| **EUCLID DEEP** | $J_{AB} \leq 26$ | 13.68 | 3.14 | 0.45 | 0.16 |
| **EUCLID DEEP** | $H_{AB} \leq 26$ | 13.68 | 3.14 | 0.45 | 0.21 |
| **WISH DEEP** | $H_{AB} \leq 27.4$ | 127 | 60 | 21 | 12 |
| **JWST NIR-CAM** | $H_{AB} \leq 29.0$ | 640 | 439 | 225 | 148 |
| **MICADO** | $H_{AB} \leq 29.9$ | 1371 | 1038 | 640 | 452 |

### Other synergies

High resolution (δz<~0.001) and high S/N Ly-α observations, coupled with deep ALMA observations of the [CII] 158μm emission line and far-IR continuum, will allow us to address a number of interesting issues. In particular, ALMA observations will allow us to determine the systemic redshift of the galaxies with unprecedented accuracy. Once precise galaxy redshifts are known, the asymmetric Ly-α profiles can be accurately modelled in terms of IGM absorption and outflows (e.g., Dayal et al. 2011), thus tightly constraining the neutral-hydrogen fraction (without being affected by uncertainties associated with giant HII regions surrounding luminous quasars at similar redshifts). Further into the future, one can envisage programmes with the Square Kilometre Array (SKA) to study the HI 21 cm gas in galaxies with E-ELT and ALMA data, and combinations of these with future X-ray missions (e.g., *ATHENA+*).

## 1.5. Requirements

### Spectral Resolution

*R*>3,000 is required to target emission lines between the OH sky lines (e.g., Navarro et al. 2010; Villanueva et al. 2012). Resolving Ly-α profiles with typical FWHM ~150-500 km/s (Tapken et al. 2007,

Kashikawa et al. 2006) requires at least $R$=4,000 to marginally sample the narrowest lines, with the optimal requirement being $R$=5,000 to separate the effect of the IGM transmission (i.e. the ionisation state of the IGM) from radiation transfer effects in the host galaxy and its close environment, such as galactic winds, known to alter the Ly-α profile (e.g., Santos 2004; Verhamme et al. 2006).

**Multiplex/Observing modes**

Current Ly-α observations reach depths of ~3-10×$10^{-18}$ erg $s^{-1}$ $cm^{-2}$ (e.g., Penterici et al. 2011; Schenker et al. 2012). Ly-α is expected to have fluxes as faint as ~$10^{-18}$ erg $s^{-1}$ $cm^{-2}$ at $z$~7.5 in galaxies with $H_{AB}$ = 30 mag, but if the possible decrease of Ly-α emission at $z$>7 is confirmed (Pentericci et al. 2011; Ono et al. 2012), the required flux depth for $m_{AB}$ = 28-30 sources would be closer to ~$10^{-19}$ erg $s^{-1}$ $cm^{-2}$. An ELT-MOS should be able to obtain a S/N~10 (per resolution element) for such fluxes in reasonable integration times (a few tens of hours) for targets down to $H_{AB}$ ~ 29 (see below).

Considering current observations with the *HST*/WFC3 (e.g., Bunker et al. 2010; Bouwens et al. 2011; Wilkins et al. 2011; Lorenzoni et al. 2011, 2013; Grazian et al. 2012, Oesch et al. 2012; Ellis et al. 2013; McLure et al. 2013) and integrating the sources with $H_{AB}$ ≤ 30 over 6.5 ≤ z ≤ 9.5, the minimum surface density of LBGs is found to be of order ~10 $arcmin^{-2}$ (Fig. 2); i.e., in this redshift range the number of potential targets for emission-line detection within a 7(10) arcmin diameter patrol field is at least ~300(800) galaxies. At $R$>3,000, more than half of the *J*- and *H*-bands are free of OH sky lines (e.g., Navarro et al. 2010; Villanueva et al. 2012), which lead to potential target densities of at least ~150 (400) galaxies in emission per patrol field (single-line detection). Note that the detection of such faint objects well below the sky continuum background will require techniques such as cross-beam switching (see Yang et al. 2013) for high-multiplex, integrated-light observations. Thus, the number of target pick-offs to be deployed will need to be twice the number of science targets (i.e. the physical multiplex for the instrument design is double the scientifically-motivated multiplex, as in plans for VLT-MOONS).

Detections in the continuum or interstellar absorption-lines will be limited to the brightest objects. Simulated observations suggest that such a limit is $m_{AB}$~27 (and perhaps a magnitude deeper for 'Large Programme'-like deep observations, see below). The surface density at the faintest magnitudes is an order-of-magnitude smaller, which leads to ~15-40 targets in the E-ELT patrol field. Taking into account the spectroscopic success rate of metal lines in the rest-frame UV and extrapolating number counts from the *HST*, one indeed finds 40 sources at $z$~7 down to $J_{AB}$~28 and 20 galaxies down to $H_{AB}$~28 at $z$~8 (Evans et al. 2010). For such faint sources, very sensitive IFU observations are desirable, which calls for an AO-fed multi-IFU observing mode on the ELT-MOS. Note that both AO-corrected IFU and integrated-light (i.e. GLAO) observations will advance this field; via excellent sensitivity in the case of the former, and a large multiplex in the latter case. There is no specific requirement for simultaneous use of these modes, although we discuss the potential merits of this later in the White Paper.

**IFU spatial sampling**

All imaging observations of very high-$z$ galaxies undertaken to date suggest that early galaxies are extremely compact, with half-light radii <0.15 arcsec (Oesch et al. 2010; Grazian et al. 2012; Bouwens et al. 2012; Ono et al. 2013), resulting in full diameters <0.6 arcsec. However, these measurements are all based on *HST* imaging, which is inevitably most sensitive to high surface-brightness peaks in emission.

Most galaxies at z>2 indeed appear clumpy (see also SC3). Such clumps have typical sizes of ~1 kpc, which translates into ~180 mas above z~6. If one wants to resolve such internal structures, an IFU spatial sampling of 40 mas provide good sensitivity, while providing spatially-resolve kinematics and other properties of the clumps; spatial sampling above ~90 mas would prevent resolving the internal properties of these structures. The optimal IFU spaxel scale results from a compromise between spatial resolution and surface brightness sensitivity (see simulations in Figs 3, 4, 5, and 6).

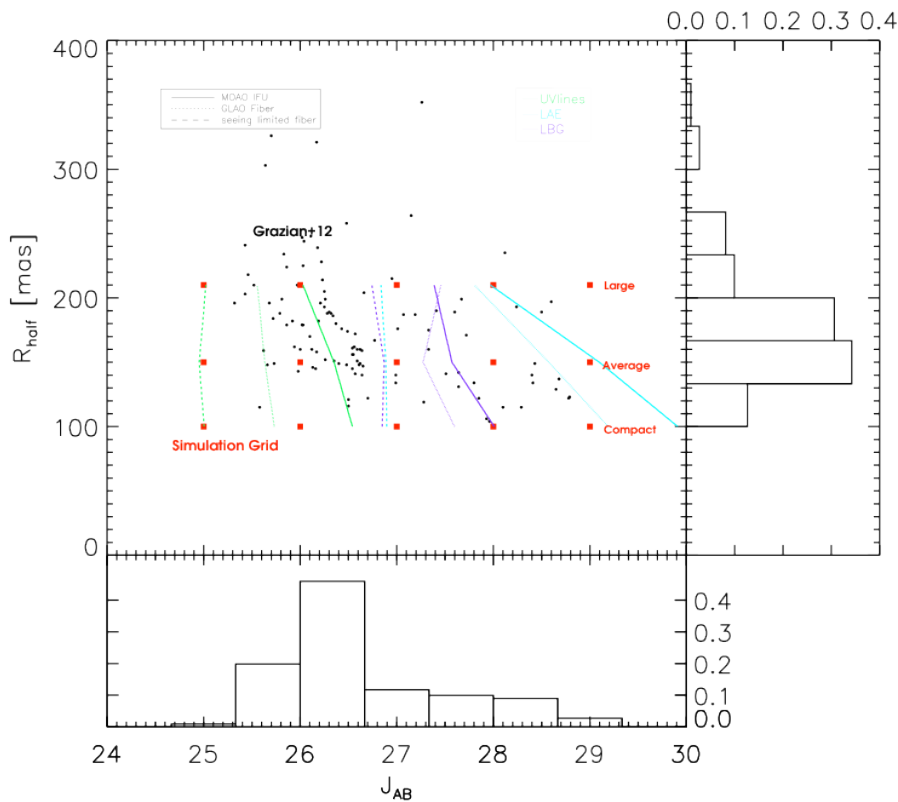

**Figure 3**: Observed magnitudes ($J_{AB}$) vs. radii ($R_{half}$) of candidate Lyman-break galaxies at z~7 using data from Grazian et al. (2012). The grid of simulated observations from Disseau et al. (2014) is shown by the red squares. The vertical coloured lines indicate the limits for which S/N = 10 is reached for the optimal spatial-pixel size or aperture size (depending on whether the observations are for multi-object AO or seeing-limited/GLAO, respectively); these are from integrations of 40 hrs for the UV absorption lines and for 10 hrs for the LAEs/LBGs.

### Expected sensitivity

We simulated spatially-resolved IFU observations of clumpy galaxies at z~7 and z~9 with $t_{int}$ = 10 and 40 hr in dark-time conditions, and MOAO corrections with ensquared energy of 30 % in 80 x 80 mas (Disseau et al. 2014). The morphology and kinematics were set-up using hydrodynamical simulations of rotating, clumpy galaxies from Bournaud et al. (2008), which were rescaled to the expected sizes and fluxes for z=7, to encompass the range of observed magnitudes and sizes (see Fig. 3): 'compact' ($R_{half}$ ~100mas), 'average' ($R_{half}$ ~150mas), and 'large' ($R_{half}$ ~220 mas). UV interstellar lines were fit using the Shapley et al. (2003) LBG empirical template, while the Ly-α emission line was fit as a Gaussian truncated on the blue side with equivalent widths following measurements at z~7 from Jiang et al.

(2013). The Ly-break was simulated assuming zero flux bluewards of Ly-$\alpha$. An integrated spectrum was constructed from each simulated datacube by summing the individual spectra after ordering them by decreasing S/N, and continuing until the maximum S/N was achieved (e.g., Rosales-Ortega et al. 2012).

- UV interstellar lines (Fig. 4): S/N~3-5 (per spectral/spatial element of sampling) can be reached $J_{AB}$~27 with integrations of 40 hr, depending on the IFU sampling and target size (provided the IFU spaxel scale is in the range ~60-140 mas). Limiting magnitudes for S/N~10 (per spectral resolution element) are $J_{AB}$~26-27, depending on the IFU sampling scale and the AO correction. $J_{AB}$ = 28 should be reachable in a 'large programme' with larger integrations (i.e. hundreds of hours) and/or using the stacking technique.

- Lyman Break Galaxies (LBGs, Fig. 5): S/N~5 (per spectral/spatial element of sampling) can be reached for compact targets with $J_{AB}$~28 with integrations of 10 hr, depending on the IFU sampling and target size. Limiting magnitudes for S/N~10 (per spectral resolution element) are $J_{AB}$ ~27-28, depending on the IFU sampling scale and AO correction.

- Lyman Alpha Emitters (LAEs, Fig. 6): S/N~7-13 (per spectral/spatial element of sampling) can be reached for $J_{AB}$~29 with integrations of 10 hr, depending on the IFU sampling and target size. Limiting magnitudes for S/N ~10 (per spectral resolution element) are $J_{AB}$ ~29-30, depending on the IFU sampling scale and AO correction.

Note that in the case of lensed systems, the above limits can likely be pushed deeper by ~2 magnitudes.

**Integrated-light aperture size**

GLAO or seeing-limited observations will provide PSF FWHMs in the range 0.2-0.55 arcsec at 1.0 μm, and 0.15-0.45 arcsec at 1.65 μm (as inferred using simulated ATLAS/GLAO PSFs, calculated for the Phase A studies, which assumed 0.81 arcsec integrated seeing). The S/N for point-like sources is maximized for aperture diameters ~1.45 FWHM. Compromising between the extremes of the spectrum, it should be possible to use a single 0.45 arcsec diameter aperture with losses of < 20% in S/N over the full spectral range considered. Assuming a ±0.13 arcsec positioning uncertainty (from the EVE Phase A study), this leads to an optimal aperture of ~0.6 arcsec; the same compromise leads to an aperture of ~0.8 arcsec if pure seeing is considered (i.e., no GLAO correction). These estimates are confirmed by end-to-end simulations as a function of target size and magnitude (see Figs 4 to 6). Further simulations are on-going to determine the optimal aperture for integrated-light spectroscopy (as a function of target size and magnitude).

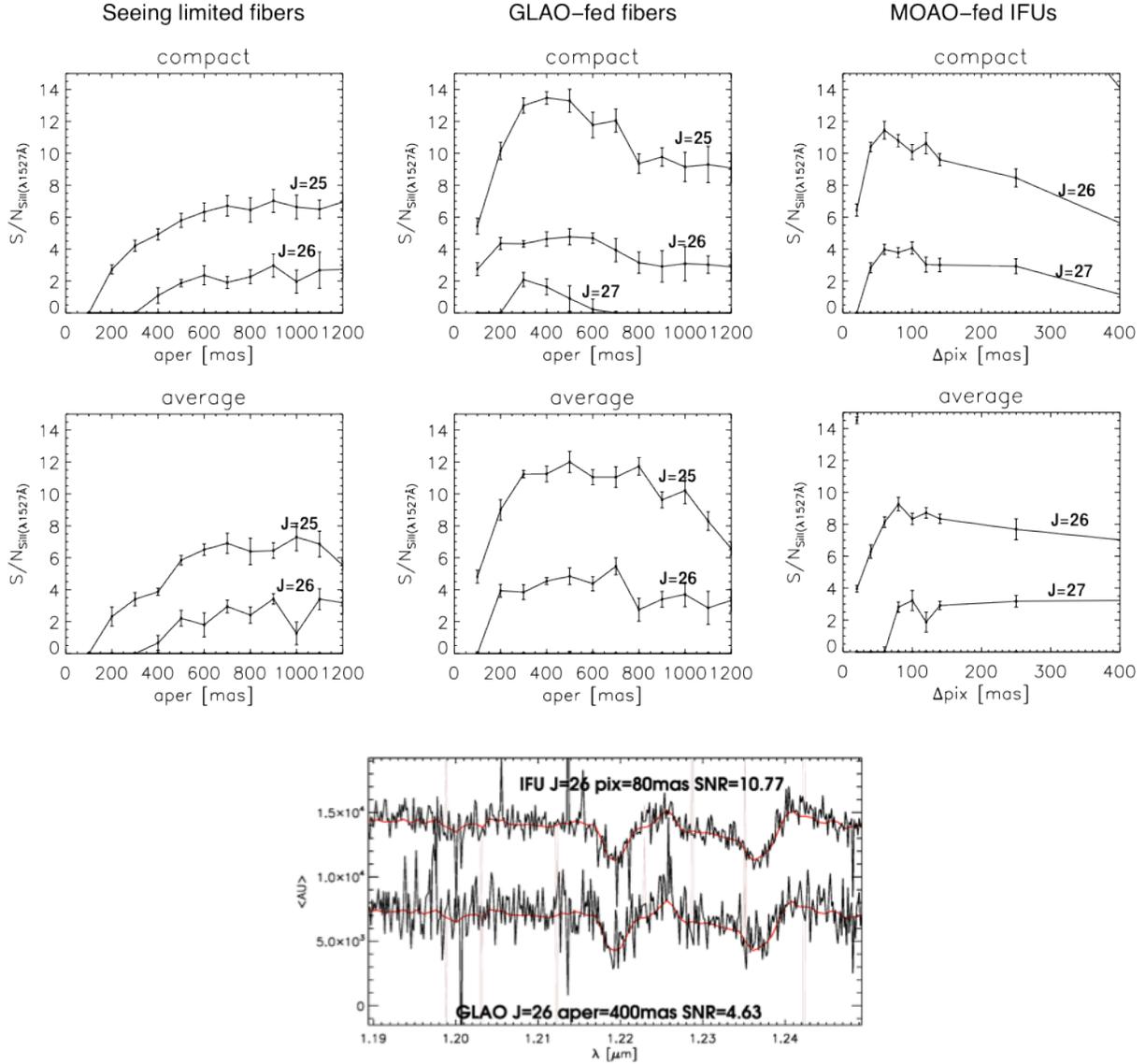

**Figure 4:** Simulated observations of UV interstellar absorption lines at $z \sim 7$ with $t_{int}$ = 40 hrs on source (Disseau et al. 2014). The two upper rows show the S/N ratios for the SiII λ1527 line from integrated spectra of compact ($R_{half}$ = 100 mas) and average ($R_{half}$ = 150 mas) sources as a function of aperture size for high-multiplex (seeing-limited/GLAO) and high-definition (MOAO) cases. Example spectra (black) and fits (red) for $J_{AB}$ = 26 mag are shown in the lower panel.

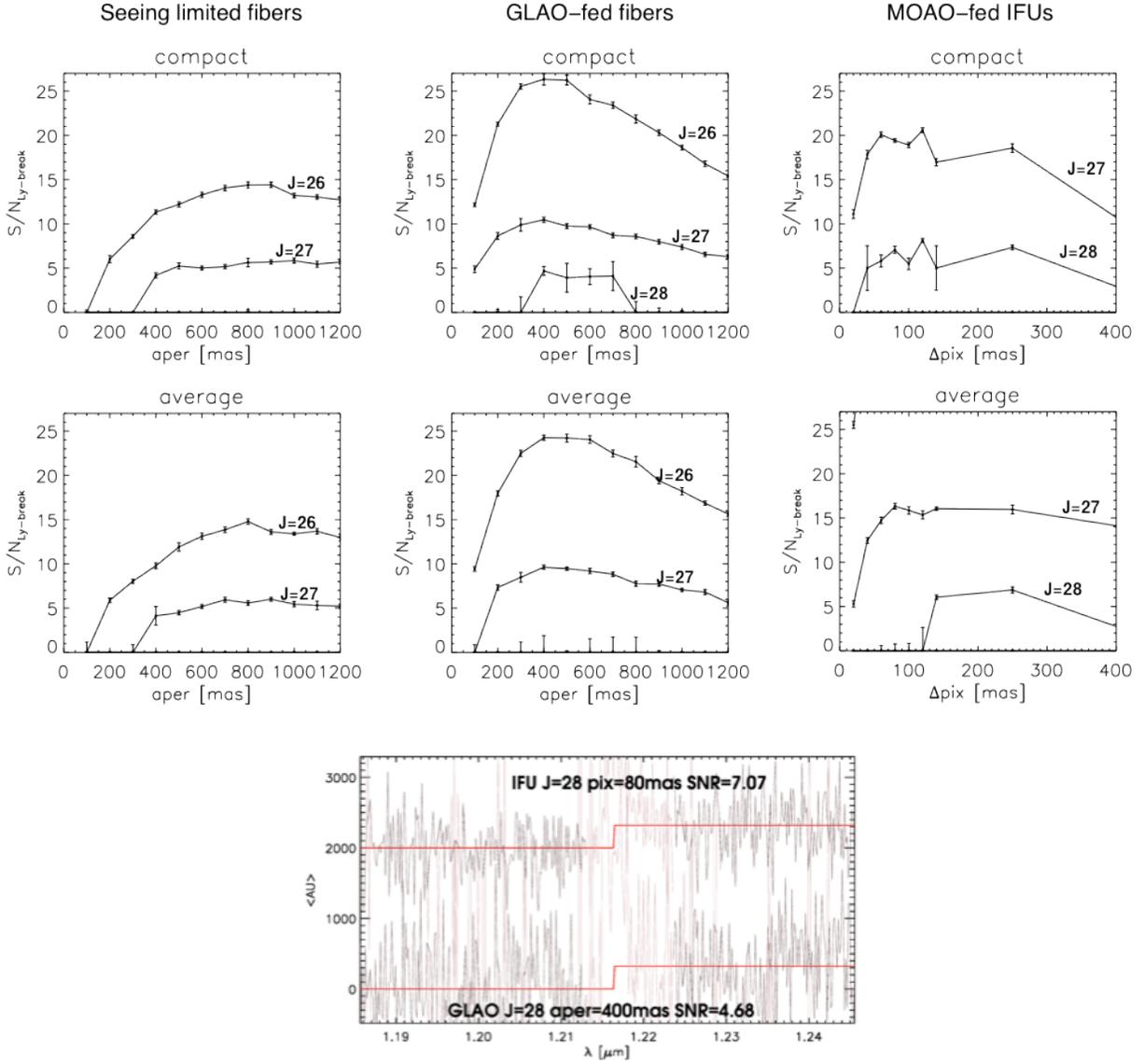

**Figure 5**: Simulated observations of Lyman-break galaxies at $z\sim9$ with $t_{int}$ = 10 hrs on source (Disseau et al. 2014). The two upper rows show the S/N ratios for the Lyman break ($\lambda912$) from integrated spectra of compact ($R_{half}$ = 100 mas) and average ($R_{half}$ = 150 mas) sources as a function of aperture size for high-multiplex (seeing-limited/GLAO) and high-definition (MOAO) cases. Example spectra (black) and fits (solid red line) for $J_{AB}$ = 28 mag are shown in the lower panel.

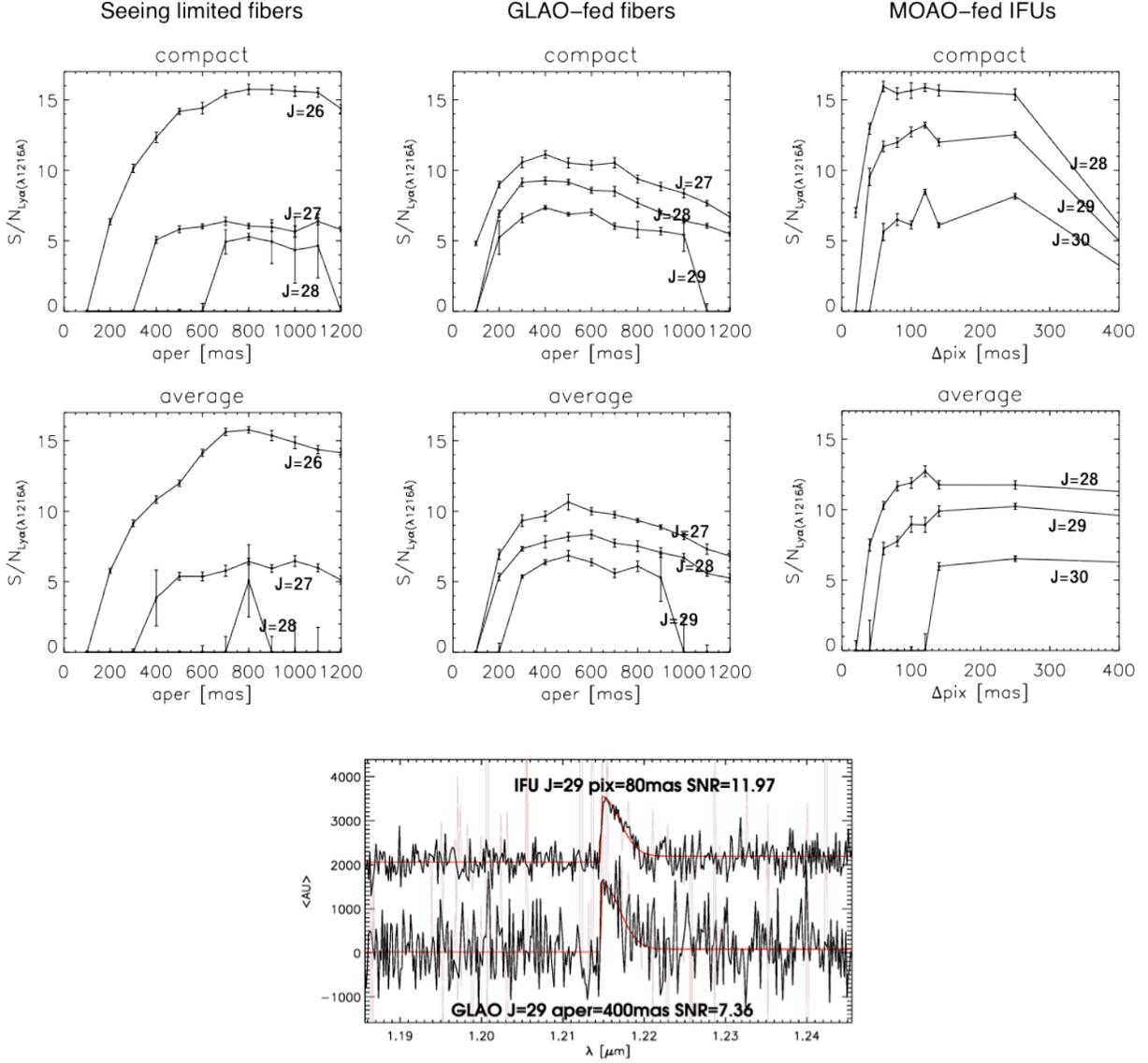

**Figure 6**: Simulated observations of Lyman-α emitters at z~9 with $t_{int}$ = 10 hrs on source (Disseau et al. 2014). The two upper rows show the S/N ratios for the Ly-α line (λ1216) from integrated spectra of compact ($R_{half}$ = 100 mas) and average ($R_{half}$ = 150 mas) sources as a function of aperture size for high-multiplex (seeing-limited/GLAO) and high-definition (MOAO) cases. Example spectra (black) and fits (red) for $J_{AB}$ = 29 mag are shown in the lower panel. In the simulations, the Ly-α equivalent width was assumed to be 10Å up to $J_{AB}$ = 27, then 20Å to $J_{AB}$ = 28, 40Å to $J_{AB}$ = 29, and 70Å to $J_{AB}$ = 30 (following Jiang et al. 2013 at lower redshifts).

## 1.6. Comparison with other facilities

In Table 2 we compare the capabilities of other potential future instruments/facilities with an ELT-MOS to provide the observations required to satisfy SC1 (taking into account spectral resolution, spatial resolution, spectral range, field-of-view/patrol field, sensitivity, and multiplex).

Table 2: Comparison of different instruments/facilities for observations toward SC1. Green cells indicate compliance, while red cells indicate clear obstacles/drawbacks. Orange cells indicate those where part of the scientific objectives may be possible, or that are non-optimal in terms of survey speed.

|  | ELT-MOS | ELT-IFU | ELT-HIRES | JWST-NIRSpec | TMT/IRIS | TMT/IRMS | TMT/IRMOS |
|---|---|---|---|---|---|---|---|
| **DETECTION OF REIONISATION SOURCES** |  | No multiplex | Too low Sensitivity (high $R$) | Limited sensitivity & patrol field | No multiplex | Limited multiplex | Limited multiplex |
| **PHYSICAL PROPERTIES** |  | No multiplex | Too low Sensitivity (high $R$) | No multiplex | No multiplex | no IFU | Limited patrol field |

# EVOLUTION OF LARGE-SCALE STRUCTURES

## 2.1. Tomography of the IGM

The gas in the IGM is revealed by the numerous hydrogen absorption lines that are seen in the spectra of quasars bluewards of the Ly-α emission from the quasar. It has been shown that the high-z IGM contains most of the baryons in the Universe and is therefore the baryonic reservoir for galaxy formation. In turn, galaxies emit ionising photons and expel metals and energy through powerful winds that determine the physical state of the gas in the IGM. This interplay of galaxies and gas is central to the field of galaxy formation. It is complex (see, e.g., the simulations in Fig. 7) and happens on scales of the order of 1 Mpc (~2 arcmin on the sky at z~2.5, using standard cosmological parameters). The main goal of this case is to reconstruct the 3D density field of the IGM at z~2.5 on these and larger scales to study its topology, its chemical properties, and to correlate the position of the galaxies with the density peaks.

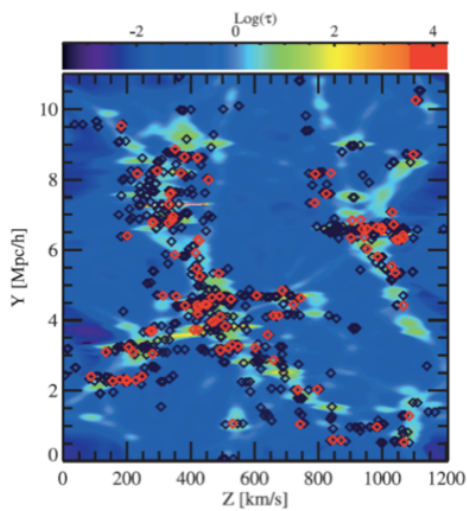

**Figure 7:** Simulations of the optical depth of neutral hydrogen compared to the location of galaxies (black and red diamonds, in which the latter have stellar masses of >$10^8$ $M_\odot$ (from Kollmeier et al. 2006).

The Ly-α forest arises from moderate density fluctuations in a warm photo-ionised IGM. The spatial distribution of the IGM is related to the distribution of dark matter and the full density field can be reconstructed using a grid of lines-of-sight (Pichon et al. 2001; Caucci et al. 2008). First results using distant star-forming galaxies to do this (rather than quasars) were recently presented by Lee et al. (2014). Fig. 8 shows the level to which the matter distribution can be recovered with 100 parallel sight-lines drawn through a 50×50×50 Mpc N-body box for which synthetic spectra were then generated. From analysis of the simulated data, it was possible to recover structures over scales of the order of the mean separation of the sight-lines. To reach S/N ≥ 8 at $R$ = 24.8 (considering the background sources as point-sources), 8-10 hr exposure would be required per field, giving an ambitious ~750 hrs for 1 deg². In support of such a survey it would also be desirable to obtain spectra of all galaxies brighter than $R_{AB}$ ~

25.5 in the same region of the sky, to determine their redshifts, identify AGN (by looking for UV emission lines), starburst systems, and to correlate all their properties with those of the IGM.

### 2.1.1. Instrument Requirements

**Multiplex**

About 900 randomly distributed targets per deg$^2$ are required to recover the matter distribution, with a spatial sampling of 0.5–2 arcmin at z~2. At z~2.5 there are ~900 LBGs per deg$^2$ (at r < 24.5) and only ~50 QSOs (at *r* < 22.5; fainter than this and the continuum emission from QSOs is contaminated by the host galaxies and so they are effectively subsumed into the LBG population). These beacons can be used to map the IGM structure between z~2.2 and 2.4 (at lower z, the response function of the spectrograph/telescope will drop dramatically as the lines move into the far blue).

Once the density field is recovered, topological tests can be applied to recover the true characteristics of the density field. The ~900 potential targets per deg$^2$ require a multiplex gain of roughly 10–15 per 40 arcmin$^2$ field, although a higher multiplex would be highly desirable to target additional galaxies in the field. In fact, the large amount of time required for IGM tomography makes it advisable to couple these observations with another project to maximize scientific return. For example, 3D spectroscopy of background (and other) galaxies would allow a more complete investigation of the structure of LBGs and interactions with their environment.

The spatial scale of the structures sets the desired survey size and source density: the total survey size correlates with the spatial scale of interest, which anti-correlates with the source density. Given this, the most promising project to be achieved with a telescope as powerful as the E-ELT is to map the high-z IGM at intermediate scales to reveal the network of filaments that connect galactic structures. The ultimate goal is to statistically associate galaxy distributions to filaments. For this, individual Ly-α absorption lines (of width about 100 km/s) should be measured.

**Wavelength coverage**

This case pushes the blueward extent of the necessary wavelength coverage, requiring observations of the Ly-α forest. Thus, coverage down to at least 0.4 μm (and with good throughput) is essential, with coverage to 0.37 μm desirable. In addition, observing the objects redwards of the Ly-α emission would allow us to normalise the continuum appropriately, study the LBGs themselves, the metals in the IGM (in particular CIV λ1550), and the quasars themselves if the IR is accessible. Thus, a continuous wavelength range of 0.37 to 0.60 μm is desirable.

**Spectral resolving power**

The optimal resolving power for such IGM tomography is set by a compromise between (1) the drop in S/N ratio per pixel as *R* increases (at fixed source magnitude and hence fixed density of sources), and (2) the drop in the ability to deblend lines as *R* decreases, especially when LBGs are used as background sources (e.g., confusion between interstellar absorption lines blueward of Ly-α and intergalactic Ly-α absorption). Between *R*~3,000 and 5,000 is an optimal trade-off for such a project (although we note that the SDSS project has used observations at *R*~2000, e.g., McDonald et al. 2006).

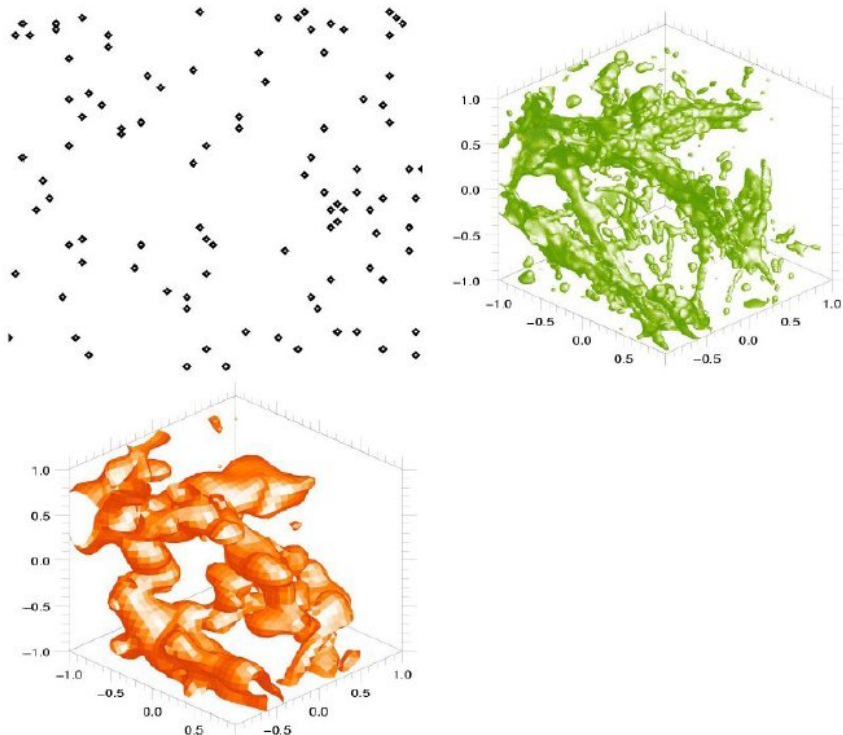

**Figure 8:** The input simulated density field is shown in the right-hand box (in green). One hundred (randomly-spaced) sight-lines to background sources are drawn through this box (upper-left). The corresponding spectra are used as the input data for the reconstruction, with the reconstructed density field shown in the lower box (in orange).

### Observing mode

As LBGs are not point sources it may be useful to use an IFU to capture the background source details.

## 2.2. The large-scale distribution of galaxies at early epochs z>2

Galaxies are thought to form through the cooling of baryonic gas within extended dark matter halos (White & Rees, 1978), which grow through subsequent mergers in a hierarchical fashion. Studies of large-scale structure provide a means to link galaxies with the underlying density field. More specifically, clustering studies allow a statistical measurement of the dark-matter halo mass in which galaxies reside (e.g. Mo & White 1996; Sheth & Tormen 1999). Thus, by studying the clustering of galaxies in the first Gyr ($z > 5$) through to the present day, we can relate the assembly of stellar mass to the build-up of dark matter halos, providing direct tests of models of galaxy formation and the evolution of large-scale structure. A major challenge is to connect the observable properties of galaxies to those of the dark-matter halos in which they are embedded. Not only do we want to understand the nature of the first galaxies, but also where they are formed (specifically, if this is in the densest peaks of the dark matter density field).

Spectroscopic surveys help us to reveal this large-scale structure of the Universe on Mpc scales. The distribution of galaxies over large scales at early epochs will enable us to measure the mass of dark matter halos and the number of dark matter satellites per halo, e.g., using halo occupation distribution models of the two-point correlation function. This will be compared/combined with estimates of dark matter from weak lensing studies. Once the relation between galaxies of various types/masses with their environment is established, we will be able to identify the contributions of nature vs. nurture on their evolution.

The spatial distribution of galaxies at $z<1.5$ is now well constrained from large surveys which have probed the dependence of galaxy clustering as a function of physical properties such as luminosity, colour, morphology, and stellar mass (Guzzo et al. 1997; Norberg et al. 2002; Zehavi et al. 2005; Le Fèvre et al. 2005; Coil et al. 2006; Li et al. 2006; Meneux et al. 2006, 2008, 2009; Pollo et al. 2006; de la Torre et al. 2010, etc). Studies using photometric redshifts have begun to probe the epoch at $z>2$ (e.g. Ouchi et al. 2004; Hartley et al. 2013), but spectroscopic studies are required to probe the full three-dimensional density field. For instance, a large (~50 Mpc) structure was identified at $z\sim5$ in the Subaru Deep Field (Shimasaku et al. 2004), but the size of the observed field was relatively small (25x45 arcmin), so subject to cosmic variance. Are such structures really common at that epoch? To date, we have very limited information on clustering at $z\sim3-5$, and then only for the most luminous systems.

In summary, current observations of high-$z$ galaxies are limited by small sky coverage (a few hundred sq. arcmin, so suffer from cosmic variance), small samples (a few hundred galaxies), and uncertainties in their distances (e.g., IR-selected studies have mainly used photometric redshifts), leading to large uncertainties in the deprojected correlation lengths. Moreover, the selection of these targets is (necessarily) biased, i.e. using colour-colour selction techniques from broad-band filters, which do not allow us to probe the bulk of the galaxy population, and so the majority of the stellar mass at $z>2$ is likely to be missed (van Dokkum et al. 2006). Estimates of galaxy clustering at $3<z<6$ are shown in Fig. 9, highlighting the uncertainties between the available data and the models. The next breakthrough requires a large spectroscopic survey of normal galaxies at $z>2$, to trace their large-scale structure, determine their dark-matter halo masses, and to relate these to physical characteristics such as their stellar masses, ages, sizes, and stellar populations.

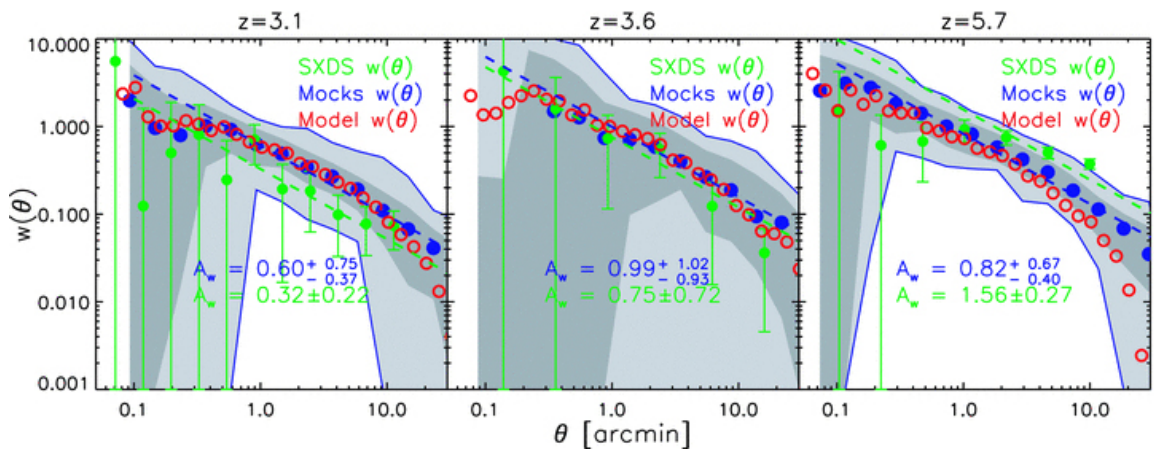

**Figure 9:** Angular correlations from SXDS mock catalogues (Orsi et al. 2008) cf. observations (green). Large uncertainties remain between observations and simulations, particularly for the poorly constrained data at small angular resolutions.

The scale of a potential survey depends on the source densities of Ly-α emitters over the range 3<z<6. Adopting theoretical models and apparent magnitudes from Le Delliou et al. (2005, 2006), emitters of $z_{AB}$ = 26 mag correspond to a Ly-α flux of a few $10^{-16}$ erg s$^{-1}$ cm$^{-2}$. This gives a source density above this flux of 1-10 Ly-α emitters per arcmin$^2$ with a typical angular size of 0.1 arcsec. If a MOS were able to observe hundreds of spectra simultaneously, a survey of 100 pointings, covering over a deg$^2$ would give tens of thousands of spectroscopic redshifts, spanning a volume of $10^7$/h$^3$ Mpc$^3$ between 3<z<6. Galaxies will be selected in the near-IR, providing an unbiased census (cf. past optical/UV selections) to study the growth of structure from an age of 1 Gyr to the present day (using the methods mentioned above). Determining the spatial distributions of galaxies will provide estimates of the typical masses of their host halos and how they are populated by the galaxies, along with estimates of $\sigma_8$, the galaxy/mass bias, and detections of the first clusters/groups present in the centre of large-scale structures. Directly linking dark matter halos to the build-up of stellar mass in the early Universe will provide a powerful test for galaxy formation models.

### 2.2.1. Instrument Requirements

**Wavelength coverage**

The diagnostic features span a broad range in wavelength, from 0.37 to 1.4 μm. Given the large source densities of targets, simultaneous coverage of this whole range is not required (as the targets can be split into sub-samples for observations).

**Spectral resolving power**

Low-dispersion of $R > 300$ is sufficient, although near-IR observations argue for $R > $ ~3,000 to adequately subtract the sky lines (and the spectra could later be rebinned).

**Multiplex**

Source densities of 1-10 per arcmin$^2$ argue for a multiplex of >400 (assuming a 40 arcmin$^2$ field).

## 2.3. Studies of galaxy clusters

Following on from the large-scale structures discussed above, we now highlight the contribution that an ELT-MOS will make to our understanding of galaxy clusters, which provide us with high-density regions in which we can probe the dynamical mechanisms governing galaxy formation and evolution.

### 2.3.1. High-redshift clusters

Rich samples of clusters and proto-clusters of galaxies have been discovered in recent years, and such discoveries can be expected to continue to grow in the coming decade. This is particularly relevant in the critical redshift range to study their assembly and the formation of their galaxy populations, at z > 1.5. These objects, selected as red or star-forming overdensities (e.g., Castellano et al. 2007, 2011; Kurk et al. 2009; Papovich et al. 2010; Tanaka et al. 2010; Andreon et al. 2009; Gobat et al. 2011; Santos et al. 2011; Stanford et al. 2012; Zeiman et al. 2012; Muzzin et al. 2013; Galametz et al. 2013; Mei et al. 2014), host the progenitors of the local cluster population, and their study is key to understanding both cluster assembly in the current cosmological model, and of the first formation/evolution phases of the early-type population.

From numerical simulations based on the concordance cosmological model, the assembly of a cluster happens by the merger of smaller halos to a main cluster. The scale at which the mass of the final clusters is distributed in progenitor halos at z~2-5 extends to 5-10 Mpc (Chiang et al. 2013; Shattow et al. 2013). Recent observations have identified overdensities of these halo progenitors (Chiang et al. 2014) but the details of their populations are beyond the reach of current spectroscopic facilities.

From IR spectroscopy (both from ground-based and space facilities), at least ~50% of cluster and proto-cluster galaxies appear to be undergoing star formation. However, some clusters at z~2 already show a well-defined red sequence, and a predominance of passive galaxies (Zeiman et al. 2012; Gobat et al. 2013; Newman et al. 2014). This variety of behaviour suggests that the early-type populations in the most massive clusters were already quenched at z>2, and that their progenitors are star-forming galaxies at z~2.5-5, where we start to detect proto-clusters with deep far-IR observations and line-emission surveys (e.g. Capak et al. 2011; Tadaki et al. 2014; Tan et al. 2014). A major contributor to this field in the coming years will be observations with *Euclid,* which are expected to detect and (spectroscopically) identify ~$10^7$ massive quiescent galaxies, including the rarest, most massive (>$4\times10^{11}$ $M_\odot$) systems at z>1.5. This will include candidate proto-clusters from identification of overdensities of early-type galaxy progenitors (which include star-forming galaxies at z~2 and are mostly star-forming galaxies at higher redshifts).

To follow-up the progenitors of early-type galaxies in primordial halos up to z~5, we need wide-field IR spectroscopy, to cover the spatial extent of the halos that will evolve to the final cluster seen in the local Universe; each of these halos will host a few up to tens of galaxies. Current scenarios of early-type galaxy formation propose both galaxy mergers and disk accretion. To understand which percentage of the early-type progenitors formed by these different modes, we will study their internal dynamics using the redshifted H$\alpha$, [OII] and [OIII] emission lines; the ratios of these will also permit us to identify AGN and to study the stellar populations in the galaxies.

### 2.3.2. Galaxy cluster dynamics

Moving closer to home, massive galaxy clusters contain an important diffuse luminous component, known as intracluster light (ICL) and comprised of stars external to galaxy halos in the intracluster environment (e.g. Arnaboldi et al. 2002, Burke et al. 2012, Guennou et al. 2012). The ICL can be used to obtain insights into the dynamical history of the clusters as stars in the external regions of galaxy halos are pushed into the intracluster environment by tidal stripping arising from galaxy interactions (e.g. Rudick et al. 2011).

Deep imaging is required to measure the elusive ICL, and recent observational programmes have been probing the spatial distribution of the ICL in clusters as a function of redshift. For example, a recent estimate of the spatial distribution of the ICL in cluster CL0024+17 (at *z* = 0.4, see Fig. 10) has been derived from deep *R*-band images obtained by the Large Binocular Camera at the LBT to derive hints on the dynamical history of the cluster. Accurate evaluation of the systematics introduced in the flat fielding and background subtraction procedures was required for these data, together with careful removal of the light from galaxy halos in the cluster centre. The average radial profile of the ICL fraction ($L_{ICL}/L_{galaxies}$) increases with decreasing radius, reaching a maximum of 40% at R~70-80 kpc. At smaller radii the profile bends, followed by a significant decrease, due to overlap of the halos of the brightest cluster galaxies.

Simple predictions in CDM scenarios show that the ratio of the total stellar mass lost by stripping compared to the stellar mass present in galaxies is almost independent of the physical properties of the generic (galaxy-hosting) stripping action. The ratio mainly depends on the global cluster properties such as the total mass and DM-profile shape. Adopting a mass of M = $1.7\times10^{15}$ for CL0024+17 and a NFW DM profile (with a concentration parameter, $c_{vir}$ = 9, derived from a lensing analysis of the cluster), the predicted ICL fraction profile broadly agrees with the observational trend.

If the ICL production is due to the tidal-stripping action, we should expect a loss of mass/luminosity and fewer faint dwarfs in the cluster core compared to the outer regions. Hints of this effect are seen in the bending of the faint end of the galaxy luminosity function approaching the cluster centre. Again taking the example of CL0024+17, the emissivity 'lost', as shown by the observed LF bending, appears in broad agreement with the measured integrated ICL fraction (23%) in the core. This suggests that the stripping activity, bending the galaxy counts at the faint end, can quantitatively explain the observed ICL fraction. In the example of CL0024+17, any measure of spectral features from the ICL and dwarf galaxies in the cluster would provide considerable insights into to the cluster dynamics. In this respect, an ELT-MOS will be essential to probe the tidal and merging dynamics in cluster cores up to *z* ~1.5.

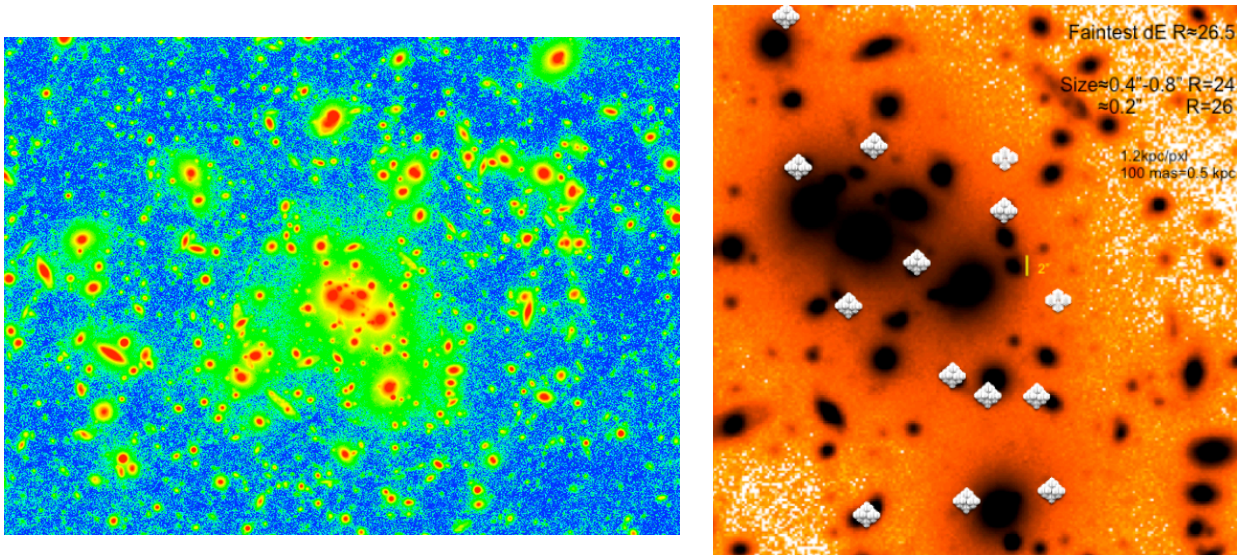

**Figure 10:** *Left:* Core 3×2.4 arcmin of the CL0024+17 cluster (*z* = 0.4; see Giallongo et al. 2014). *Right:* LBT *R*-band image of the central part of CL0024+17, with illustrative ELT-MOS targets to probe the dynamics of the intracluster light (ICL) in the cluster. The image was obtained in 0.7" seeing and the pixel scale is 1.2 kpc/pixel (~0.23").

### 2.3.3. Instrument requirements

**Multiplex & spatial resolution**

To give a sense of the scales involved, the characteristic cluster radius ($R_{200}$, Hansen et al. 2005) at z~1 is of order 7 arcmin in projection on the sky. In the specific example of a cluster at z~0.5, the lower-mass galaxies we want to observe (*R* ~24 to 26 mag) will be moving in the cluster core such that tidal forces will partially or fully remove their DM halos – this will destroy their stellar components at r < 0.33 $rv_{ir}$, which will spread out in the intracluster volume and contribute to the ICL. A multiplex of 50-100 within the E-ELT patrol field will enable the study of the velocity distribution of these dwarfs as a function of distance from the cluster centre. Integrated-light, GLAO observations (at spatial scales of ~200-400 mas) will be sufficient for this case. Observations at finer spatial resolution (~50 mas) would enable investigation of the internal halo dynamics of the intermediate-luminosity galaxies in the clusters (e.g., *R* = 23-24 mag, at *z* ~0.5); this would reveal any signs of tidal interactions when smaller dwarfs get through to the larger galaxy halos. A lower multiplex of ~15-20 would be sufficient for this case (with similar IFU sizes as discussed for SC3).

**Spectral resolving power**

*R* > ~3,000 is required to provide sufficient velocity resolution and to observe the relevant emission lines robustly between the sky lines.

**Wavelength coverage**

The primary diagnostic lines of interest at low- and intermediate-*z* are the strong Balmer absorption features. Thus, coverage of the 0.6-1.0 μm region in the optical is required (i.e. at the finer spatial scales of ~50 mas, correction in the *I*-band is required). To study the more distant galaxy population up to z~5, we need observations of the Hα, [OII] and [OIII] emission lines. Specifically, this will be enabled by coverage to the longward end of the *K*-band, enabling studies of Hα up to z~2.5, and observations of proto-clusters at up to z~5-6 (via [OII] emission). Identification of passive galaxies will require coverage of the (rest-frame) 4000Å break, and the main Fe and Mg absorption lines.

# MASS ASSEMBLY OF GALAXIES THROUGH COSMIC TIMES

## 3.1. Spatially-resolved spectroscopy of high-z emission line galaxies

The availability of IFUs on 8-10 m class telescopes has heralded a new era of galaxy studies. IFU spectroscopy is now routinely deriving the spatially-resolved kinematics of massive, emission-line, distant galaxies (with stellar masses ≳$10^{10}$ M$_\odot$), from z~0.4 to 3, see reviews by Glazebrook (2013) and Conselice (2014). These studies have allowed us to establish that a significant fraction of high-z galaxies are not dynamically relaxed (e.g., Hammer et al. 2009; Förster-Schreiber et al. 2009; Epinat et al. 2012), and that a large fraction show unusually large velocity dispersions compared to local counterparts (e.g., Puech et al. 2006; Law et al. 2007; Epinat et al. 2009). Spatially-resolved data have proved to be essential to study accurately the evolution of the Tully-Fisher relation between velocity and mass (Puech et al. 2008, 2010; Crescri et al. 2009, Gnerucci et al. 2011; Vergani et al. 2012), and the evolution of the angular momentum of rotating disks (Puech et al. 2007; Bouché et al. 2007).

Despite all these efforts, we are still lacking a clear and coherent picture of galaxy evolution, and different and perhaps conflicting pictures have emerged. On the one hand, at z ≳1.5, numerical simulations have suggested that cold-gas streams from inter-galactic filaments could feed galaxies with fresh gas (e.g., Dekel et al. 2009; Agertz et al. 2009; Ceverino et al. 2009; Keres et al. 2009), which could explain their very high SFRs and clumpiness (e.g., Genzel et al. 2008; Bournaud et al. 2009). On the other hand, a few Gyr later (z~0.6), IFU observations have suggested that major mergers could play a more significant role, in agreement with predictions from semi-empirical Λ-CDM models (Hopkins et al. 2010; Puech et al. 2012). A significant fraction of local spirals could have rebuilt their disk following a gas-rich major merger at z<2, as supported by numerical models (e.g., Barnes 2002; Robertson et al. 2006; Hopkins et al. 2009), observations of individual objects (Hammer et al. 2009; Puech et al. 2009), and semi-empirical models (Hopkins et al. 2010b). So, is there a transition epoch where the main driver for mass assembly shifts from cold gas streams to major mergers around z ~0.8-1.4 (Puech 2010; Contini et al. 2012)? Were cold streams overestimated in numerical simulations (e.g., Nelson et al. 2013)? Are clumps surviving long enough to provide a viable channel for bulge formation (Genel et al. 2012, Hopkins et al. 2012; Newman et al. 2012; Bournaud et al. 2014)?

What we currently know is too fragmentary to tell us exactly how galaxies have assembled their mass. Above z~1.5-2 only the most massive, high-surface-brightness systems are within the grasp of current facilities. In addition, most of these samples have been assembled (necessarily) from a collection of selection criteria, or from rest-frame UV luminosities, which can be significantly affected by star formation episodes. Whether these samples are representative of the early Universe remains uncertain, but models of galaxy formation and evolution rely on analytic/empirical prescriptions, with inputs such as metallicity, angular momentum, and the spatial distribution of the gas taken from the existing observations (e.g., Hopkins et al. 2009; Dutton et al. 2012). This probably explains why the processes by which disk galaxies assembled their mass remain much debated and, in particular, the importance of minor vs. major mergers (e.g. Ownsworth et al. 2014), as well as the accretion of hot vs. cold gas.

As a result, there are a wide range of opinions in the community regarding what processes drive galaxy evolution, and making progress in this area will require dissection of a relatively large sample of mass-selected, emission-line galaxies to map their physical and chemical properties. Building a census of mass assembly and star formation in distant galaxies will require the deepest, high-resolution near-IR imaging available (e.g. *HST*/WFC3, *JWST*) combined with an ELT-MOS. More specifically, with such a sample, it will be possible to:

- Infer the dynamical state of the observed galaxies, via the combination of spatially-resolved kinematics from the bright H$\alpha$ $\lambda$6563 or [OII] $\lambda\lambda$3726,3729 emission lines, and deep near-IR imaging from *HST*/*JWST*. Such a combination is essential to determine if a galaxy is a rotating disk (cf. morpho-kinematic models of a regular rotating disk, e.g. Neichel et al. 2008), and to study internal galaxy structures (e.g., Forster-Schreiber et al. 2011; Law et al. 2012).

- Constrain the properties and dynamics of the stellar populations and ISM of the galaxies by comparing deep spectroscopy and SEDs (derived from deep multi-band imaging from the *JWST*, *Euclid*, or MICADO) with stellar population models. Further constraints on the source of ionization in the ISM gas will be derived using emission-line widths from 3D data (see Lehnert et al. 2009). The relative kinematics of the stellar and gaseous phase will be compared through potential shifts between absorption and emission lines, i.e., in/outflows will be detected and quantified (e.g., Rodrigues et al. 2012). [OII] line ratios will be used to derive electron densities (see Puech et al. 2006b), which will also constrain gas concentrations due to star formation or in/outflows.

- Derive metallicity gradients (using the $R_{23}$ index which employs the H$\alpha$, H$\beta$, [OII], and [OIII] emission lines) to investigate past gas accretion or mergers in distant galaxies (Cresci et al. 2010; Queyrel et al. 2012) as employed in the Local Universe (Lagos et al. 2012). This will enable us to constrain their star-formation histories and past interactions between the galaxies and the IGM.

- Characterize the evolution of the fundamental scaling relations between velocity, mass, and size. We will derive the integrated properties of all galaxies, including:
    - Stellar masses, from both rest-frame colours and *K*-band luminosities (Bell et al. 2003) and SED fitting (e.g., Papovich et al. 2001);
    - Star-formation rates (combining UV and IR luminosities, e.g. Hammer et al. 2005);
    - Gas fractions, by inverting the Schmidt-Kennicutt law between star formation and gas densities; (e.g., Law et al. 2009; Puech et al. 2010);
    - Gas metallicities, using deep integrated spectroscopy and the $R_{23}$ index.

These will be used to study the evolution of scaling relations between mass and velocity (the stellar mass and baryonic Tully-Fisher Relations, Puech et al. 2010; Cresci et al. 2009; Vergani et al. 2012), mass and metallicity (the stellar mass-metallicity relation, e.g., Rodrigues et al. 2008, Mannucci et al. 2009, as well as exploring the baryonic mass-metallicity relation), and between specific angular momentum and stellar/baryonic mass (Puech et al. 2007; Bouche et al. 2007).

- Use all these constraints to model all the galaxies in the sample, using hydro-dynamical models (e.g., Hammer et al. 2009; Hopkins et al. 2013, Perret et al. 2014). This step is crucial to disentangle what is driving star formation in each object and guiding us in discriminating between the possible disk origins. Such models can be used to derive the evolution of the merger rate as a function of redshift within a factor two-to-three in accuracy compared to theoretical predictions (Puech et al. 2012). Statistics will allow us to assess different possible mechanisms and investigate the cause of the star formation and mass build-up. The results from such a survey, such as the average gas fraction or the merger rate (e.g., Rodrigues et al. 2012), are invaluable constraints for semi-empirical models of galaxy evolution (e.g., Hopkins et al. 2009, 2010).

Progressing our understanding of galaxy formation and evolution requires such observations of substantial samples of galaxies, over a large enough volume to rule out field-to-field biases ('cosmic variance') and a large range of mass and redshift – only then will we be able to test the models rigorously with representative samples and reliable observables. The capabilities of the E-ELT will provide, for the first time, the potential for spatially-resolved observations of unbiased and unprecedented samples of high-z galaxies. Assembling a spectroscopic survey of hundreds of spatially-resolved high-z galaxies remains a strong scientific motivation for the E-ELT, and can only be obtained efficiently with a multi-IFU capability.

## 3.2. The puzzling role of high-z dwarf galaxies in galaxy evolution

At smaller stellar masses, dwarf galaxies play a key role in galaxy formation and evolution. In hierarchical models, they are thought to be the first structures to form in the Universe (Dekel & Silk 1986) and are believed to have an important contribution to the reionization process (Bouwens et al 2012). However, to date, their study has been biased to the local universe or in clusters.

The $\Lambda$-CDM model predicts that galaxies are assembled over cosmic time through hierarchical merging of progressively more massive sub-units (e.g., White & Rees 1978), possibly combined with massive accretion of cold gas (e.g., Keres et al. 2009, although see Nelson et al. 2013). It is therefore expected that the fraction of dwarf galaxies among the whole galaxy population should increase as a function of redshift (Khockfar et al. 2007), which appears to be reflected in a steepening of the faint-end slope of the galaxy luminosity function. (e.g., Ryan et al. 2007, Reddy & Steidel 2009, Ilbert et al. 2013, Muzzin et al. 2013, Mortlock et al. 2015).

While the deepest *HST* images have revealed a plethora of intrinsically faint galaxies at $1 \leq z \leq 3$ (Ryan et al. 2007) their faint magnitudes ($m_{AB}$ ~ 24-27) prevent spectroscopy with current facilities. Beyond understanding the overall role of dwarf galaxies in the bigger picture of the evolution of mass and star-formation in the Universe, there are also some more specific questions that can only be addressed with samples of sub-L* galaxies observed with an ELT-MOS, as illustrated by the following three examples.

### 3.2.1. The impact of low-surface-brightness galaxies (LSBGs) at high-z

LSBGs have typical surface brightnesses of $\mu_0(B) > 22$ mag. arcsec$^{-2}$ or $\mu_0(R) > 20.7$ mag. arcsec$^{-2}$ (Zhong et al. 2008) and represent 9% of the baryonic mass and a third of the HI mass density in the local Universe (Minchin et al. 2004). However, their effect on the distant Universe is a complete mystery. They are gas rich and have average stellar masses comparable to dwarfs, such that they may contribute significantly to the increase of the number density of dwarfs at high-z. Since, the gas fraction appears to increase rapidly with z, at a rate of 4% per Gyr from z = 0 to 2 (Rodrigues et al. 2012), LSBGs might include a significant (even dominant?) fraction of the Universal baryonic content at high-z. With large-multiplex, integrated-light observations, it will be possible to investigate their integrated ISM properties (dust, metal abundances, star-formation rates), while complementary IFU observations will provide their ionised gas fractions by inverting the Kennicutt-Schmidt law.

### 3.2.2. HII galaxies as a probe of the curvature of the Universe ($\Lambda$)

With careful target selection (i.e., a nebular component that dominates the continuum flux and with Gaussian emission components with $EW_0(H\beta) \geq 50$Å), HII galaxies show a remarkably tight correlation between the luminosity of recombination lines and the velocity dispersion of the ionised gas, e.g., $L(H\beta)$ vs. $\sigma(H\beta)$, over four dex in luminosity (see Melnick et al. 1988, Bordalo & Telles, 2011). This arises when a starburst fully dominates its host galaxy by the correlation between ionising photons and the turbulence of the ionised gas; Melnick, Terlevich & Terlevich (2000) demonstrated its validity up to z~3. Employing this correlation, Chavez et al. (2012) have used HII galaxies to provide an independent estimate of $H_0$ in the local Universe. Indeed, due to their brightness, HII galaxies at large redshifts (e.g., z~1.5-2.4) could provide larger observational samples than those of distant type Ia supernovae, at epochs for which the curvature effect is most pronounced. This could provide a complementary method to evaluate the dark energy equation-of-state at larger redshifts than currently probed by type Ia supernovae (Plionis et al. 2011). Deep exposures will be required to reach S/N~5 in the continuum of HII galaxies with $J_{AB}$ ~ 27 using integrated spectroscopy (requiring total exposures of up to ~40 hr);

### 3.2.3. The origin and mass assembly of dwarf galaxies

A signature of primordial dwarfs could be the peculiar properties of IZw18-type galaxies, whose spectra show prominent [OIII] 5007Å and H$\alpha$ equivalent widths (>>1000Å) that severely contaminate the continuum, impacting on the shape of their SEDs. Deep near-IR imaging may provide targets down to $J_{AB} \geq 29$ and detection of their prominent emission lines will require MOS observations with the E-ELT, enabling determination of the evolution of the number density of primordial galaxies from z = 0-1.5 (or up to z = 2.4). Deep spectroscopy of these systems will provide accurate redshifts and emission-line fluxes and, from comparisons with advanced population-synthesis models, we will be able to investigate their star-formation histories. The latter will allow us to map the stellar mass assembly of these remote systems (e.g. Rodríguez-Muñoz et al., 2015), to compare with predictions of galaxy evolution models of dwarfs at different epochs.

Moreover, it has been claimed that most present-day dwarfs could have a tidal origin (Okazaki & Taniguchi, 2000). With an ELT-MOS this hypothesis can be probed directly at z = 0.8-2.4, the era at which most of the mergers would have occurred (e.g., Puech et al. 2012). The 3D locations in space of the dwarfs could be obtained relative to massive galaxies observed in a major-merger process (in a sample drawn from ELT-MOS IFU observations). Their locations are expected to follow the tidal tails that can be modelled at the in-situ merger (e.g. Hammer et al. 2009), but this can only be tested via ELT (integrated-light) spectroscopy. It is not clear whether such a mechanism would explain the observed density of dwarfs, thought it would be fascinating to explore them in the distant, gas-rich Universe. Another motivation for such a study would be to assess the number density of satellite galaxies and compare it to the expected number of halos from the $\Lambda$CDM theory, from z = 2.4 to the present day. With target magnitudes of $J_{AB}$ ~ 26 (27), an exposure time of 17 hr (assuming GLAO) would provide S/N~8 (3) per continuum pixel.

## 3.3. Emission-line galaxies as a benchmark for the epoch of reionisation

Ly-$\alpha$ emission from galaxies is one of the most useful diagnostics available to us to explore the high-z Universe (see SC1). At the largest redshifts there are few useful alternatives and even *JWST* will need to rely on it at z>6.6; beyond this NIRSpec will lose access to H$\alpha$, while MIRI has no multiplex capability. Even if H$\beta$ (from space) and/or CIII] 1908 (from the ground) might provide supporting information, it is important to understand the escape fraction of Ly-$\alpha$ from galaxies, and its dependence on their luminosities, as this provides us with clues to the reionisation process.

Combining Ly-$\alpha$ and H$\alpha$ measurements is particularly attractive as this provides the cleanest method to estimate the Ly-$\alpha$ escape fraction ($f_{esc}$) without additional assumptions (and with only the unknown dust extinction affecting H$\alpha$). The H$\alpha$ line provides a well-understood tracer of the instantaneous star-formation rate, and is less model (and extinction) dependent than the near-UV continuum and other emission lines; it is also not subject to the same radiative-transfer effects which influence Ly-$\alpha$. Indeed, the resonant nature of Ly-$\alpha$ makes it susceptible to scattering at very modest densities of neutral hydrogen (and therefore a problem in star-forming galaxies). Thus, the radiative transfer of Ly-$\alpha$ photons through a galaxy is a complex process, influenced by the density, distribution, kinematics and dust content of the ISM.

Studies of UV absorption lines with *HST* have highlighted the importance of the neutral gas kinematics (Kunth et al. 1998; Wofford et al. 2013), but which parameters directly affect Ly-$\alpha$ remain unknown. By combining Ly-$\alpha$ and H$\alpha$ fluxes from VLT observations, Hayes et al. (2010) measured an average $f_{esc}$ of only 5% at z=2, providing a benchmark for studies at higher redshifts and galaxy-formation models (e.g. Le Delliou et al. 2006). However, extrapolating these results to larger redshifts is not possible as $f_{esc}$, as inferred from other methods, evolves with redshift (Hayes et al. 2011). Estimates of $f_{esc}$ for individual objects can vary with orders of magnitude for a given extinction (Hayes et al. 2010), but empirical estimates, as a function of the host galaxy parameters, would enable a better understanding of the observed Ly-$\alpha$ strength.

At z=2-2.6 (ages of 3.3 to 2.6 Gyr) both lines are accessible by ground-based observations and, when combined with other rest-frame UV features (e.g. SiII, CII) and optical lines ([OII], H$\beta$, [OIII], [NII]), the fundamental ISM properties that regulate $f_{esc}$ can be determined (i.e., the neutral and ionised gas kinematics, covering fractions, dust content, metallicity, and mass). Such estimates of $f_{esc}$ as a function of luminosity and physical parameters would provide valuable input to models of galaxy formation, and provide an empirical reference point for observations of Ly-$\alpha$ at larger redshifts (see SC1). Such studies can be extended to z=3.9 (age of 1.6 Gyr) by using H$\beta$ instead of H$\alpha$ (and using H$\gamma$ to estimate the reddening), albeit with reduced precision on the dust and metallicity distributions.

Spatially-resolved observations (from 1.1-2.4$\mu$m) are required to probe the rest-frame optical emission lines ([OII] through to H$\alpha$), as well as in the range of 0.38-0.7$\mu$m to probe the rest-frame UV lines (Ly-$\alpha$, HeII 1640, CIII] 1908, and SiII/CII/SiIV ISM absorption lines).

- Near-IR observations will provide information on the gas dynamics to investigate feedback effects (ionised ISM outflows), the presence of large-scale disturbances (from mergers/interactions), and estimates of dynamical masses. Analysis of the emission lines will reveal the presence/absence of secondary components and/or broad wings from outflows. Comparisons of the kinematics of lines with different excitation levels (e.g. H$\alpha$ or H$\beta$ vs [OIII]) would also help reveal outflows of hot gas.

- Optical observations: the resonant scattering leads to the effect that galaxies appear significantly larger at Ly-$\alpha$ than in the UV continuum (Östlin et al. 2009; Steidel et al. 2011; Hayes et al. 2011, 2013). Thus, if a constant-sized aperture (as a function of wavelength) is used, large aperture corrections for Ly-$\alpha$ are required. Alternatively, a larger aperture can be used, e.g., 2x2 arcsec, sampling 17x17 kpc at z=2.5, comparable to *HST* studies of low-z systems (Hayes et al. 2013). Note that the diffuse nature of the emissions means that high-order AO correction is not required. Optical observations will also be used to probe the kinematics of the neutral ISM, analogous to what is currently done at low-z using *HST*-COS (mapping the covering fraction and column-density distribution of neutral gas, as a function of velocity, along the line-of-sight using transitions of SiII). When combined with analysis of the Ly-$\alpha$ line profile, this provides vital information on the Ly-$\alpha$ radiative transfer (cf. models such as those from Schaerer et al. 2011) and the ISM structure. High-ionisation lines such as SiIV can be used to map the hot gas in a similar way. The far-UV also contains photospheric features that are sensitive to winds from hot stars, providing a direct measure of feedback and better age determinations (Wofford et al 2013).

With at least eight parameters (metallicity, dust attenuation, global kinematics, luminosity/SFR, specific SFR, stellar population age, neutral ISM kinematics and covering fraction), we require a minimum of 256 galaxies for such H$\alpha$ and Ly-$\alpha$ studies (i.e. 28 galaxies, to allow two independent observations per quantity). There will be some overlap between the H$\alpha$ and Ly-$\alpha$ samples, but at least 400 galaxies will be studied. These will be selected from narrow-band imaging in several luminosity bins, ranging from L* to the detection limit of deep, narrow-band surveys. The goal would be to have 1000 sources for both lines, employing existing deep *HST* imaging (e.g., HUDF/GOODS-S, Abell 1689, Frontier Fields), which provide direct probes of the leakage of ionising photons. Using H$\alpha$-selected galaxies is superior to LBG selection because it is insensitive to model uncertainties that relate the near-UV flux and star-formation rate to the ionising output, as it is the same photons that leak out and give rise to H$\alpha$. Observations of CIII] 1908 may also provide an alternative diagnostic at z>7, although this line has not been studied systematically to date and its strength relies on unknown physical properties. Similarly, observations of HeII 1640 will also provide insights into the metal-poor/high-mass stellar content (e.g. Schaerer, 2003).

## 3.4. Stellar populations in high-z galaxies: absorption-line spectroscopy

The emergence of the massive end of the Hubble sequence remains an open issue. Little is known about the early evolution of massive galaxies and in particular about the relative impact of in-situ (star formation) vs. external events (mergers, large-scale environment) in their subsequent mass and size growth. In particular, the period spanning 1<z<3 has been established as a key epoch that defines the properties of the massive galaxy population. Indeed, both the cosmic star-formation rate and the accretion rate onto galactic black holes peaks in this era (e.g. Chapman et al. 2005; Hopkins & Beacom 2006), resulting in roughly half of all the stellar mass we see today (e.g., Bernardi et al. 2003; Dickinson et al. 2003; Rudnick et al. 2003; Fontana et al. 2006; Borch et al. 2006). Understanding their formation and subsequent evolution is a key objective of modern observational cosmology and a central science case for upcoming missions and facilities such as *Euclid* and *JWST*.

### 3.4.1. From the blue cloud to the red sequence

Galaxies present a clear bimodality in the color-magnitude (or stellar mass) plane from low- to high-redshift (z~3; see, e.g., Baldry et al. 2004; Willmer et al. 2006). Also, the mass function appears to assemble continuously from z~4-5 to 0 (e.g., Perez-Gonzalez et al. 2008). This suggests that there is an on-going transition from the so-called 'blue cloud' of star-forming galaxies to the 'red sequence' of passive galaxies, which must happen rapidly at all epochs. While we seem to know many of the processes that might contribute to this transition (merging, AGN/stellar feedback, cold gas accretion, disk instabilities), we still lack of a comprehensive view of the relative importance of each process at different cosmic epochs.

Martin et al. (2007) found that galaxies at z~0.1 typically make this transition in <1 Gyr, suggesting that an external process, and not simple exhaustion of the gas reservoir, was probably causing the quenching of star formation. At z~0.8, Gonçalves et al. (2012) found that the quenching galaxies are more massive, and that they transition even faster. Both these results suggest that there is a downsizing of quenching with time, with the massive end of the red sequence forming earlier and more rapidly. Over the important 1<z<3 period, massive galaxies are seen to systematically shut off their star formation and then build-up the red sequence (e.g. Muzzin et al. 2013), but we do not know which mechanisms and on which time-scales this quenching occurs, nor its impact on chemical evolution.

### 3.4.2. Mass and size growth in the red sequence

The presence of a population of massive compact galaxies at z>1 has also been established (Daddi et al. 2005, Huertas-Company et al. 2013). These formed the bulk of their stellar content in very short timescales at very early epochs. Whether this primordial population evolved to match the properties of today's massive elliptical galaxies and contributed to the growth of the red sequence, or simply remained a marginal population of relics, is an open debate in the literature (e.g., Carollo et al. 2013; Poggianti et al. 2013). Answering these questions requires a detailed analysis of the progenitors of these present-day (~$10^{11} M_\odot$) galaxies, both in terms of stellar populations and environment. Current deep, near-IR *HST* surveys such as CANDELS are starting to find such progenitors at z~3-4, which appear to be very compact but actively forming stars (e.g., Barro et al. 2013, 2014), suggesting that disk instabilities or mergers could make them compact before quenching. Deep spectroscopy (S/N>20) is now required to put reliable constraints on their star-formation histories and chemical enrichment, hence on the physical processes driving their early formation, which cannot be obtained from SED fitting given the existing degeneracy between models (Barro et al. 2014). Such spectroscopy is clearly out of reach of the current 8-10m class facilities with reasonable exposure times (Belli et al. 2014). Only an ELT-MOS will allow us to obtain accurate physical parameters in a significant sample of massive galaxies during their early phases of evolution.

The ELT-MOS will not only probe the internal structure of these objects but also their close outskirts, giving key information regarding their early assembly (and, in particular, during the quenching phase). If mergers are truly playing a major role during this phase, we expect to find faint companions around these massive progenitors, for which lower S/N spectra could be obtained (in a single observation) using a relatively large patrol field. The typical magnitude of a massive quiescent galaxy at z~2 is $m_{AB}$>21, thus absorption-line studies require deep spectroscopy in the near-IR. Even with long exposures, the current generation of 8-10m telescopes can only provide marginal S/N for even the brightest such galaxies at z~2 (e.g. Kriek et al. 2009; Toft et al. 2012). To observe fainter, less massive galaxies, we require the sensitivity of the ELTs, equipped with high-throughput, multi-object spectrographs (to deliver large samples in a realistic amount of observing time). Seeing-limited or GLAO observations are sufficient for this case (with the latter reducing the sky background in the integration aperture somewhat). Thus, with a detection limit for absorption-line spectroscopy of $m_{AB}$ ~ 27, it would be possible to observe galaxies some two-to-three magnitudes fainter than L* at z = 2 and 3, respectively (Gabasch et al. 2004, 2006).

### 3.4.3. Deconstructing the stellar populations

The study of galaxy evolution has been traditionally based on two complementary approaches. The so-called 'archaeological' approach uses the information encoded in the stellar populations of present-day galaxies to infer their past star-formation history. The other method studies a census of galaxies at progressively earlier cosmic epochs, allowing us to trace the evolution of galaxies as a population back in time. However, in the latter case, our fundamental ignorance of the actual genealogy of galaxies makes it hard to distinguish among different evolutionary paths. The application of the archaeological approach at different epochs can shed light on this problem by constraining possible evolutionary paths and removing some of the degeneracy between star-formation histories and halo/galaxy assembly histories. Complementing the star-formation histories with details on the chemical enrichment of stars and gas is essential to provide clues about the exchange of gas between the galaxy, its halo, and the surrounding environment, as well as the feedback process.

Such detailed information at z>1 can only be achieved with deep spectroscopy to obtain reliable measurements of the absorption features in the stellar continuum in the (rest-frame) UV-optical, which requires S/N ≥20 (per rest-frame Å). Diagnostic features include the 4000Å break ($D_{4000}$) and the higher-order Balmer lines (H$\gamma$, H$\delta$ and H$\beta$) to constrain the age and recent SFH, with metallic features (including iron and a variety of $\alpha$-elements) providing key constraints on chemical enrichment and its typical timescales (via the $\alpha$/Fe ratio). Moreover, the great diversity of star-formation and chemical-enrichment histories at high-z requires large observational samples of thousands of galaxies, which span the full range of galaxy masses (down to ~$10^{9-11}$ M$_\odot$).

While the E-ELT can provide the necessary S/N in a few hours, observing thousands of galaxies one by one would be prohibitive and a MOS is required. The target density within the E-ELT patrol field (e.g., >1,000 galaxies with $H_{AB}$ < 23.5 at z>1.5) will allow optimal observations with a multiplexing of >100.

## 3.5. Finding the targets

### Emission-line galaxies: IFU and integrated-light observations

A minimum requirement is to probe the stellar mass function down to at least sub-M* galaxies at z=4 (see below), which have typical $m_{AB}$ ~25-26 (Puech et al. 2010). Selecting these targets requires pre-imaging and initial redshift determinations which are complete to at least this depth, with enough targets (i.e., ≥100 at z~4) to fill several mass/redshift bins with a substantial number of galaxies. *HST*/WFC3 data in the GOODS field were investigated to estimate the source densities down to $H_{AB}$ = 26, between z=3.5 and 4.5 (using photometric redshifts from Dahlen et al. 2010), finding a density of order 5 arcmin$^{-2}$.

Accounting for a spectroscopic success rate of ~50% (because we need emission lines between the OH sky lines, Evans et al. 2010), this leads to an effective density of observable targets of ~2.5 arcmin$^{-2}$. This roughly matches the expected densities of star-forming galaxies at z~4 down to $M_{stellar}$ > $10^8$ M$_\odot$ from Muzzin et al. (2013, their Table 2), i.e., deep enough into the stellar mass function at z~4. We also note that comparable source densities are found for $M_{stellar}$ > $10^9$ M$_\odot$ at z = 2-3 (e.g., from *HST*-CANDELS imaging); a MOS Large Programme would provide unprecedented insights into the kinematics and (resolved) star-formation rates of galaxies at this key evolutionary epoch.

Deep imaging from the *Euclid* mission, scheduled for launch in 2020, should be ideal for building such samples. The above source density translates into ~175,000 targets in the two Euclid deep fields (40 deg$^2$, assuming that only half of it will be in the Southern hemisphere), of which ~100 are expected to lie within a given ELT-MOS patrol field. Observing all the *Euclid* candidates at z~4 would therefore require ~1750 MOS pointings (assuming an effective multiplex of ~100), which would observe all the necessary targets (not to mention that *JWST*-NIRCam, that will also provide a substantial additional number of faint targets, see SC1). For galaxies at the highest redshifts (out to z~5.6), one needs to go down to $m_{AB}$ ~27-28; the GOODS-WFC3 data suggest that we would have ~130-150 galaxies between z=4.5 and 5.6 down to $H_{AB}$ = 27-28, i.e., 30-50% additional targets.

Slit-less spectroscopy from *Euclid* should provide redshifts for a fraction of these galaxies. However, *Euclid* is ultimately only a 1.2m telescope and therefore will not reach the required depth; in addition, slit-less spectroscopy is usually biased towards high surface-brightness objects. *JWST*-NIRSpec should alleviate part of this issue (see SC1), while current or next generation facilities should not go much deeper than $m_{AB}$ ~25 (UKIDSS-UDS, UltraVISTA, VLT-MOONS Deep Survey). This directly calls for integrated-light observations with a high-multiplex mode on an ELT-MOS. This mode would provide near-IR selected (as a proxy for stellar mass) redshift surveys complete down to at least J/H~26-27 mag. Considering all potential targets in the GOODS-WFC3 catalogue with z≥2 and H≤28 results in ~2,400 objects, hence a density ~17 arcmin$^{-2}$ (after applying a 50% spectroscopic success rate). Thus in the notional MOS patrol field one would have 680 targets, which calls for a multiplex of a few hundred in a high-multiplex mode.

### Integrated spectroscopy of dwarf galaxies

From the GOODS-WFC3 data (see Fig. 11), the number of high-z dwarfs already identified with $M_{stellar}$ ≤ $3\times10^9$ $M_\odot$ (i.e., masses comparable to the LMC or lower) with $J_{AB}$ ≤ 26 (≤ 27) within the notional 40 arcmin$^2$ patrol field is found to be ~800 (1750). This already gives enough targets to initiate such a programme, and more targets will be identified over this redshift range with the new VLT-MOONS instrument. Going deeper to $J_{AB}$ < 29 using the 3D-*HST* catalogue from Skelton et al. (2014) for the GOODS-S field, the source densities are even larger (of order 5000 galaxies per 40 arcmin$^2$ field).

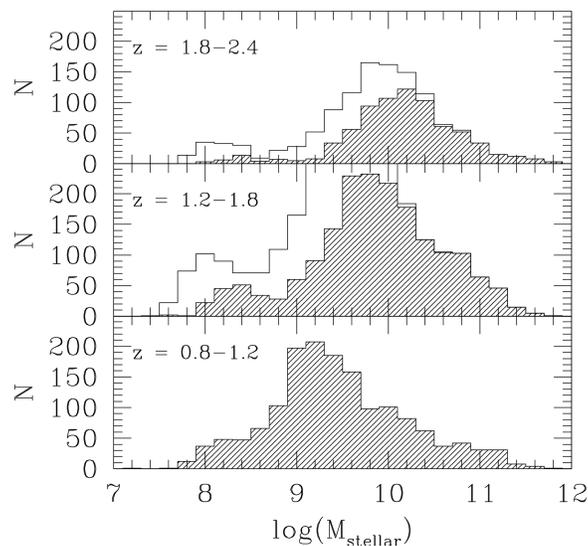

**Figure 11:** Distribution of stellar masses for galaxies selected by *HST*-WFC3 in the GOODS-S field within an area of 6.8'×10'. Photometric redshifts are derived from Dahlen et al. (2010; private comm.), absolute magnitudes are determined from an interpolation method (Hammer et al. 2001), and stellar masses are determined from the results of Bell et al. (2003). Shaded (and non-shaded in the upper panels) histograms show galaxies with $J_{AB}$ ≤ 26 (≤ 27) with 1,460 (3,100) dwarfs, with $M_{stellar}$ ≤ $3\times10^9$ $M_\odot$.

**Emission-line galaxies and Ly-α escape fractions**

For Hα narrow-band selected sources at z=2.2, a flux of $3.6 \times 10^{-18}$ erg s$^{-1}$ cm$^{-2}$ corresponds to a star-formation rate of 1 M$_\odot$/yr (while L* corresponds to 100 times brighter fluxes). This is approximately the limiting depth of Hα narrow-band surveys at the VLT; the current deepest observations are from Hayes et al. (2010) with HAWK-I giving a flux limit for Hα-selected sources of $6.8 \times 10^{-18}$ erg s$^{-1}$ cm$^{-2}$ in 60ks). By aiming at a Ly-α flux of $1 \times 10^{-18}$ erg s$^{-1}$ cm$^{-2}$ we will reach $f_{esc}$ of 3% assuming no Hα dust attenuation. If we assume an average extinction in Hα of a factor of two (Hayes et al. 2010) we would reach $f_{esc}$ = 1.5%. For Ly-α, the narrow-band flux limit is almost the same as in Hα. Targets this faint could be found with VLT-FORS. With Hα spectroscopy going down to $1 \times 10^{-18}$ erg s$^{-1}$ cm$^{-2}$, one can reach sources with $f_{esc}$ < 40%, assuming unobscured Hα (a Lyα source with $f_{esc}$ = 100% would have an Hα flux of $4 \times 10^{-19}$ erg s$^{-1}$ cm$^{-2}$). Hβ is expected to be approx. three times fainter, thus with an H-band flux limit of $7 \times 10^{-19}$ erg s$^{-1}$ cm$^{-2}$, all but the faintest sources can be probed. The [OIII] lines are generally brighter than Hβ, and would allow estimates of the metallicity. Thus, the goal is to obtain estimates for the physical properties of galaxies for star-formation rates down to 0.5 M$_\odot$/yr.

## 3.6. Synergies with other facilities

Kinematic studies of the ionized gas at high-z have attempted to investigate the origin of the gas in these galaxies, by examining their velocity structure (Law et al. 2009; Főrster-Schreiber et al. 2009; Newman et al. 2013). However, these studies suffer from a combination of poorer resolution and surface brightness dimming, which are intrinsic to high-redshift observations. Observational effects can lead to misinterpretation of the data, making galaxies appear more 'disk-like' than in reality (Basu-Zych et al. 2007; Overzier et al. 2010, 2011; Gonçalves et al. 2010).

In this area the combination of ALMA and an ELT-MOS should be very fruitful. In particular, ALMA will allow for resolved mapping of the molecular gas in the observed samples, while an ELT-MOS will allow us to map line emission to very low surface densities. With high spatial resolution (of order 80 mas) from the E-ELT, both facilities will probe similar physical scales of ~200 pc for low-redshift galaxies, i.e. on comparable scales to individual giant molecular clouds (e.g. Bolatto et al. 2008). The expected sensitivity of an ELT-MOS will enable us to probe down to SFR surface densities of ~$10^{-2}$ M$_\odot$ yr$^{-1}$ kpc$^{-2}$, which is the lower limit for the linear regime between gas and molecular gas densities. In other words, an ELT-MOS will allow us to investigate the resolved Schmidt-Kennicutt relation at scales of ~200 pc, covering a dynamic range of four orders of magnitude, between $10^{-2}$ and $10^2$ M$_\odot$ yr$^{-1}$ kpc$^{-2}$, in a single galaxy.

On a longer timescale, the SKA will also provide measurements of HI gas mass and its distribution. This will allow us to understand the origins of the cold gas and satellite galaxies, and how the internal galaxy properties are impacted by the properties of their close environment.

## 3.7. Requirements

**Observing modes**

The above cases require both an AO-corrected, multi-IFU mode to spatially resolve the properties of distant emission-line galaxies, and a single-object, large-multiplex, near-IR mode to measure the integrated properties of absorption-line and dwarf galaxies. Spatially-resolved spectroscopy in the optical (with a minimum FOV of 2"x2") is desirable to determine Ly-$\alpha$ escape fraction, although with no specific AO requirement. Alternatively, large-multiplex, single-aperture spectroscopy could be used at the cost of aperture effects from the extended Ly-$\alpha$ emission (provided that narrowband/tunable-filter imaging is available). There is no specific requirement for simultaneous use of all these modes.

**Wavelength coverage**

The goal of the IFU observations is to target optical emission lines redshifted into the near-IR (see Fig. 12). An effective redshift limit to such studies is given by the [OII] emission line leaving the *K*-band above z~5.6. However, *K*-band observations will require significantly larger exposure times than at shorter wavelengths because of the increased thermal background (Puech et al. 2010, see Fig. 15). Without coverage of the *K*-band, observations would be limited to z~4 (*H*-band). To estimate age and metal abundances of the stellar populations in galaxies, measurements are required of absorption lines which span a range of rest-frame wavelengths, from 3700-5500 Å, which are redshifted into the near-IR. Building on the work by Gallazzi et al. (2005, 2012), we identify three main cases:

- Age determinations, with little information on metal abundances: this requires observations of the Balmer features (and optionally the $D_{4000}$ break), and a minimal set of metal absorption lines to lift the age-metallicity degeneracy.

- Resolving age and iron abundances: requires the Balmer features (and optionally the $D_{4000}$ break), combined with at least one iron feature.

- Recovering age, total metallicity and $\alpha$/Fe ratio: requires Balmer features (and optionally the $D_{4000}$ break), and a set of features which are sensitive to both iron and $\alpha$-elements.

In Fig. 13, we show the redshift ranges that can be investigated for these three cases (in blue, green and red, respectively), using observations over three wavelength ranges: *YJ*, *YJH*, and *YJHK* (top-to-bottom panels, respectively). The solid panels show ranges where the $D_{4000}$ break is also observable – this key feature provides more robust age determinations (and, consequently, also better metal-abundance determinations). Coverage of only the *YJ*-bands does not really exploit the potential of a 40m-class telescope, limiting such studies to galaxies at z<1.5-2 (already feasible with current facilities, although with much longer exposure times). The addition of the *H*-band allows us estimate ages and iron abundances up to z~3, and full metallicity determination up to z=2.3-2.5. Addition of the *K*-band would enable us to go significantly beyond z~3, allowing us to probe the evolution of the stellar content of galaxies at the epoch of their birth, which is one of the key goals for the E-ELT. On the other hand, we note the *K*-band observations are more challenging due to the high thermal background, and comment that such observations might be better served by *JWST*-NIRSpec. In summary, the essential requirements are access to the *YJ*- and *H*-bands (not simultaneously, but one band per observation), with the *K*-band as optimal/desirable. Determining the Ly-α escape fraction requires observations of Hα redshifted into the *K*-band, and optical (spatially-extended) spectroscopy is required over 0.385-0.70 μm (with down to 0.37μm desirable).

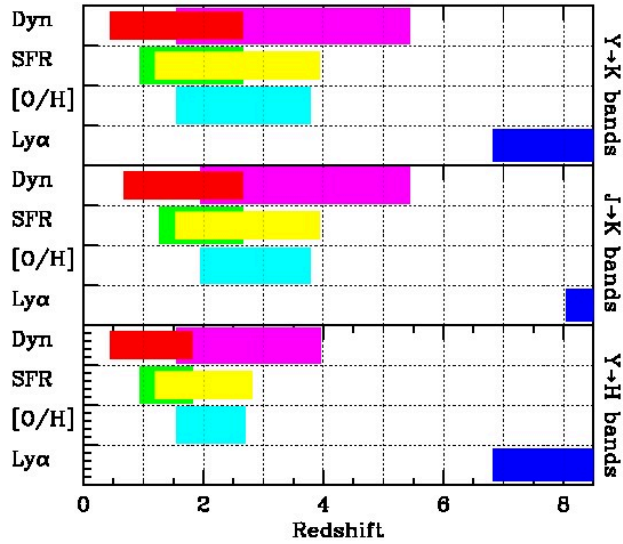

**Figure 12:** Physical quantities determined from emission lines as a function of redshift and different combinations of wavelength coverage. Rest-frame lines considered: Hα λ6563 and [OII] λλ3726, 3729 for dynamics (red and pink, respectively); Hα and Hβ λ4861, or Hβ and Hγ λ4341 for dust-corrected star-formation rate (green and yellow); Hα, Hβ, [OII], and [OIII] λλ4959, 5007 for dust-corrected $R_{23}$-metallicity (cyan). For comparison, the blue areas show the redshifts for which Ly-α can be targetted (see SC1). With IFU observations, all these properties can be spatially-resolved in the target galaxies.

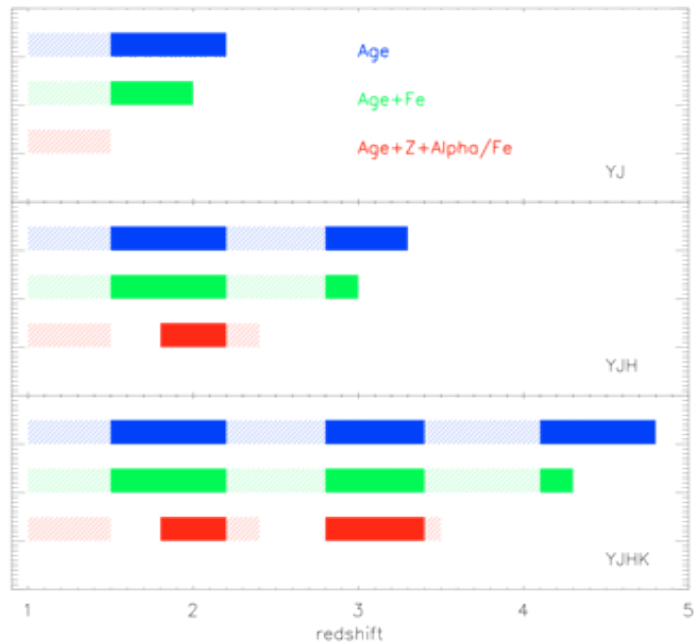

**Figure 13:** Physical quantities that can be determined from absorption lines as a function of redshift and different combinations of wavelength coverage. Solid panels represent windows where the Balmer break ($D_{4000}$) is observable (the hatched regions show where this feature is unavailable). These quantities will be estimated for large samples of galaxies from integrated-light spectroscopy.

## Multiplex

Preliminary estimates of the number of emission-line galaxies of interest for IFU observations in the ~7 arcmin diameter patrol field were estimated to be tens to ~100 galaxies, taking into account the spectroscopic success rate of measuring several emission lines between OH sky lines (Evans et al. 2010). Using results of the Design Reference Mission of the E-ELT, a survey of ~160 galaxies, spanning $2 \leq z \leq 5.6$ in three redshift and mass bins, would require ~90 nights of observations (including overheads) with a multiplex of 20 IFUs (see Puech et al. 2010). A similar survey limited to z≤4 (i.e. omitting the *K*-band) would provide observations of ~240 galaxies in ~12.5 nights. In contrast, the latter sample with a single IFU instrument would require ~250 nights, needing 4-5 years of observations (assuming standard operations of sharing between instruments and programmes).

For the large-multiplex, integrated-light mode, there are hundreds of potential targets in an E-ELT patrol field given the observed source densities. For instance, there are ~800 (1750) potential satellite/dwarf galaxies with $J_{AB} \leq 26$ ($\leq 27$) within a E-ELT patrol field. The number of absorption-line galaxies in the same patrol field is found to be >1,000 for galaxies with $H_{AB} \geq 23.5$ and z>1.5. Thus, both kinds of targets require a multiplexing ≥100 for optimal observations.

For ISM absorption-line studies in Hα and Ly-α emitters at z~2, S/N=10 in 10 hrs equates to a point-source sensitivity of approximately $m_{AB} = 26$ ($2 \times 10^{-19}$ erg s$^{-1}$ cm$^{-2}$ Å$^{-1}$). For a Ly-α emitter with a rest-frame EW~20 Å this leads to $1 \times 10^{-17}$ erg s$^{-1}$ cm$^{-2}$. The number of sources this bright per patrol field is > 200.

## Aperture size of high-multplex (integrated-light) mode

Given the small size of dwarf galaxies, the aperture for the single-object mode should maximise the S/N ratio for point-like sources in GLAO conditions, and was shown to be ~0.6'' in the relevant wavelength range (see SC1), or 0.8" in seeing-limited conditions. For absorption-line spectroscopy, Wuyts et al. (2011) found that the most extended galaxies at z~2 have $R_{eff}$ ~3-4 kpc, which translates into 0.35-0.5" on-sky. A 0.8" diameter aperture will cover a radius of 3.4 kpc, collecting ~50% of the light; similar figures apply to lower redshifts 1<z≤1.5. Larger apertures are inadequate for compact (quiescent) galaxies at z~2, which have $R_{eff}$ ~1 kpc (= 0.12") as apertures of >0.8" would lead to a prohibitive background contribution. Conversely, apertures of <0.8" result in large aperture losses for the most extended galaxies.

In conclusion, the ideal aperture size for this case is ~0.8", as a compromise to target a large variety of galaxies with single apertures. As an alternative, an aperture of ~0.6" diameter, in agreement with other target requirements, would optimize the S/N for the small-size/low-luminosity part of the sample, with multiplexed IFUs used for larger galaxies, in synergy with the case for the spatially-resolved spectroscopy.

## IFU Spatial resolution

Most z>1 sources reveal clumpy morphologies (e.g., Elmegreen et al. 2005, 2007), which were proposed to be important steps of bulge formation (e.g., Bournaud et al. 2008). Such clumps are typically 1 kpc large, i.e., ~140 mas at z=4. Resolving these structures would therefore require at least ~70 mas/spaxel. However, AO-assisted programmes on the VLT are currently assembling significant z~2 samples that will shed light on these issues (Mancini et al. 2011; Newman et al. 2012, 2013), not to mention that HARMONI will be particularly well-suited to resolve these clumps at the diffraction limit of the E-ELT.

We therefore argue that instead of optimising the spatial sampling of an ELT-MOS to resolve clump instabilities, a better compromise would be to favour survey speed by balancing spatial sampling with surface brightness detection at the global galaxy scale. Indeed, the next relevant spatial scale for studying distant galaxy structures is the diameter of the target galaxies, which is the scale at which large-scale motions are imprinted and which traces the dynamical nature of galaxies (e.g., Puech et al. 2008, see Fig. 14). Given the size of high-z galaxies, this argues for spatial sampling in the range 50-75 mas (e.g. Puech et al. 2010 and references therein). More specifically, a ~75 mas pixel scale would allow us to resolve sub-M* galaxies at z~4, or equivalently M* galaxies at z~5.6 (Puech et al. 2010), while larger pixel scales would be too limiting on our ability to probe the mass functions at the largest redshifts (see Table 3). Simulations show that the required ensquared energy in 2x2 spatial pixels is ~30% (Puech et al. 2008; see Fig. 14).

**Table 3:** Evolution of the stellar-phase size as a function of redshift (z*)* and stellar mass (expressed in fractions of M*, i.e., the knee of the galaxy stellar mass function, at a given redshift); the quoted sizes are the expected half-light radii in the *K*-band in milliarcseconds (see Puech et al. 2010, and references therein). Boxes in green/red are those for which 75 mas spatial sampling does/does not provide (at least) two pixels per half-light radius, as required to resolve the galaxies of given mass and redshift, respectively.

| z | 0.1M*(z) | 0.5M*(z) | M*(z) | 5M*(z) | 10M*(z) |
|---|---|---|---|---|---|
| 2 | 170 | 300 | 380 | 670 | 850 |
| 4 | 80 | 150 | 190 | 330 | 430 |
| 5.6 | 70 | 130 | 160 | 280 | 350 |

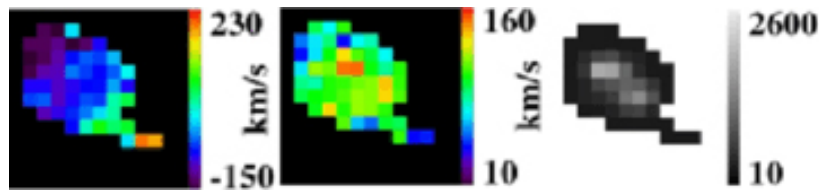

**Figure 14:** Simulated kinematics of a major merger at *z*~4 observed with an IFU with 75 mas/pixel. *From left-to-right*: velocity field, velocity dispersion map, and [OII] emission-line map, from which the kinematics was derived. Total integration time was 24 hr and the ensquared energy in 150x150 mas was 34%. The object diameter is 0.8" (~5.5 kpc). The peaks corresponding to two progenitors can clearly be seen in the emission-line and velocity-dispersion maps. Fitting the velocity field also reveals non-circular motions, characteristic of perturbations induced by galaxy (major-)mergers (see Puech et al. 2008).

**Spectral resolving power**

The requirement for absorption-line analysis and optimal separation between absorption and emission lines is a resolution of ~2-3Å rest-frame (i.e., $R$~2500, equivalent to a velocity dispersion of ~50km/s). In practise, $R$=4,000-5,000 is needed to resolve the brightest OH sky lines and identify absorption and emission lines between them; this also allows resolution of the [OII] doublet in the most distant galaxies. Observations of H II galaxies, reaching the continuum level and avoiding contamination by stellar light, while investigating Gaussian H$\beta$ profiles (asymmetric vs. multiple, or rotationally-contaminated profiles) argues for $R\geq$5,000 (with $R$~10,000 desirable).

**Field-of-view of each IFU**

This is defined by the size of the galaxies on the sky and the need for good sky subtraction (e.g., A-B-B-A dithers within the IFU) and entails an IFU size of order 2"x2". Only the closest and most massive/luminous galaxies would require specific offset sky measurements but the relatively large collecting power of the E-ELT should limit the time penalty to reasonable total observing times of no more than a few hours depending on mass, size, and redshift.

**Expected Sensitivity**

In the frame of the ESO DRM, a large range of simulated observations of distant emission-line galaxies were conducted to determine the achieved depth with an IFU pixel scale of 50 mas and $R$~5,000 (Puech et al. 2010, see Fig. 15 below), i.e., very similar to what is envisaged for the ELT-MOS. These results suggested that integrated magnitudes of $m_{AB}$ ~28 and emission-line fluxes of a few $10^{-20}$ erg s$^{-1}$ cm$^{-2}$ should be reached within ~24 hr of integration time (with an average S/N per spaxel of 5). This will allow us to probe the galaxy stellar mass function down to a few tenths of M* over the redshift range z=2 to 4. Observations at z>5 will be strongly limited by the telescope/instrument thermal background.

## 3.8. Comparison with other facilities

In Table 4 we compare the capabilities of other potential future instruments/facilities with an ELT-MOS to provide the observations required to satisfy SC3 (taking into account spectral resolution, spatial resolution, spectral range, field-of-view/patrol field, sensitivity, and multiplex).

Table 4: Comparison of different instruments/facilities for observations toward SC3.

|  | ELT-MOS | ELT-IFU | ELT-HIRES | JWST-NIRSpec | TMT/IRIS | TMT/IRMS | TMT/IRMOS |
|---|---|---|---|---|---|---|---|
| SPATIALLY-RESOLVED SPECTROSCOPY |  | No multiplex | Too low Sensitivity (high $R$) | No multiplex | No multiplex | No IFU | Limited patrol field |
| DWARFS AND LSBs |  | No multiplex | Too low Sensitivity (high $R$) | Limited patrol field | No multiplex | Limited multiplex & fov | Limited multiplex |
| INTEGRATED STELLAR POPULATIONS |  | No multiplex | Too low Sensitivity (high $R$) | Limited patrol field | No multiplex | Limited multiplex & fov | Limited multiplex |

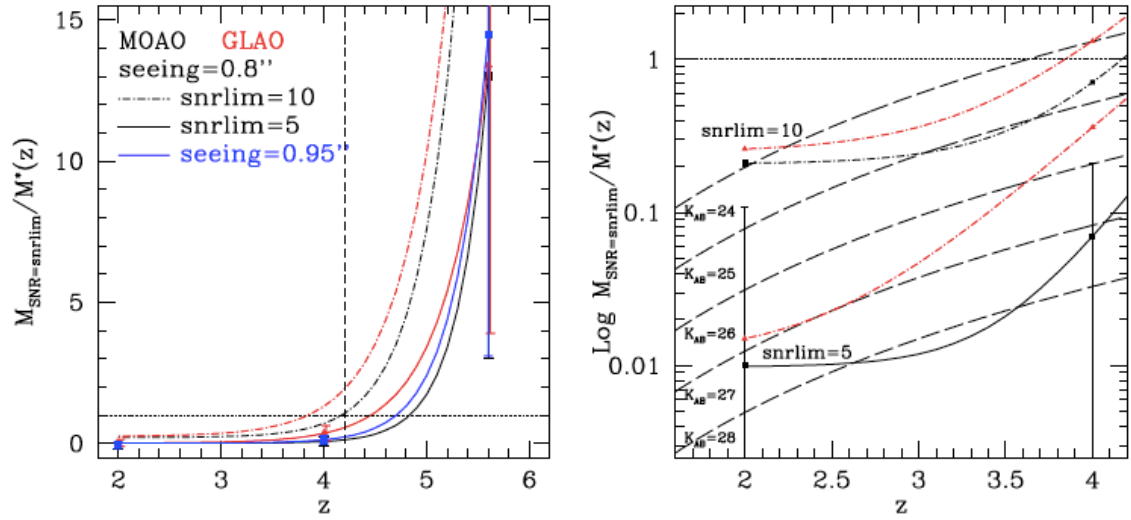

**Figure 15:** *Left:* Stellar mass limit that can be reached at a S/N limits of 5 and 10, as a function of redshift (z), AO correction, and seeing; error-bars represent the range of the mass limits obtained when considering a wide range of mopho-kinetic templates. The vertical dashed line indicates the limit above which the [OII] line is redshifted into the *K*-band. *Right:* Expanded version of the same plot, over the range 2 < *z* < 4; long-dashed lines are iso-magnitude curves. For clarity, only a subset of the results in main panel are plotted, and the range of limiting mass is only shown for the MOAO, 0.8" seeing case. (Taken from Puech et al. 2010.)

# AGN/GALAXY CO-EVOLUTION & AGN FEEDBACK

One of the most important and still largely unsolved questions in cosmology is the combined evolution and growth of galaxies and their central super-massive black holes (SMBHs). A key question is how galaxy and SMBH growth is self-regulated by the feedback processes thought to be associated to active galactic nuclei (AGN) and SNe-driven outflows, which can heat up the host galaxy ISM and expel it from the galaxy.

Powerful, AGN-driven outflows are expected to play a major role in galaxy evolution (e.g. Croton et al. 2006; Menci et al. 2008; Zubovas & King 2012). From galaxy-evolution models, the amount of energy deposited into the ISM by such outflows is large enough to blow away most of the reservoir of cold gas, thus terminating galaxy growth and creating a population of 'red-and-dead', gas-poor galaxies. Observational evidence of such self-regulation processes in action has been long sought for.

Outflows have been widely observed in AGN, and especially in luminous quasars, from X-rays through to UV and optical/near-IR spectroscopy. AGN-driven outflows have historically been detected in the ionized gas component by observation of broad absorption-line systems (BALs), seen in 10-40% of quasars (Weymann et al. 1991; Dai et al. 2008). BALs trace the fast winds of the ionized medium, with velocities of up to a few 10,000 km/s seen in both high- and low-ionization optical/UV lines (e.g. Reeves et al. 2009; Moe et al. 2009). The spatial extent of the BAL winds is still a matter of debate (see Borguet et al. 2013, and references therein). Moreover, faster outflows – 'ultra-fast outflows' (UFOs) – have been found with velocities of up to a fraction of the speed of light. These UFOs are traced by absorption lines in the 6-8 keV range (see Tombesi et al. 2013, and references therein), and trace energetic nuclear winds (with $E_{kin}$ ~a few percent of $L_{bol}$) which likely originate close to the accretion disks of SMBHs, on scales of a few pc.

Outflows have also been detected via the observation of broad, blueward-skewed profiles of [OIII] $\lambda$5007 (Zamanov et al. 2002; Aoki et al. 2005). These profiles cannot be ascribed to gravitational motions but are thought to reflect the kinematics of energetic outflows. Zhang et al. (2011) analyzed a large sample of radio-quiet, type-I AGN at z<0.8, finding significant broad [OIII] components in most objects. This finding suggests that outflowing ionized material is present in most AGN (e.g., Villar-Martin et al. 2011; Oh et al. 2013).

Spatially-resolved optical/near-IR/sub-mm spectroscopy of about two dozen AGN has found broad (>1,000 km/s) molecular and (forbidden) ionized emission lines, extended over kpc scales. For example, the IFU observations of a luminous quasar at z = 2.4 shown in Fig. 16 (from Cano-Diaz et al. 2012). The main diagnostic tools in these studies have been the [OIII] broad emission components, the H$\alpha$ emission, and the broad sodium absorption lines. (The latter of these is the most demanding as it requires the detection of absorption lines in the spectra of the host galaxy.)

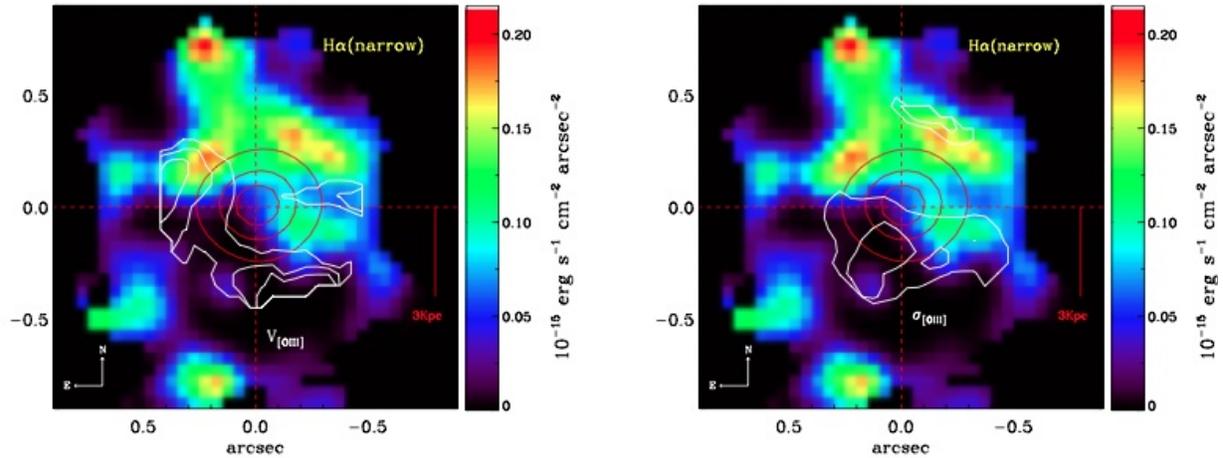

**Figure 16:** VLT-SINFONI integral-field spectroscopy of a luminous quasar at z = 2.4 from Cano-Diaz et al. (2012). Star formation (as traced by Hα) appears suppressed in the SE region where the strongest outflow is traced by [OIII] emission.

In some cases it has been possible to constrain the morphology of the outflow and its size, thus providing estimates of the mass-outflow rate, $M_{out}$, and the outflow kinetic energy $E_{kin}$. $M_{out}$ is often > 100-1000 $M_\odot$/yr, i.e. larger than the star-formation rate in the host galaxies, suggesting that these outflows could be the long sought cause of 'quenching' of star formation, e.g., see Fabian (20120, as well as the discussion in Sect. 3.3). For example, in the case shown in Fig. 16, there is evidence that the SFR (as determined from the narrow Hα emission) is smaller in the regions affected by the [OIII] outflow, providing the first direct evidence of star-formation quenching due to an AGN-driven outflow.

In summary, we have strong statistical evidence that massive, AGN-driven outflows are common and that they affect the host galaxies of most AGN. However, to date we have only been able to measure the physical and geometrical parameters of the outflows in about two dozen cases (the size of the outflow and its mass-outflow rate are the most important parameters), as shown in the redshift-luminosity plane in Fig. 17. Bright sources (i.e., nearby AGN or extremely luminous, high-z AGN) have been observed with the latest instrumentation, such as AO-assisted spectrographs on 8-10m class telescopes, the ALMA Early Science observations, and using the Plateau de Bure Interferometer. However, the bulk of the AGN population remains unexplored. In particular, normal L* AGN at z = 1-3, i.e., the peak of both AGN and galaxy evolution, are effectively impossible to study with present faciltiies. In the coming years the full capabilities of ALMA will hugely expand the discovery space for molecular outflows; the E-ELT will provide strong synergies with ALMA via a similar gain for studies of outflows of ionized/neutral gas.

The typical bolometric luminosity of faint AGN accessible to the E-ELT will be ~$10^{44}$-$10^{45}$ ergs s$^{-1}$, which is close to or even below L* at z = 1-3. The sensitivity of an ELT-MOS would therefore also allow additional diagnostics of outflows, such as: mapping [NeV] broad emission components and/or mapping Al III broad absorption-lines, or other UV absorption troughs; these would open-up the search and characterisation of outflows of ionized gas at redshifts significantly larger than those we can access with current facilities, as illustrated by Fig. 17.

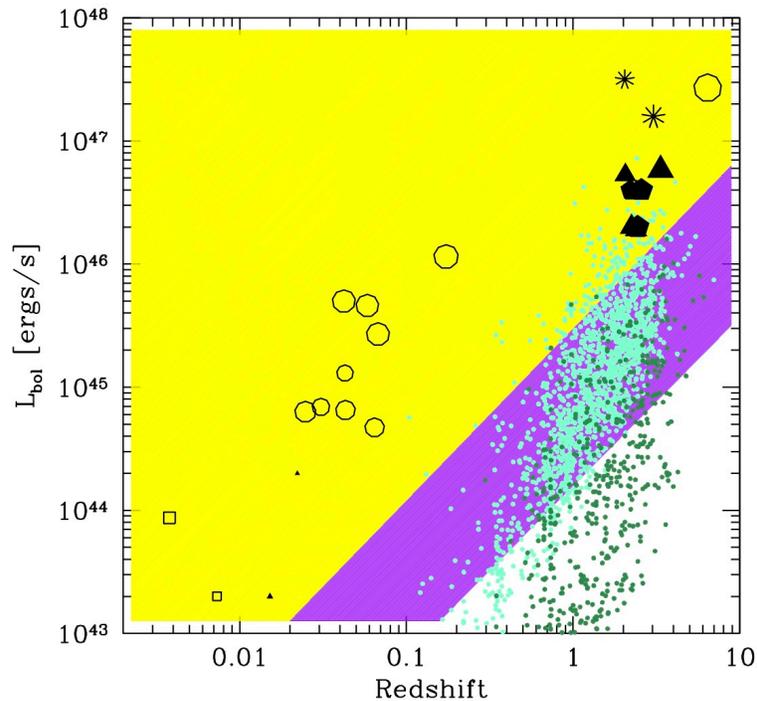

**Figure 17:** The $L_{bol}$-redshift plane for AGN with measured outflow properties from the literature (black points). AGN from the *Chandra*-COSMOS and CDFS surveys are indicated by the cyan and green points, respectively. The yellow area indicates the region of parameter space accessible with current facilities – the E-ELT will enable observations in the region highlighted in purple, providing significant new parameter space in the study of AGN outflows.

We are at the beginning of quantitative study of AGN/galaxy interactions through their powerful outflows. Observations with the E-ELT will enable us to answer questions such as:

- What are the masses, mass-flow rates and kinetic energies of the observed outflows?
- What are the best tracers to obtain reliable estimates of the outflow properties?
- What are the outflow morphologies?
- SNe are also expected to drive energetic outflows. If so, what is the relation between AGN- and SNe-driven winds? Can we disentangle the two processes? Which is dominant? (And in which type of galaxy?)
- Is there any positive feedback induced by outflows?
- How common are massive outflows at the peak epoch of AGN and galaxy assembly (z = 1-3)?
- In which galaxy/AGN types do massive outflows occur? And in which phase of the evolution of a galaxy? And how long does the active feedback phase last?
- How are molecular outflows linked to ionized outflows?
- How do outflow properties (masses, mass-loss rates, sizes) correlate with other AGN/galaxy properties such as: bolometric luminosity, obscuration, Eddington ratio, SMBH mass, star-formation rate in the host galaxy, and age of the stellar populations?

To address these issues we need to characterize outflows in large samples of AGN over cosmic time, up to the peak of AGN/galaxy activity and beyond (into the era of galaxy and AGN formation). We also need a comprehensive census of the normal AGN population, not only data on the most luminous examples. An ELT-MOS will enable such a census, providing a self-consistent picture of the impact of outflows on SMBH growth and of the regulation of star formation in the host galaxies.

## 4.1. Instrument requirements

An ELT-MOS is expected to play a crucial role in the study of AGN outflows and feedback. AO-corrected IFUs will be necessary to reliably map the outflows in the ionized and neutral gas, to constrain their geometries and measure their sizes and to determine their mass outflow rates and kinetic energies.

**Example observation**

The peak of the [OIII] equivalent width distribution is ~13Å (Risaliti et al. 2011). The median of the [OIII] EW ratio of broad-to-total emission is 0.6, with a semi-interquartile of 0.11 (Zhang et al. 2011). The correlation between the [OIII] EW and AGN luminosity is quite weak (see Zhang et al.) and so we neglect it the following. Our goal is to achieve a relatively good S/N ratio for broad [OIII] lines with rest-frame EWs of order ~10Å; at z=2 this corresponds to an EW of 30Å in the observer's frame. The line S/N can be roughly estimated from the continuum S/N, multiplied by the ratio of the line EW and the line FWHM. The FWHM of broad [OIII] lines observed in some quasars and radio galaxies are ~ 50-100Å (Nesvadba et al 2006, 2008; Cano-Diaz et al. 2012; Harrison et al. 2012), which means that we need to obtain S/N ~10 for the continuum to obtain S/N ~5 for the [OIII] line. We consider $H_{AB}$ = 23 for our calculations and approximate the continuum emission with a black body with T = 105 K (the temperature of the blue bump which dominates in the rest-frame optical). Assuming a seeing of 0.4" (i.e., with GLAO correction), a S/N ~10 for the 1600Å continuum (in an area of 50mas radius) is obtained in an exposure of ~40hrs.

**Spectral resolving power**

$R$ > 3,000 is required to reduce the contamination from the OH sky lines, while also being sufficient to separate the broad and narrow components in the emission lines.

**Multiplex**

Today we are limited to AGN and galaxies brighter than $H_{AB}$ ~19-20. The huge increase in sensitivity provided by the E-ELT, coupled with AO correction, will enable studies of AGN outflows in objects as faint as $H_{AB}$ ~23-24. The AGN spatial density at $H_{AB}$~20 and $H_{AB}$ ~24 is 1.5 and 10 per 50 arcmin$^2$ field. This means that current facilities such as VLT-KMOS can only be fully exploited for over-dense fields, but as we go to larger distances/fainter magnitudes, we can observe ~10 per E-ELT pointing.

**Spatial resolution**

The extended outflows observed in both molecular and ionized gas in local AGN and ULIRGS have ~kpc sizes (Rupke & Veilleaux 2013; Feruglio et al. 2010, 2013; Cicone et al. 2013). The goal is to spatially-resolve and characterise outflows at z = 2-4. At z = 2, 3, and 4, an on-sky aperture of 1" corresponds to 8.5, 7.8, and 7 kpc, respectively. Thus, spatial resolution of ~100 mas is essential, with 50-70 mas desireable (giving better constraints on outflow geometry).

**IFU field-of-view**

Typical target sizes for AGN (at z = 2-4) are ~1-10 kpc, so an IFU of ~2"×2" will provide full coverage of the galaxies and their nearby environment (particularly as some objects may be mergers). In many cases such a field will also be large enough to provide local estimates of the sky background.

## 4.2. Additional AGN-related cases

Spectroscopy of large samples of high-z AGN can also be used to investigate important topics such as:

1. Scenarios for the formation of the black hole seeds, which will eventually grow up to form the SMBHs seen in most galaxy bulges. Two main scenarios have been proposed to date: the monolithic collapse of rare ~$10^5$ $M_\odot$ gas clouds to black holes, or an early generation of more common ~100 $M_\odot$ BHs produced by the SNe from the first (Pop III) stars. The two scenarios predict different slopes for the SMBH mass functions at z>6, with the lighter, seed black holes producing a steeper SMBH mass function. Since accretion at high-z is likely to occur at, or close to, its Eddington luminosity, spectroscopic identifications of low-to moderate luminosity AGN at z>6 will be vital to distinguish between the two scenarios.

2. The evolution of the AGN duty-cycle vs. redshift can help in disentangling different AGN triggering mechanisms; two main mechanisms have been suggested: galaxy encounters and internal galaxy dynamics. The first predicts an increase in the AGN duty-cycle by a factor of 10-20 from z~0 to z~4-5 followed by a sharp decrease at higher redshift. The peak at z=4 to 5 is due to the decrease of the rate of galaxy interaction above the redshift at which groups formed (z~4), with its decrease toward lower redshifts due to high relative velocities/low densities (Fiore et al. 2012). An accurate measure of the AGN duty cycle at z>4-5 would provide a test of the galaxy encounter scenario.

3. As alluded to in SC1, the AGN contribution to reionization (and to the heating of the IGM and its effect on structure formation) remains unclear. The number density of luminous QSOs is sharply peaked at z=2 to 3 and rapidly decreases toward both higher and lower redshifts; these objects do not appear to be responsible for heating the IGM exactly at z=2 to 3. However, at larger redshifts the density of low-to-moderate luminosity AGN can be relatively high, and their duty cycle can reach a peak at z=4 to 5 (previous point); these objects may contribute to hydrogen ionization at such high redshifts (see Giallongo et al. 2012). Spectroscopic identification of low-to-moderate AGN at z>4 to 5 is needed to measure their luminosity function, which is crucial to assess the global AGN contribution to the high-z IGM heating.

Low-to-moderate luminosity AGN at z>4-5 (log $L_{bol}$ = 44-45) have faint near-IR counterparts, *H*=25-27 (Fiore et al. 2012). Spectroscopy of such faint objects is hardly feasible with present instrumentation. *JWST*-NIRSpec may reach such depths, but the space density of high-z AGN (0.2-2 arcmin$^{-2}$) would make it difficult to collect a large statistical sample. A multiplexed ELT-MOS, with ~100 channels over a 7×7 arcmin field would match neatly to the density of high-z AGNs, enabling efficient spectroscopic identification of large samples. Targets can be selected from existing *Chandra/Spitzer* data (or future X-ray/IR facilities).

Science case 5

# RESOLVED STELLAR POPULATIONS BEYOND THE LOCAL GROUP

Discoveries of disrupted satellite galaxies have demonstrated that our evolutionary picture of the Milky Way is far from complete (Majewski et al. 2003; Belokurov et al. 2006), let alone our understanding of galaxies elsewhere in the Universe. Deep imaging from ground-based telescopes and the HST has yielded colour-magnitude diagrams (CMDs) with unprecedented fidelity, providing new and exciting views of the outer regions of galaxies beyond the Milky Way (e.g. Barker et al. 2011; Bernard et al. 2012; Weisz et al. 2014). From comparisons with stellar evolutionary models, these data enable us to explore the star-formation and chemical-enrichment histories of the targeted regions, providing a probe of the past evolution and, in particular, the merger/interaction histories of these external galaxies. New constraints on galaxy formation models have also been provided by evidence for larger-scale structure and alignment of the galaxies in the Milky Way (Pawlowski et al. 2012) and Andromeda groups (Ibata et al. 2013).

Although photometric methods are immensely powerful when applied to extragalactic stellar populations, only via precise chemical abundances and stellar kinematics can we break the age-metallicity degeneracy, while also disentangling the populations associated with different structures, i.e. follow-up spectroscopy is required. The Calcium II Triplet (CaT, spanning 0.85-0.87 μm) has become a ubiquitous diagnostic of stellar metallicities and kinematics in nearby galaxies (e.g. Tolstoy et al. 2004; Battaglia et al. 2008; Starkenburg et al. 2010). However, 8-10m class telescopes are already at their limits in pursuit of spectra of the evolved populations in external galaxies, e.g. Keck-DEIMOS observations in M31 struggled to yield useful S/N below the tip of the red giant branch (RGB) at $I > 21.5$ (Chapman et al. 2006; Collins et al. 2011; Dorman et al. 2012). Equally, observations of massive O-type stars struggle to reach beyond 1 Mpc (e.g., Tramper et al. 2011, 2014; Hartoog et al. 2012). To move beyond the Local Group – to investigate whether similar processes are at work in other large galaxies, and what role environment and galaxy morphology have on galaxy evolution – we need the sensitivity of the E-ELT.

With its vast primary aperture and excellent angular resolution, the E-ELT will be the facility to unlock spectroscopy of evolved stellar populations in the broad range of galaxies in the Local Volume, from the edge of the Local Group, out to Mpc distances. This will bring a wealth of new and exciting target galaxies within our grasp, spanning a broader range of galaxy morphologies, star-formation histories and metallicities than those available to us at present. These observations can then be used to confront theoretical models to provide a unique view of galaxy assembly and evolution. There are many compelling and ground-breaking target galaxies for stellar spectroscopy of individual resolved stars with the E-ELT including the spiral dominated Sculptor 'Group' at 2-4 Mpc and the M83/NGC5128 (Centaurus A) grouping at ~4-5 Mpc.

Beyond these targets, the E-ELT will be essential for spectroscopy of stellar populations in more distant galaxies. As we move to larger distances, observations of the evolved stellar populations will move from 'resolved' to 'semi-resolved' in terms of spatial resolution (and building a bridge toward the integrated-light studies discussed in Sect. 3.4). Nonetheless, there are a wide range of science questions to be addressed in targets such as NGC 1291 (8 Mpc), NGC 4594 (the Sombrero, ~10 Mpc), NGC 3379 (~11 Mpc) and systems in the Virgo Cluster, to address questions related to their large-scale structure (e.g. Gadotti & Sanchez-Janssen, 2012, and briefly expanded on in Sect. 5.5.5). Additional motivation to reach targets at >10 Mpc is provided by the need for detailed results in systems at the higher end of the galaxy mass distribution (e.g. M87), which would enable us to investigate the formation of bulges in disk galaxies over the full mass spectrum.

## 5.1. Spectroscopy of evolved stars in the Sculptor Galaxies

The top priority case in this area is a 'Large Programme'-like survey of the evolved populations in the Sculptor Group, comprising five spiral galaxies (NGC 55, NGC 247, NGC 253, NGC 300 and NGC 7793) and numerous dwarf irregulars. Distance estimates over the past decade have revealed that this 'group' is actually two distinct components (e.g. Karachentsev et al. 2003), at approximately 1.9 Mpc (NGC 55 & 300) and 3.6-3.9 Mpc (NGC 247, 253 & 7793). These five galaxies represent the most immediate opportunity to study the star-formation history and mass assembly of spirals beyond the limited sample available at present, i.e. the Milky Way, M31 and M33. Their masses are in the range 1.5-8 $\times$ 10$^{10}$ M$_\odot$, putting them on a par with M33 – it is exactly these late-type, low-mass, small bulge (or even bulge-less) spirals that theoretical N-body/semi-analytic simulations struggle the most to reproduce (D'Onghia & Burkert, 2004).

Two of the Sculptor galaxies are shown in Fig. 18 – these are large extended galaxies in which there is a wide range of stellar densities/crowding. A key point to note is that there is already substantial deep imaging available of such galaxies from, e.g., the *HST*, VLT, and Magellan; i.e. we already have catalogues of potential targets, but lack the facilities to obtain spectra with adequate S/N. Additionally, some relevant parts of the southern sky will also be observed with *Euclid,* which will also yield good samples for E-ELT follow-up, especially at large galactocentric radii.

A HARMONI-like instrument will be well suited to spectroscopy of stars in individual dense regions in external galaxies (and the Milky Way), but the larger samples needed to explore entire galaxy populations will require an ELT-MOS. To recover the star-formation histories and structures in the Sculptor galaxies we advocate a two-fold approach (depending on the target source densities), as discussed in the next two sections.

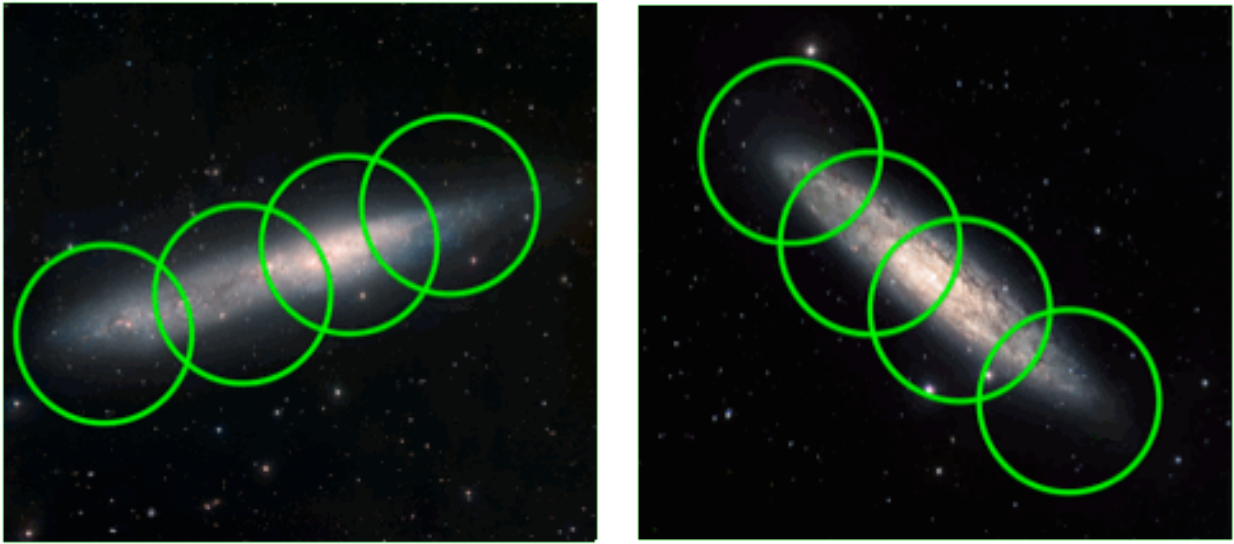

**Figure 18:** Illustrative 7' diameter ELT-MOS pointings in NGC 55 at 1.9 Mpc (*left*) and NGC 253 at 3.6 Mpc (*right*).

## 5.2. Requirements: Halos of external galaxies

### Spatial resolution/image quality

The outer regions of external galaxies can provide some of the most dramatic evidence about their past star formation and evolutionary histories. Wide-field surveys have revealed the presence of extended structures (e.g. Crnojevic et al. 2013; Greggio et al. 2014), while the low stellar densities have enabled deep *HST* photometry reaching back to early epochs (e.g. Gogarten et al. 2009; Bernard et al. 2012). The objective is to observe hundreds of stars per galaxy but, given the low source densities at large galactocentric distances, high-performance AO is not required and the GLAO correction enabled by M4 and M5 of the telescope will provide acceptable image quality.

### Wavelength coverage

We assume the use of the CaT as a metallicity diagnostic, as used by the *Gaia* mission and in many studies in the Galaxy and other galaxies of the Local Group. At large galactocentric distances we would expect relatively metal-poor populations, but an important characteristic of the CaT is that its three absorption lines are relatively strong, meaning that its relationship to metallicity, [Fe/H], is robust over a large range (e.g. Cole et al. 2004; Carrera et al. 2007), with the calibration from Starkenburg et al. (2010) valid to metallicities as low as [Fe/H] ~ −4. Other strong spectral features such as the Mg I b triplet or the G-band may also be used as additional diagnostics.

### Spectral resolving power

Battaglia et al. (2008) demonstrated that, with careful calibration and S/N ≥ 20 Å$^{-1}$, metallicities obtained from the low-resolution mode of FLAMES-Giraffe ($R$ ~ 6,500) agree with estimates from the high-resolution mode ($R$ ~ 20,000). In practise, $R \geq 5{,}000$ is sufficient, given adequate S/N (≥ 20).

## 5.3. Requirements: Disk regions of external galaxies

**Wavelength coverage**

While the CaT is a widely-used spectral feature for studies of stellar populations, the AO correction will be more effective at longer wavelengths, and it is worth considering other diagnostics. Observations in the *J*-band (at *R* of a few thousand) were demonstrated by B. Davies et al. (2010) to provide robust estimates of metallicities of red supergiants (RSGs). The 1.15-1.22 μm region includes absorption lines from Mg, Si, Ti, and Fe so, while the lines are not as strong as those in the CaT, they provide direct estimates of metallicity. The potential of this region for extragalactic observations of both RGB stars and RSGs with the E-ELT was further investigated by Evans et al. (2011). They concluded that a continuum S/N > 50 (per two-pixel resolution element) was required to recover simulated input metallicities to within 0.1 dex, sufficient for many extragalactic applications. On-sky tests of the *J*-band methods are underway via VLT-XShooter and VLT-KMOS observations of RSGs in the Magellanic Clouds and beyond. Once the additional problems of real observations are taken into account (e.g. telluric subtraction), analysis of these data suggests that S/N of ~100 is required for good metallicity estimates (Gazak et al. 2014).

**Spatial resolution/image quality**

This becomes a significant factor in the denser regions of external galaxies, and therefore makes stronger demands on the AO performance. For instance, the spatial resolution delivered by *HST*-ACS and WFC3 observations is more than a factor of three finer than the sampling provided by the GLAO correction of the E-ELT – to successfully follow-up such observations (for example), will require additional correction. The WEBSIM tool developed by Puech et al. (2008) was used to simulate performances for *J*-band stellar spectroscopy with MOAO PSFs (see Evans et al. 2011). The results in the left-hand panel of Fig. 19 show the mean S/N (and standard deviation) from ten simulations for individual stars with *J* = 22.75 and 23.75. The results in the right-hand panel of Fig. 19 shown the S/N recovered from two configurations of natural guide stars (NGS), which were representative of the two likely extremes of MOAO performance from different asterisms.

In sheer survey speed and sensitivity for one pointing, the maximum efficiency for such point-source observations would be achieved with spatial sampling of 20 mas; marginally longer exposures would be required for spatial sampling of 30-40 mas. The maximum spatial sampling acceptable is ~75 mas – equivalent to critically-sampled optical images from *HST*, and short-wavelength near-IR images from *JWST*-NIRCam. With 75 mas spatial pixels the exposure time to reach the same S/N as the 37.5 mas pixels adopted in the EAGLE study would need to be 2-2.5 times longer. However, if the same number of detector pixels were used by each IFU, coarser sampling would enable IFUs with an area twice as large (leading to observations of more stars per IFU). Thus, in terms of survey speed for a given number of stars across the full extent of an external galaxy, either would be acceptable (and in fields where the spatial resolution is more critical, individual pointings with HARMONI could be obtained).

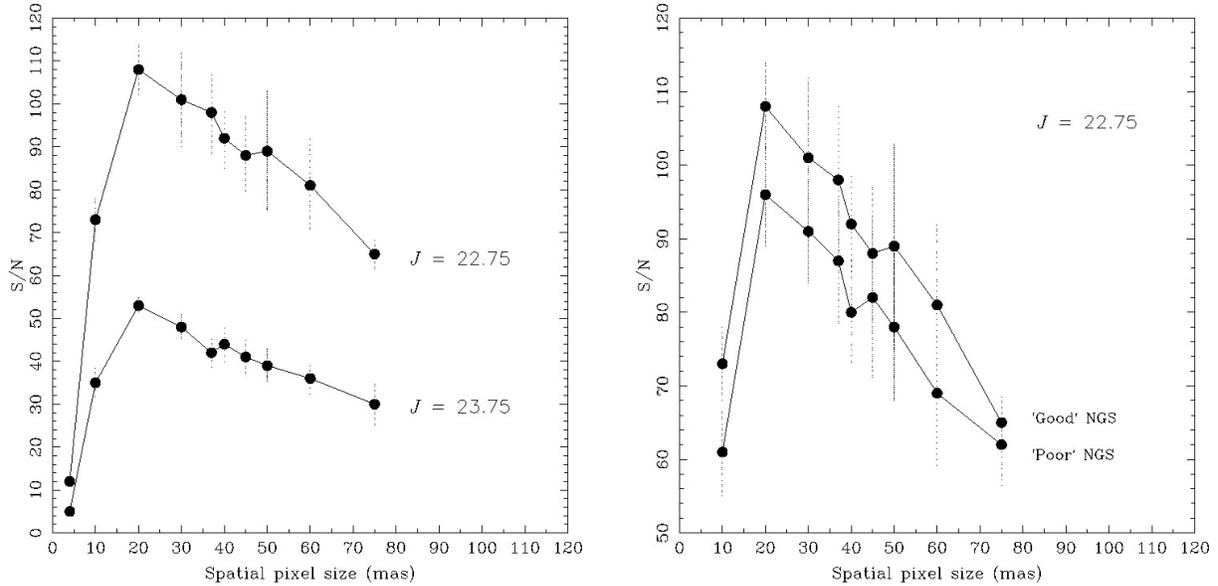

**Figure 19:** Simulated performances (S/N per 2-pixel spectral-resolution element) for individual stars (with MOAO correction). *Left-hand panel:* Illustrative performances for targets with $J$ = 22.75 and 23.75 for MOAO with a 'good' configuration of natural guide stars (NGS). *Right-hand panel:* Comparison of performances for $J$ = 22.75 with 'Good' and 'Poor' NGS configurations (note the different range plotted in S/N cf. the left-hand panel).

**Multiplex**

While the objective is to compile spectroscopy of individual stars, the extended spatial coverage provided by IFUs is an attractive means to obtain large samples. IFUs with fields on the sky of ≥1.0"x1.0" would provide adequate spatial pixels for good background subtraction combined with multiple stars per IFU. Indeed, with potentially tens of stars per IFU, the effective multiplex can be large. We require samples of >>100 stars per galaxy, to adequately sample their different populations and sub-structures (thus requiring multiple E-ELT pointings across each galaxy, e.g. Fig. 18). To assemble ~1,000 stars per target galaxy (within a few pointings) suggests a multiplex in the range of 10 to 20 IFUs. Of course, individual fields in selected galaxies could (and will) be observed with HARMONI, but to sample the full range of structures would entail a Large Programme per galaxy (which is unrealistic given the other demands on such an instrument and the E-ELT overall). An ELT-MOS could address the science goals in a few pointings and within a Large Programme could investigate all of the major Sculptor galaxies (not to mention observations in galaxies further away, e.g. Cen A, M83, etc).

## 5.4. Comparison of J-band and CaT performances

Evans et al. (2011) presented WEBSIM simulations of MOAO observations of cool stars using both the CaT and the *J*-band diagnostics. We have also simulated new CaT observations with coarser spatial sampling from GLAO using a version of WEBSIM for simulations of EVE performances, adopting similar parameters as Evans et al. (2011) to enable comparisons: a total throughput of the telescope of 80%, an instrumental throughput (including the detectors) of 35%, a low read-out noise of 2e$^-$/pixel, and an exposure time of 20×1800s. Other inputs were a spatial sampling of 0.3" (of a 0.9" aperture), $R = 5,000$, and one of the *I*-band GLAO point-spread functions (PSFs) from Neichel et al. (2008). The PSF included a ring of nine NGS at a diameter of 7' – the configuration of guide stars (both natural and laser) will be somewhat different to this, but this PSF serves as a first-order test of the likely performances in the GLAO case (excluding the known limitations in the simulations).

Taking the results from Table 4 of Evans et al. (2011) a S/N ~ 100 is achieved for *J* ~ 22.5, depending on the AO asterism and observing conditions/zenith distance; this sensitivity estimate also takes into account the known limitations in the simulated MOAO performances. From ten GLAO simulation runs to calculate CaT spectra we find a continuum S/N = 21 ± 3 (per pixel) for *I* = 23.5, sufficient to recover the metallicity (cf. Battaglia et al. 2008). The intrinsic *I* − *J* colours for RGB stars are ~0.75 mag so, to first order, the GLAO observations of the CaT and the MOAO observations in the J-band provide sufficient S/N to recover the metallicity of a given RGB star – i.e. they are roughly competitive for a given target (ignoring the effects of extinction and crowding for now).

MOS spectroscopy on the E-ELT will provide the capability to determine stellar metallicities and investigate the history of chemical enrichment out to Mpc distances (for stars near the tip of the RGB, as shown in Fig. 20) and, in the case of the AO-corrected *J*-band observations, out to tens of Mpc for RSGs. The latter case will open-up a huge number of external galaxies, spanning all morphological types, for direct study of *present-day* stellar abundances and calibration of the mass-metallicity relation, as shown in Fig. 21. Although not quite so luminous, we also note that AGB stars share many common spectral features with RSGs, and have longer lifetimes, so this population in external galaxies will also provide complementary estimates of abundances.

We propose that GLAO observations of the CaT will be sufficient for observations in the sparse regions of external galaxies such as those in Sculptor, and that MOAO observations in the *J*-band are well suited for investigation of the main bodies of the target galaxies. Provision of both of these modes is highly complementary, probing different regions and populations (thus providing good sampling of each spatial and kinematic feature). This combination of modes will give a complete view of each galaxy, which is required to confront models of star-formation histories and past interactions. This provides a relatively 'clean' split in the required wavelength coverage for the two modes at ~1μm although, pending more detailed results of the *J*-band methods, extension of the wavelength coverage of the IFUs to 0.8μm is retained as a goal.

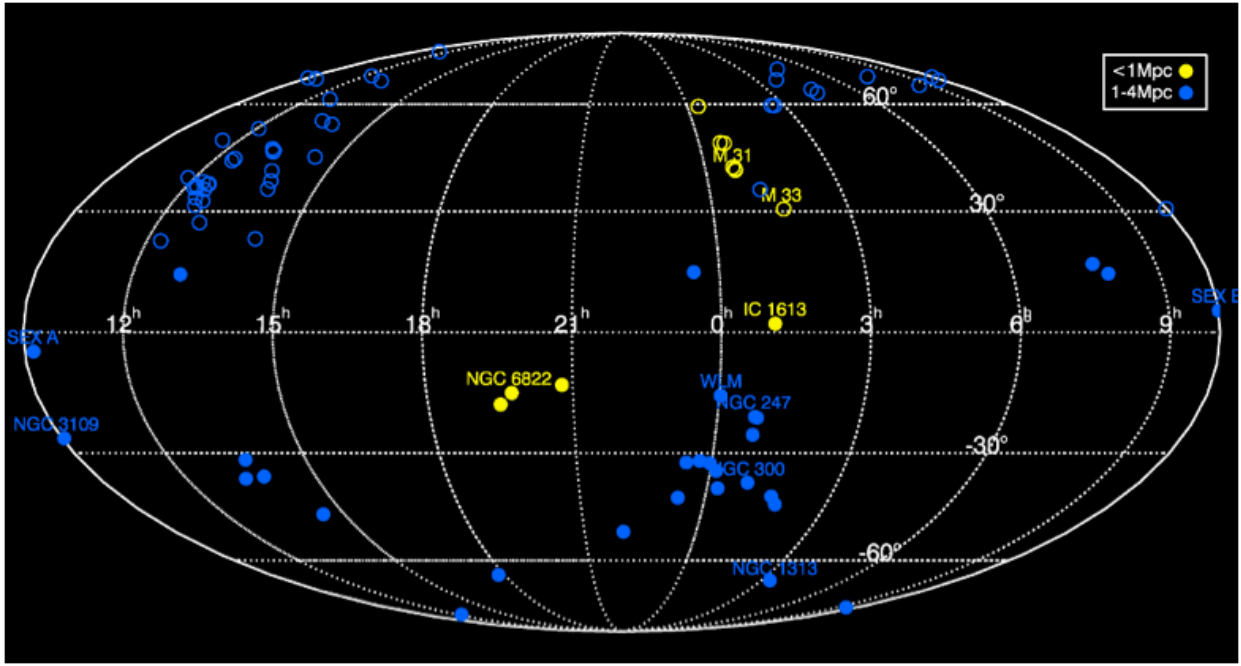

**Figure 20:** MOS spectroscopy of evolved red giants with the ELT will be possible out to distances of several Mpc, opening-up a wide range of external galaxies for direct abundance determinations. Galaxies with δ ≤ 20°, i.e. those observable from Cerro Armazones at reasonable altitudes (≥45°), are indicated by the closed symbols.

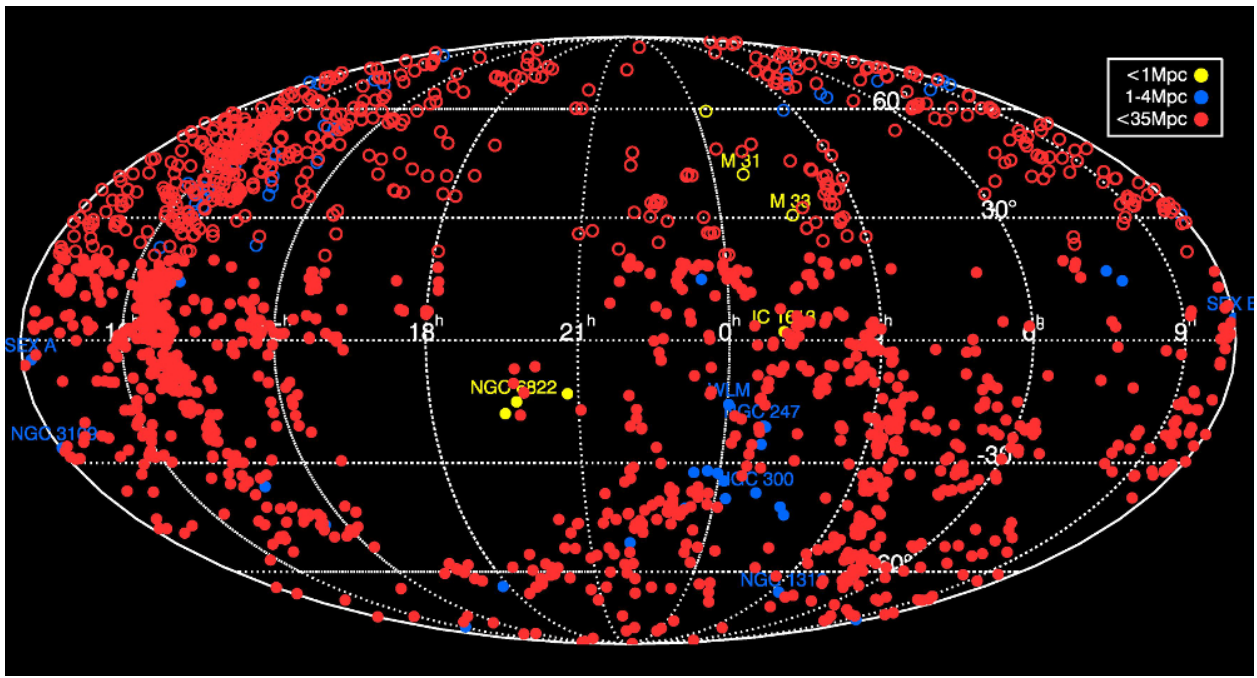

**Figure 21:** MOS spectroscopy of RSGs with the ELT will open-up a huge range of external galaxies for abundance studies. Galaxies with δ ≤ 20°, i.e. observable from Cerro Armazones at reasonable altitudes (≥45°), are indicated by the closed symbols.

## 5.5. Other cases relating to stellar populations

### 5.5.1. Extragalactic star clusters

The E-ELT will open up observations of individual star clusters out to distances of tens of Mpc, bringing a representative sample of galaxy types within reach of detailed stellar population studies. Clusters provide discrete sampling points in space and time, and trace environments ranging from starbursts and galactic disks to ancient spheroidal stellar populations. With internal velocity broadenings of only 5-10 km s$^{-1}$, their integrated light can be observed at far greater spectral resolution than that of galaxies, and detailed abundances can be determined for a wide range of Fe-peak, $\alpha$, neutron capture and light elements. Taking full advantage of this small velocity broadening requires a spectral resolving power of $R \geq 20000$.

With the current generation of 8-10 m class telescopes, kinematic information has been obtained for hundreds of ancient globular clusters around giant elliptical galaxies. For instance, 700 clusters in NGC 1399 (Schuberth et al. 2010), 737 clusters in M87 (Strader et al. 2011), and 563 clusters in NGC 5128 (Woodley et al. 2010). However, only with the E-ELT will it be possible to measure detailed chemical abundances in these clusters and to investigate their relationship with the kinematics. This will provide crucial information about the chemistry of substructure (streams, shells, etc.) identified in the kinematics, that may have arisen from accreted dwarf galaxies or mergers (e.g. Romanowsky et al. 2012).

The richest globular cluster populations are found around massive central elliptical galaxies such as M87 (in Virgo) and NGC 1399 (Fornax). Within a 10 arcmin radius, M87 hosts about 130 globular clusters brighter than $V = 21.5$ ($M_V = -9.5$) and 360 clusters brighter than $V = 22$ (Tamura et al. 2006a, 2006b). With E-ELT spectroscopy it will be possible to reach well below the turnover of the globular cluster luminosity function (at $V \sim 20.5$) in the nearest largest elliptical, NGC 5128; such a depth corresponds to cluster masses of $< 10^5$ $M_\odot$. More than 500 clusters brighter than $V = 21$ will be observable in NGC 5128 (more than 1000 with $V < 22$), distributed over roughly a square degree.

Similarly, young massive star clusters (YMCs) can provide information about stellar populations in the disks of spiral galaxies and other star-forming environments, such as mergers and starbursts (e.g. Larsen et al. 2006; 2008). Typical surface densities of YMCs in spiral galaxies are 0.5-5 kpc$^{-2}$ for $M_V < -8$ (0.2-2 kpc$^{-2}$ for $M_V < -9$). A large spiral with an area of ~100 kpc$^2$ thus hosts 50-500 YMCs with $M_V < -8$ (20-200 with $M_V < -9$). At distances of 5-15 Mpc, where a $M_V = -9$ limit equates to $V = 19.5$-22, most YMCs will be found within a field of radius of 2-5 arcmin. Such clusters will have masses in the of range several $10^4$ $M_\odot$ to about $10^6$ $M_\odot$, with the lower limit depending strongly on age. A nearby example is M83, with >100 YMCs with $M_V < -9$ known from ground-based imaging (Larsen & Richtler 1999).

### 5.5.2. Starburst and nuclear activity in merging galaxies

Somewhat farther away, the nearest major merger NGC4038/39 ('The Antennae') at ~22 Mpc hosts a rich population of young 'super-star clusters' (SSCs), which will provide rich targets for an ELT-MOS. The direct interaction of the two galaxies in the Antennae which produced the SSCs results in intense star formation that is typical of ultra-luminous infrared galaxies (ULIRGs), with star-formation rates of up to ~1000$M_\odot$/yr. These SSCs are thought to play a major role in the structural and chemical evolution of their host galaxies. Their stellar populations are coeval (typically spanning a burst of ~10 Myr), so within a given galaxy they provide us with diagnostics of the fine temporal evolution, over a wide range of stellar mass and luminosity. General laws such as the mass function and the Kennicutt-Schmidt law can be investigated, as well as whether SSCs are the progenitors of globular clusters and how stable they are with time.

Considering a ULIRG at 140 Mpc (the median distance in the catalogue from Haan et al. 2011), a SSC with a diameter of 20pc would have an angular size of ~30 mas. Probing such scales is beyond the capabilities of AO on 8-10m telescopes (or *HST* for that matter), but AO-corrected observations with the E-ELT will open-up studies of SSCs in distant systems. The goal here is not to resolve the SSCs themselves, but to be able to separate clusters in regions where they are concentrated, thus avoiding confusion which can bias the inferred properties towards high mass/luminosity values (e.g. Randriamanakoto et al., 2013)

The aim for E-ELT observations is to determine the extinction-corrected luminosities, stellar contents, ages, and dust and molecular content for SSCs across a sample of ULIRGs. This will provide distributions of properties such as luminosities and mass functions, to investigate if they diverge from star-formation relations in the nearby Universe (see, e.g., Bastian, 2008). We can also use SSCs as tracers of physical properties and merger activity in starburst galaxies. To beat the confusion of SSCs in a close ULIRG we need to be able to resolve distances of ~100pc, which leads to a spatial resolution requirement of better than 400 mas. If we then consider an example case of Arp 299 projected to z=0.1 (d=400 Mpc), this confusion limit corresponds to ~50mas, as shown in Fig. 22.

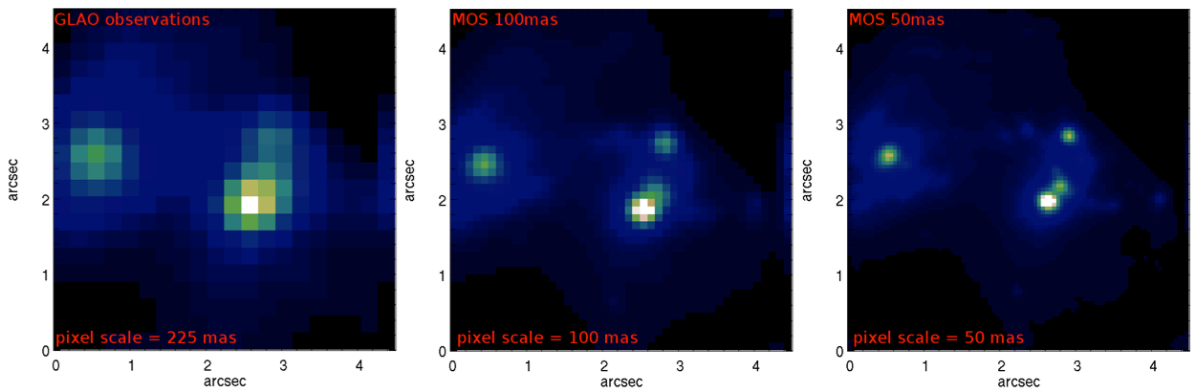

**Figure 22:** Simulated observation of ULIRG at z=0.1 for three different spatial-sampling scales.

The spectral coverage required is 0.6-2.5μm. In particular, *K*-band observations will provide insights into the molecular gas content (via the HII lines, e.g. at 2.12μm), which is an important quantity as it is related to the gas-to-star conversion efficiency and the age of the SSC. The Brγ line (at 2.165μm) will provide an estimate of the number of ionising photons, linked to the number of high-mass (OB-type) stars in the cluster. Lastly, observations of [FeII] in the *H*-band at 1.64μm provides constraints on the SNe rate and age, as well as the stellar mass function. In the (red) optical important diagnostics are provided by the Hα profile (age) and the CaII triplet/Paschen lines (between 0.85-1.0μm), which trace the populations of cool and luminous supergiants, respectively. The spectral resolving power requirements are not particularly demanding ($R$~2000) but, in practise, $R$>5000 is required to minimise blending with the atmospheric OH lines. Assuming a typical SSC luminosity of $H$=17 at the distance of Arp 299, this translates into $H$=23.4 for the same object at z=0.1. Thus, a S/N>10 should be realised by a total integration time of ~15hrs.

### 5.5.3. Testing the invariance of the extragalactic IMF

Recent results have suggested that massive early-type galaxies (ETGs, i.e. ellipticals with large velocity dispersions) may have 'bottom heavy' IMFs, i.e., they contain too many low-mass stars (<1 $M_\odot$) relative to what is seen locally. This evidence has come from dynamical measurements of the mass-to-light ratio of ETGs (e.g., Treu et al. 2010; Cappellari et al. 2012) and via observations of spectral features that are dominated by low-mass, high surface-gravity stars (e.g., Conroy & van Dokkum 2012). MOS observations with an ELT would be ideal to test the latter of these, using diagnostic features in the red-optical or near-IR. This science case may be complimentary to investigations of the IMF with near-IR imaging with MICADO. Specific features of interest include the Na I doublet (8183/8195Å) and the Wing-Ford FeH band (spanning ~9850-10200Å), and only modest spectral resolution is required (~5-8Å is sufficient) but at high S/N (~200). We propose three types of studies:

1. To quantify the radial variation of the spectral features, i.e., if the inferred IMF differs as a function of radius in galaxies. The inner regions of ETGs are thought to have formed in massive bursts of star formation (resulting in a large velocity dispersion), while the outer parts were thought to be accreted later as smaller dwarf galaxies via minor mergers. Quantifying any radial differences in the mass function will be an important constraint on galaxy assembly models (and multiple observations at fixed radii will enable summing of spectra to increase the S/N).

2. Spatially-resolved (AO-corrected) spectroscopy would enable studies at scales of ~20-50 pc in nearby ETGs (e.g., in the Virgo Cluster). We can then construct surface-brightness fluctuation maps on small scales (e.g., Conroy et al. 2013). If the apparent variations are truly caused by differences in the IMF, and not by unknown features in bright red giants (which dominate the total light of ETGs in the optical/near-IR), then the strength of the spectral features should be equivalent in both 'bright' and 'faint' spatial pixels.

3. MOS observations combined with the sensitivity of the E-ELT will allow study of individual globular clusters and the host galaxy at the same time, using the same methods. This will test if the globulars also show IMF variations and will enable comparisons with the population of the host galaxy – conclusively testing whether globulars are drawn from the same parent populations as the stellar content of the host galaxy.

Each of these cases would provide valuable input to models of star formation, as the IMF is the key prediction which is compared to observations – if the IMF were truly demonstrated to vary, it would have a drastic impact on our understanding of galaxy evolution.

### 5.5.4. Cepheids and the extragalactic distance scale

Cepheids are a well-known primary distance indicator as they are used to calibrate the extragalactic distance scale (and, in turn, the Hubble constant $H_0$) through their Period-Luminosity (PL) relations. Several factors prevent us from obtaining an accuracy better than a few percent on $H_0$, in particular the metallicity dependence of the PL relations (Freedman et al. 2012). For PL-relations in the *V*-band no agreement has been reached between models and observations (Romaniello et al. 2008), while the metallicity dependence is probably marginal for PL-relations in the *K*-band (Bono et al. 2010). It seems promising to use Period-Wesenheit (PW) relations instead where the Wesenheit function has been constructed to be reddening-free. The zero-point of the $PW_{VI}$ relation appears to be strongly dependent on metallicity (Storm et al. 2011) but this preliminary result is based on only six Cepheids in the SMC.

There are numerous observations of Galactic Cepheids, but they have a limited range of metallicities and periods ([Fe/H] > –0.5 dex, P < 90 d). [Fe/H] goes down to –0.9 dex for Cepheids in the Magellanic Clouds, but only a few dozens of stars have direct metallicity measurements. This should improve with the next generation of MOS instruments on existing telescopes (e.g. VLT-MOONS) but there are no direct metallicity measurements for the Cepheids discovered in external galaxies at greater distances. Their metallicity is therefore determined from emission lines of HII regions or planetary nebulae, assuming that they have the same metallicity at a given galactocentric distance and a linear correlation between [O/H] and [Fe/H].

Only an ELT-MOS will have the sensitivity required to determine the metallicity of individual Cepheids in each galaxy where they have been detected, particularly in the young, metal-poor systems that are mostly beyond the grasp of current and future facilities. While current methods require high-resolution in the visible, it is likely (but not firmly established yet) that the metallicity of Cepheids can also be determined from low-resolution observations in the *J*-Band, as for RSGs (see Sect. 5.3; Davies et al. 2010; Evans et al. 2011).

### 5.5.5. Structural components in external galaxies

With the E-ELT we will be able to extend our observations out to even larger distances to address questions relating to the complex structures of galaxies. Resolved low-z disk galaxies display a wealth of substructures, such as different bulge families (classical, disk-like, box/peanut, e.g., Athanassoula 2005; Gadotti 2009), inner disks, spiral arms, bars, rings, lenses and tidal features (e.g., Laurikainen et al. 2005; Gadotti 2008). All such structures hold clues to the formation and evolution of galaxies. To extract this information one needs kinematical, chemical and structural properties from the different components (both their stellar and gas constituents), requiring IFU capabilities.

A desirable patrol field, given the angular size of potential target galaxies, is at least 5×5 arcmin. The relatively large angular size of the galaxies also implies multiplex capabilities are required, i.e. a few tens of IFUs (covering several arcsec$^2$ each) that can be distributed within the field covered by the galaxy, targeting particular points of interest.

As the structural components can be compact and blended together in an apparently seamless fashion, high spatial-resolution such as provided by GLAO is a minimum requirement. This also means that the IFUs must be able to be placed close to each other when necessary, with minimum separations of no more than a few arcsec. In this context, the availability of simultaneous GLAO and MOAO-IFU observations would also be beneficial. The image quality from GLAO correction will be sufficient in less dense regions, where structural information can be derived from other imaging facilities. In more crowded regions, where different substructures overlap, MOAO-corrected IFU observations will be required.

However, the true major leap forward in this field is in disentangling the stellar populations of the different components. Studies of resolved stellar populations in the central regions of the Milky Way have provided a wealth of information and have revealed the great complexity in the formation and evolution of the central components in our own galaxy. The goal is to attain a comparable understanding in the information available for external galaxies, hence the requirement for high spatial resolution from MOAO.

The primary diagnostic feature of note is the CaT, but given that many of the target substructures are in the central regions of galaxies, coverage into the near-IR (at least to the *J*-band) will minimise the impact of dust absorption. The range in kinematical properties – such as the stellar velocity dispersion between the different components – is often only a few tens of km/s, therefore requiring a spectral resolving power of $R$~10,000-20,000 (e.g., Gadotti & de Souza 2005).

# GALAXY ARCHAEOLOGY

The Universe that emerged from the hot and dense phase after the Big Bang had an extremely simple chemical composition: stable isotopes of hydrogen and helium, with traces of $^7$Li. As discussed in SC1, the source of the photons which reionised the early Universe is still unclear. A leading contender is thought to be the ionising UV photons from the first stars which formed from the primordial gas (so-called 'Population III' stars). A consequence of the extremely low metallicity is that this first generation of stars is expected to be comprised of very massive stars (e.g. Bromm & Larson, 2004) – metallic species are thought to play an essential role in cooling while the natal gas undergoes collapse and, without these vital cooling agents, fragmentation of the gas is probably inhibited, leading to a unique and relatively massive first generation of stars. The nuclear reactions at work in the stellar interiors of this first generation of stars will then start manufacturing metals, with the result that the pristine ISM will be 'chemically polluted' within a few stellar generations as the massive stars explode as supernovae.

This theoretical paradigm requires observational confirmation, e.g., searching for emission features from (rest-frame) He II 1640Å as mentioned in SC1. In particular, we want to know the form of the IMF for this first generation of stars, i.e. what mass range it spanned?  While we cannot constrain the massive end directly, by comparing chemical abundances for the low-mass survivors from this era with the yields predicted by models with variable IMFs, we can investigate if there is evidence for a `top heavy' IMF, i.e. were all the first stars very massive, or were lower-mass stars also formed from the primordial gas? Moreover, recent numerical simulations suggest that the distribution of possible stellar masses in this epoch may be much broader than previously thought, extending down to $\leq 1 M_\odot$ (e.g., Clark et al. 2011; Greif et al. 2011). If stars with $M \leq \sim 0.8 M_\odot$ were actually formed a few hundred Myr after the Big Bang, they should still be shining today as their lifetime is larger than the Hubble time.

We can constrain the issues of star-formation in very metal-poor environments by analysing the metallicity distribution functions (MDFs) of stellar populations in the local Universe. One of the best tracers to determine these MDFs are stars at the main-sequence turn-off (MSTO), which are: numerous and relatively uncontaminated in terms of other populations, easy to select on the basis of their colours, and are the brightest population among the long-lived, chemically-unmixed stars, i.e. their chemical abundances are relatively unaltered since their formation. To advance this field we need the sensitivity of the E-ELT, combined with both a large multiplex (to compile large samples of stars) and sufficient spectral resolving power for detailed chemical analysis ($R \geq 20{,}000$). The relative proximity of the targets means that the GLAO correction provided by the E-ELT is generally sufficient in terms of the required angular resolution.

## 6.1. Metal-poor stars

### 6.1.1. The Galaxy

Significant effort has been invested over the past few decades in searching for primordial stars in the Galaxy. These stars are the long-lived descendants from the earliest stellar generations, and will have formed from a (near-)pristine ISM, which would have only been weakly enriched in metals from the first supernovae. Their atmospheres therefore give us a fossil record of the ISM from which they were formed, corresponding to redshifts of $z \geq 10$. Having a direct tracer of chemical abundances at such an early time can provide fundamental constraints on the properties of the first generations of stars, as well as giving indirect information on their masses and ionising feedback (of interest in the context of reionisation).

Until recently, the most metal-poor stars known were four giants and one turn-off binary star with $-4.1 \leq$ [Fe/H] $\leq -3.7$ (Norris et al. 2000; François et al. 2003; Cayrel et al. 2004; Gonzalez Hernandez et al. 2008), and the deepest survey searching for metal-poor stars was the Hamburg-ESO survey (HES, Christlieb et al. 2008), going down to $V = 16$. The MDF derived from the HES has a vertical drop at [M/H] $= -3.5$ (Schörck et al. 2009), in excellent agreement with predictions that a 'critical metallicity' of [M/H] $= -3.5$ is required for the formation of low-mass stars. Above this value, cooling from the fine-structure lines of ionised carbon and neutral oxygen is thought to be sufficient for the formation of low-mass stars (e.g. Bromm & Loeb, 2003; Frebel, Johnson & Bromm, 2007). The discovery of metal-poor stars with strong C and O enhancements provided support to this theory (Christlieb et al. 2002; Frebel et al. 2005; Norris et al. 2007), but the discovery that the extremely metal-poor star SDSS J102915+172927 is not strongly enhanced in C or O has challenged this model (Caffau et al. 2011). Indeed, an alternative scenario predicts that dust cooling plays the dominant role and argues for a lower critical metallicity of [M/H] $\sim -5$ (e.g., Schneider et al. 2006; Dopcke et al. 2011).

Work is now underway to improve our understanding of the low-metallicity tail of the MDF – major progress can be expected in the coming years using data from, e.g., the Sloan Digital Sky Survey (SDSS, York et al. 2000) and the SkyMapper telescope (Keller et al. 2007), as evidenced by the dramatic recent finding of a star with [Fe/H] $< -7.1$ (Keller et al. 2014) and new samples of very metal-poor stars (e.g. Norris et al. 2013; Cohen et al. 2013; Roederer et al. 2014).

### 6.1.2. The Galactic bulge

The stellar populations of the Galactic bulge are a template for studies of ellipticals and bulges of spirals. The formation history of the bulge can give hints on proto-galaxy counterparts observed at high-redshift, providing strong motivation for the detailed study of this component of our Galaxy. The formation of our bulge is still a controversial issue, and it is probably a combination of a pseudo-bulge population, and an old, spheroidal, true bulge. Studies of the chemical compositions and kinematics of its stellar populations will be the key to disentangle its formation mechanism(s).

Due to fast chemical enrichment at early times, the lower end of the MDF is as high as [Fe/H]= −1.1 (Hill et al. 2011), with small numbers of stars with lower metallicities (González et al. 2013). This moderately metal-poor stellar population is probably the oldest component of our Galaxy and its kinematics are compatible with that of an old spheroid, whereas the more metal-rich population is compatible with a bar, suggesting differing formation scenarios (Babusiaux et al. 2010). Current observations are limited to giant stars, but observations of dwarfs will be important to disentangle the complex mix of stellar populations in the bulge. A few tens of dwarfs have been studied so far in the bulge via microlensing techniques which, while providing some first interesting results, are limited to small numbers of stars (e.g. Bensby et al. 2013).

Typical MSTO magnitudes of stars in the bulge are: $V_{MSTO}$ = 19.5 for Baade's Window, $V_{MSTO}$ = 20.8 for the metal-rich globular cluster NGC 6528 (Ortolani et al. 1995), and $V_{MSTO}$ = 20.4 for the metal-poor globular NGC 6522 (Piotto et al. 2002; Barbuy et al. 2009). A few clusters are brighter than these examples, but the majority are fainter.

VLT-FLAMES spectroscopy at $R$~20,000 in the optical can reach $V$ = 17 in about 2 hrs for a reasonable S/N. In the near-IR it is possible to reach two-to-three magnitudes deeper profiting from the fact that the SEDs of these stars peak at longer wavelengths. However, the near-IR lines available do not include some key elements: lines of FeI and FeII of varied excitation potential are only found in the optical, as well as the lithium line, and many lines from heavy elements. It is interesting to note that Bensby et al. (2010) observed lithium in a metal-poor Bulge dwarf (also brightened by a microlensing event), finding it to be compatible with the abundance of the Spite plateau in the Galactic halo. It is difficult to draw strong conclusions from this result, yet the discovery of a Spite plateau (see next section) in the metal-poor bulge population may have far-reaching cosmological implications.

Only an ELT-MOS will be able to obtain spectroscopy of dwarf and turn-off stars in the Bulge, allowing us to compile statistically-significant samples to address specific problems such as the existence of the Spite plateau and/or the evolution of the heavy elements in this important Galactic component.

### 6.1.3. The galaxies of the Local Group

It is important to know if the star-formation process in other galaxies resembles that in the Galaxy (see review by Tolstoy, Hill & Tosi, 2009). Interestingly, the MDFs of the dwarf spheroidal galaxies obtained to date are significantly different from that in the Galaxy (see Fig. 23); the metal-poor tail of the Galactic MDF is significantly more populated than in the dwarf spheroidals, but it is not clear why this is the case.

The MDFs available for nearby external galaxies have been obtained from [Fe/H] measurements in giant stars, generally employing the CaT calibration (e.g., Battaglia et al. 2006, Helmi et al. 2006). The improved CaT calibration from Starkenburg et al. (2010) is valid over the range −4 ≤ [Fe/H] ≤ −0.5, and indicated that the numbers of extremely metal-poor stars in these external galaxies were actually not as low as thought previously. This was confirmed by precise measurements from high-resolution VLT-UVES observations of five metal-poor candidates in three dwarf spheroids (Fornax, Sculptor, and Sextans), finding [Fe/H] ≤ −3 in each case (Tafelmeyer et al. 2010).

Another interesting aspect of this field is related to the ground-breaking discovery by Spite & Spite (1982) that Galactic metal-poor stars at the MSTO have a constant Li abundance irrespective of their metallicity or effective temperature (the so-called 'Spite Plateau'). The most straightforward interpretation of this result is that the observed Li is primordial, i.e. produced during the first three minutes of the Universe. As mentioned earlier, in those earliest moments only nuclei of deuterium, two stable He isotopes ($^3$He and $^4$He), and $^7$Li were synthesized. Their abundances depend on the baryon-to-photon ratio, thus on the baryonic density of the Universe. In principle, the Spite Plateau allows us to determine the baryon-to-photon ratio, which can not be deduced from first principles.

This interpretation of the Spite Plateau is seriously challenged by the measurement of the baryonic density, with unprecedented precision, from the fluctuations of the cosmic-microwave background by WMAP (Spergel et al. 2007). However, we have indications that the plateau is universal. For instance, Monaco et al. (2010) observed the Spite Plateau in ω Cen, which is generally considered to be the nucleus of a disrupted satellite galaxy. It is of paramount importance to verify if this is the case in other Local Group galaxies, which possess metal-poor populations, but spectroscopy of extragalactic stars at the MSTO requires the sensitivity of the E-ELT.

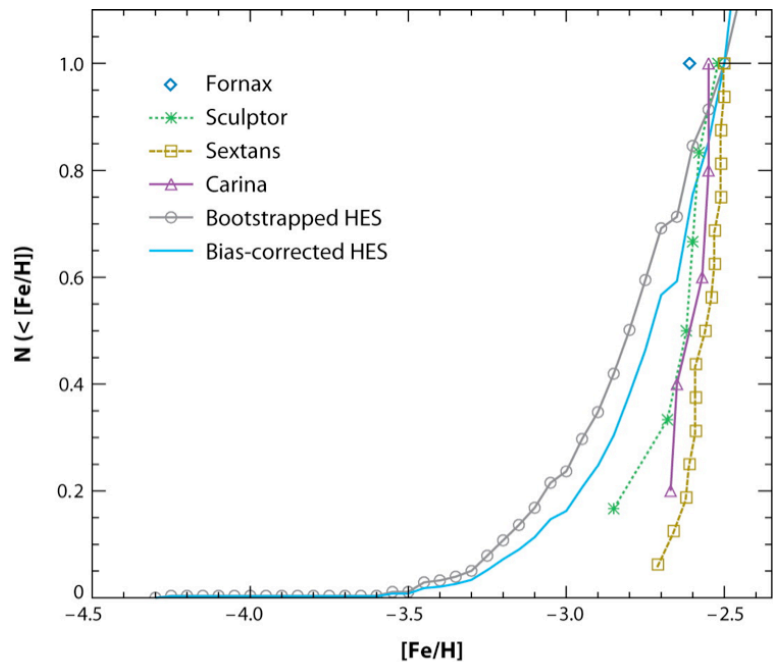

Figure 23: Galactic metallicity-distribution functions (MDFs) from the Hamburg-ESO Survey (HES; bias-corrected results are from Schőrck et al. 2009) compared to MDFs for four dwarf spheroidals from the DART survey (Helmi et al. 2006); this is Fig. 10 from Tolstoy, Hill & Tosi (2009). The figure changes somewhat using the CaT recalibration from Starkenburg et al. (2010), but significant differences remain between the Galaxy and its dwarf satellites.

### 6.1.4. The ultra-faint dwarf galaxies: fossils of the early Universe

The ultra-faint dwarf galaxies (UFDs) that have been discovered recently around the Milky Way are the smallest, faintest, most metal-poor and dark matter (DM) dominated galaxies that we know of. Their properties point to some of them having concluded their star formation and chemical evolution much earlier in the history of the Universe than the other galaxy types, and they are now essentially fossils, perhaps of the reionization epoch (e.g., see the results from the *HST* program by Brown et al. 2012). The existence of such extreme objects has raised several fundamental questions, including forcing us to rethink the concept of what a 'galaxy' is, and what distinguishes it from a stellar cluster (e.g. Willman & Strader, 2012).

With their enormous dynamical mass-to-light ratios of up to 1000s $(M/L)_\odot$, e.g. Simon & Geha (2007), UFDs are potentially the best targets for deciphering the nature of DM by detecting the signal of annihilation/decay of candidate particles (e.g. Essig et al. 2009) and are also excellent places to test for the possible presence of cusps in the DM density profile, as predicted by $\Lambda$CDM. Interestingly, these very low luminosity dwarfs appear to display similar DM masses in their central regions as the brighter 'classical' dwarf galaxies (e.g. Strigari et al. 2007), raising the question of whether there is a mass threshold for galaxy formation (i.e., a common mass scale). As the most metal-poor galactic systems (<[Fe/H]> down to –2.8, see McConnachie 2012) the UFDs are also natural probes of chemical enrichment in the early Universe and provide us with targets in which we can maximize our chances to find the most pristine stars (e.g. Frebel & Bromm, 2012; Frebel, Simon & Kirby 2014).

Because of their intrinsic faintness, there are simply not enough stars in the UFDs which are sufficiently bright to be followed-up with intermediate-resolution spectroscopy using 8-10m class telescopes. Therefore our best estimates of their DM content and determinations of their metallicity properties (e.g. MDF) are derived from only a handful of stars. Not only is the DM content of UFDs highly uncertain, but determination of the DM density profile cannot even be attempted applying the most precise techniques available (see Breddels et al. 2013; Breddels & Helmi 2013), as this would require accurate line-of-sight velocities for of the order of 500-1000 stars.

## 6.2. Goals for the E-ELT

One the key questions in the Local Group is whether the galaxies were formed from the early primordial gas, or if they formed from gas that had already been partially enriched, thus providing a metallicity floor. This has important cosmological implications. In the standard hierarchical scenario the first structures to form are dwarf galaxies which subsequently merge to form larger structures like the Galaxy and other massive disk galaxies. If the MDFs of Local Group dwarfs display a clear metallicity floor, then either the hierarchical galaxy formation model is wrong, or the present-day 'surviving' dwarf galaxies formed later (in which scenario, they are not the relics of the primordial dwarfs). In this sense, the Local Group poses one of the most stringent tests of the hierarchical structure formation paradigm of cosmology.

The current limiting factor in extragalactic studies is that only giant stars are bright enough to have high-quality, high-resolution spectroscopy with an 8-m class telescope (e.g. giant-branch stars in 'nearby' dwarf spheroidals have $V \leq 18$). Unfortunately, the statistics available from the analysis of extragalactic RGB stars is not sufficient to determine the metal-poor tail of the MDF robustly in their host galaxies. If the fraction of stars below [Fe/H] ~ –4 is $10^{-4}$, then at least $10^4$ stars need to be observed to find just one. There are simply not enough giant-branch stars in most of the Local Group dwarf galaxies to sample these rare populations, and the MDFs determined to date are likely to be severely incomplete in their metal-poor tails. As mentioned earlier, the best stars to obtain a well-sampled MDF are those at the MSTO, and to observe these faint stars in extragalactic systems we need the sensitivity of the E-ELT.

We therefore require large samples of stars at the MSTO, in multiple galaxies (to further improve the statistics). The *V*-band magnitudes of stars at the MSTO in nearby galaxies are summarised in Table 5; thus we require high-resolution, absorption-line spectroscopy down to *V*~25. An additional and important piece of information will be carbon abundances of these metal-poor populations – if they are carbon-enhanced this would indicate that the star-formation paradigm advocated by Bromm & Loeb (2004) is dominating the process, whereas (mainly) carbon-normal abundances would favour star formation driven by dust cooling and fragmentation. Carbon abundances can be determined for MSTO stars from observations (at $R \geq 10,000$) of the G-band feature (at 0.43 μm).

Similarly for the UFDs, we require intermediate-resolution spectroscopy of stars as faint as *I*~25. The sensitivity of the E-ELT, combined with a MOS, would enable us to compile samples of 150-2000 stars for each of the currently known UFDs, with distance moduli of 17-21 mag and *V*-band surface brightnesses of ~28-29 mag./arcsec$^2$. Our understanding of all of the known systems would be substantially advanced by determination of metallicity properties while, for systems where ~1000 or more stars are within reach, we will be able to provide accurate determinations of the DM content with which to confront theoretical predictions.

Table 5: *V*-band magnitudes of the main-sequence turn-off (MSTO) in illustrative nearby southern galaxies.

| Galaxy | $V_{MSTO}$ (mag) |
|---|---|
| **SAGITTARIUS DWARF** | 22.4 |
| **LMC** | 23.9 |
| **SMC** | 24.5 |
| **SCULPTOR DWARF** | 24.8 |
| **SEXTANS DWARF** | 24.8 |
| **CARINA DWARF** | 25.0 |
| **FORNAX DWARF** | 25.7 |

## 6.3. Instrument requirements

Observations are required of stars at the MSTO in Local Group galaxies down to *V*~25. Effective temperatures of each target will be derived from the photometry used to select candidates, as well as from fits to the wings of the Hα line at 0.656 μm (giving sufficient precision of ~ ±200 K). Stellar gravities will also be derived from the photometry, but can be further constrained from spectroscopy.

**Spectral resolving power**

High-resolution spectroscopy is required at $R \geq 20{,}000$ (optimal), $R \geq 15{,}000$ (essential) to determine accurate chemical abundances/metallicities, but only selected wavelength regions are required. In the UFDs a resolving power of $R > 5{,}000$ (over the 0.8-0.9 μm region) will provide sufficient precision for line-of-sight velocities (to ±3-4 km/s) to compare with dynamical models. With current spectral synthesis techniques a resolving power of $R \geq 6{,}500$ will enable metallicity estimates for these very metal-poor stars.

**Wavelength coverage**

The optimum optical regions for high-resolution spectroscopy are well understood from the extensive work already completed with, e.g., VLT-FLAMES and VLT-UVES. Abundance estimates are required for: Fe, α-elements (e.g. Mg, Ca), C (from the G-band), N (from the A-X CN band), Li, Ba, Eu, and Sr. Thus, the optimal wavelength ranges required are 0.38-0.52 μm and 0.64-0.676 μm, with an essential requirement on the bluewards range of 0.41-0.46 μm. If these are obtained simultaneously then that would save by a factor of two in total observing time. In the UFDs, coverage of 0.65-0.90 μm will suffice for the diagnostic lines required for metallicity estimates.

Science case 7

# GALACTIC CENTRE SCIENCE

One of the most compelling astrophysical laboratories available to us is the central parsec of our own Galaxy, providing us with a unique view of the processes at the heart of a massive galaxy. Indeed, one of the most spectacular results of the past decade was arguably the observed orbits of stars around Sgr A*, the central black-hole (Ghez et al. 2005; Eisenhauer et al. 2005). Studies of these 'S' stars in the central arcsecond (~0.04 pc) have since been used to provide the most accurate estimates of the distance to the GC to date (8 kpc) and the mass of the central compact object ($4\times10^6$ $M_\odot$). However, studies in this region remain challenging even with 8-m class facilities, primarily because the line-of-sight extinction is so high ($A_V = 30$, $A_K = 3$). We now briefly consider the contribution of the E-ELT to GC studies, focussing on cases where we are currently limited by sensitivity rather than spatial resolution.

## 7.1. The parsec-scale central stellar cluster

Beyond the central arcsecond there are ~100 known massive (OB-/WR-type) stars in the inner 0.5 pc, with past debate as to whether they are structured in one or two discs (see, e.g., Lu et al. 2010 cf. Paumard et al. 2006). Two scenarios have been proposed for the formation of these disc(s), either in-situ star formation (in an accretion disk orbiting Sgr A*) or formation in an offset massive star-cluster that later migrated onto Sgr A* and dissolved in the central parsec. Observations of fainter (lower-mass) stars would provide a larger sample with which to investigate this important cluster but require the sensitivity of the E-ELT. HARMONI observations will provide spectroscopy of the central regions with a relatively small number of pointings, but observations of the stellar populations at larger distances from the centre will be a vital complement to our understanding of the region. Such observations further from the core are particularly important as the two models predict different surface density and mass profiles at larger radii. Deep ELT-MOS observations of regions in the central arcmininute would provide unique constraints on these models.

## 7.2. Gas content of the regions around Sgr A*

Moving further out, the torus-like Circumnuclear Disk (CND, with inner and outer radii of ~2 and 7 pc, respectively) is also of interest. It is relatively clumpy in appearance and has been suggested as a potential stellar nursery; interestingly it has a significant but low inclination to the Galactic plane (~10°) and is somewhat reminiscent of an accretion disk. Although it may be currently quiescent in terms of star formation, the CND is thought to be an important ingredient of the GC region – it may later evolve into a real accretion disc, unleashing an AGN-like event in the Milky Way and forming a new disk of stars. Several teams are currently looking for star-formation sites or self-gravitating nodes within the CND, and resolving the dynamics of its structures would provide vital understanding of its properties and likely future.

The hole within the CND is occupied by an ensemble of gas and dust clouds (Sgr A West), observable in ionised lines (e.g., H I, He I) and which, when viewed together, look like a three-arm spiral. This 'Minispiral' is actually a 3-d structure comprised of dusty, thick and extended clouds which each orbit Sgr A* in their own planes. The northern arm of Sgr A West has been suggested as a clump of the CND which has fallen inwards after a shock of some sort (perhaps following the supernova explosion that created the nearby Sgr A East shell). The advantage of near-IR compared to radio observations is that it enables detections of stars and molecular (H2) and ionised (HI) gas in the same data, from which we can study their respective interactions. Spatially-extended spectroscopy from IFUs would provide the required mapping of selected pointings in the CND, and an ELT-MOS programme could potentially observe the large majority of the neutral-ionised region (as illustrated in Fig. 24). Assuming a field of ~2"x2" for each IFU, this would require >>100 pointings with HARMONI to map the CND but, depending on the multiplex, MOS observations would be significantly faster.

## 7.3. Stellar populations in the inner Galaxy

The stellar populations beyond the central region shown in Fig. 24 comprise massive star clusters (e.g. the Arches and Quintuplet clusters), apparently isolated massive-stars, other gas clouds, and X-ray sources (Mauerhan et al. 2010), which will all benefit from spectroscopic follow-up from ELT-MOS programmes. For example, high signal-to-noise observations of the massive stars will provide important insights into their physical properties, binarity, and dynamics. Moreover, we note that the *Chandra* observations from Mauerhan et al. detected >6000 X-ray sources; a thorough and efficient census of these sources to identify their natures would require a high-multiplex mode of an ELT-MOS.

Current predictions (e.g., Yusef-Zadeh et al. 2009) suggest that the presence of a bar-like structure is crucial in order to sustain the high star-formation rate seen in the Nuclear Bulge. It has been proposed that such a structure would drag gas from the inner disk into the Nuclear Bulge (Athanassoula 1992; Kim et al. 2011). However, the young populations such as LBVs in the Quintuplet cluster (Najarro et al. 2009), nitrogen-rich W-R stars in the Arches cluster (Martins et al. 2008), and red supergiants in the local field population (Davies et al. 2009a) all appear to have solar iron abundances. Moreover, red supergiant clusters at the near-end of the Galactic Bar have sub-solar abundances (Davies et al. 2009b). In contrast, Cepheids in the inner disk have supersolar abundances (by up to +0.5 dex, e.g., Genovali et al. 2013). As these tracers have similar ages (of a few to a few tens of Myrs), but very different metallicities, the mechanism(s) feeding the Bar and the Nuclear Bulge with gas could well be more complex than currently assumed.

## 7.4. The innermost region

Spectroscopy of the central Sgr A* region is important, both for characterising the stellar orbits of the central 'S' stars, and for monitoring of flares from the central source (during which its brightness can increase by up to several magnitudes in the *K*-band). HARMONI and MICADO will likely be the ELT instruments of choice for spectroscopic/astrometric studies of the 'S' stars, but monitoring of the activity of Sgr A* could be provided by parallel observations with an ELT-MOS while undertaking the observations discussed above.

## 7.5. Instrument requirements

ELT-MOS studies of the stellar populations and gas dynamics beyond the innermost region around Sgr A* place some general requirements on the instrument:

### Spectral resolving power

The essential requirement is $R \geq 5{,}000$, with a goal of $R \sim 10{,}000$ (to enable detailed quantitative analysis of the massive stars in the region).

### Wavelength coverage

Coverage of the *H*- and *K*-bands is essential for the relevant diagnostic lines in the near-IR, as well as to penetrate the significant line-of-sight extinction toward the GC. Specifically, the *K*-band is much richer than the *H*-band in diagnostic lines, and includes lines to characterise massive stars (ionised H, He, N, C), cool stars (CO bandheads), and both ionised (H, He, Fe) and molecular gas (H2). (Indeed, inclusion of the *K*-band would also enable studies of the embedded young stellar populations in star-forming regions in other parts of the Galaxy, the Magellanic Clouds and beyond.)

### Target configurations

The desire to map the CND (and the stellar populations to some extent) requires IFU spectroscopy. The IFUs should cover the largest possible total field-of-view and should also be able to be closely clustered (i.e. enable contiguous mapping with a small number of dither positions).

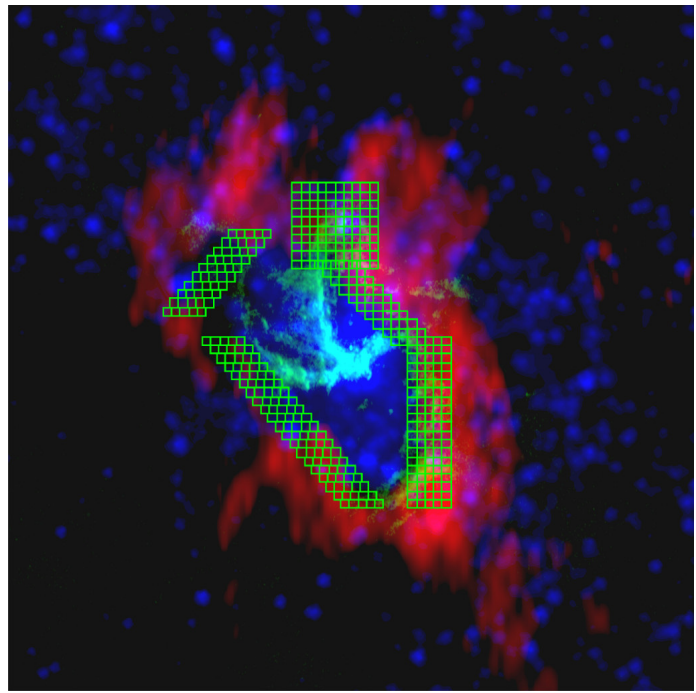

**Figure 24:** Example ELT-MOS fields to observe the interface between the circumnuclear disk and the minispiral near the Galactic Centre. Each IFU footprint is 2"x2". Image courtesy of NRAO/AUI (Credit: F. Yusef-Zadeh, M. Wright, S. Stolovy).

Science case 8# PLANET FORMATION IN DIFFERENT ENVIRONMENTS

The complicated process of planet formation is still not well understood. The two most commonly discussed processes are core-accretion and disk-instability scenarios. In the first, a solid core forms over a time-scale of at least 1 Myr. If the core is massive enough (5-10 $M_{Earth}$), hydrogen and helium is accreted from the disk, ultimately leading to the formation of a gas giant planet. Alternatively, planets may also form directly via an instability of the disk; this process is very fast but requires a massive disk to work. The task ahead is to understand which physical parameters are most relevant to the formation process(es). For example, if planets form quickly via disk instability, the mass ratio of the star to the disk would be important, but the lifetime of the disk would be less relevant. In contrast, the lifetime of the disk and the abundance of heavy elements in the disk are both important for the core-accretion scenario.

Observationally, the probability for a star to have a massive planet appears to depend on both the mass and metallicity of the star. The correlation with metallicity has been interpreted as evidence for the core-accretion scenario. Simplistically, the cores consist of heavy elements and so larger abundances in disks should make it easier to form a core. However, recent ALMA observations of proto-planetary disks have shown that they are complicated structures with local density enhancements and regions with varying sizes of dust grains (e.g. van der Marel et al. 2013). Additionally, VLTI observations have shown that the outer regions of disks have equal amounts of pyroxene and olivine, while the inner regions are dominated by olivine (van Boekel et al. 2004); such variations likely play a role in the formation of the cores. Thus, not only the global properties of disks but also their detailed structure are thought to be crucial factors in planet formation.

Other external factors are also thought to contribute to planet formation. For instance, the majority of stars are thought to form in stellar clusters, and studies have shown that cluster environments can have strong effects on the evolution of proto-planetary disks and on the planets after formation. There are two important environmental effects in this regard:

- The intensive X-ray and extreme UV (EUV) radiation from hot stars in the clusters;
- Close (dynamical) encounters with stars.

## 8.1. The effects of EUV radiation in clusters on the disk

The EUV radiation in clusters will usually be dominated by the radiation from the most massive star of the cluster (a consequence of the steep gradient in the initial mass function). Notably, the UV radiation from the most massive star impinging on circumstellar disks can often dominate over the radiation fields produced by their respective central stars (Holden et al. 2010). This EUV radiation can photo-evaporate the disks, i.e., the most massive star of the cluster can strongly influence planet formation in the region (see left-hand panel of Fig. 25, from Armitage, 2000), with predicted disk lifetimes in clusters with >300 stars of ≤$10^6$ yrs (right-hand panel of Fig. 25).

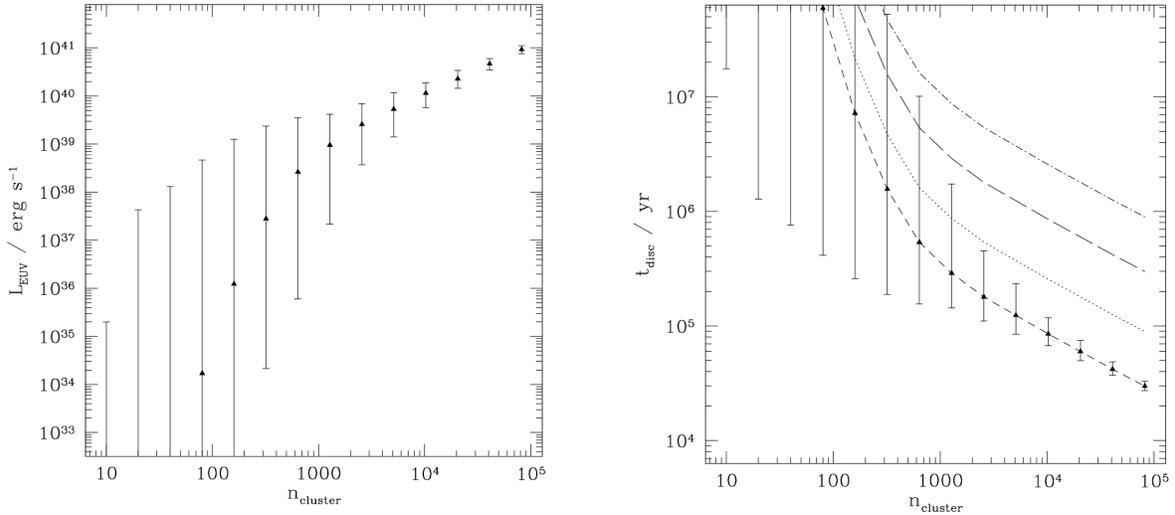

Figure 25: *Left:* Median-integrated EUV luminosities for clusters with $n_{cluster}$ stars; error bars show the range enclosing 90% of the distribution of luminosities. *Right:* Disk lifetimes as a function of $n_{cluster}$ stars. (Both from Armitage, 2000.)

Going one step further, from consideration of nuclear star-clusters, starburst clusters in M83, compact high-z galaxies, and some Galactic globular clusters, Thompson (2013) concluded that, during formation, their temperatures likely exceeded the ice-line temperature ($T_{Ice}$ ~150-170 K) for a timescale comparable to that for planet formation, i.e. giant planets should not form in these systems. Thus, Thompson suggested planet-search programmes in NGC 6366, 6440, 6441, and 6388 to test this hypothesis. If stellar density is more important than metallicity, the globular cluster NGC 6366 ([Fe/H] = −0.82) should yield detections, whereas the relatively metal-rich globular clusters NGC 6440, 6441, and 6388 should be devoid of giant planets because of their high stellar densities.

### 8.1.1. Does a dense environment help rather than inhibit planet formation?

It is thought that ~50% of stars form in clusters/associations with more than ~1750 stars, so we might also expect that 50% of planet-hosting stars should arise from such clusters (e.g. the Orion cluster in Fig. 26). However, the disk lifetimes in such clusters are only $10^5$-$10^6$ yrs (e.g. Fatuzzo & Adams 2008, and right-hand panel of Fig. 25). One could hypothesize that all planet-hosting stars have formed in small clusters, but the conclusion that the Sun formed in a cluster which contained ~1200 stars argues otherwise (Gounelle & Meynet, 2012; Pfalzner et al. 2013). Equally, we do not see evidence for low-mass stars forming first in clusters, i.e. the massive stars forming after planet formation – for example, the small age spread for stars in the Scorpius-Centaurus OB assocation (Preibisch & Zinnecker, 2007).

Thus, it is something of a mystery how planets can form in such an environment. However, Throop & Bally (2005) and Mitchell & Stewart (2010) have pointed out that photo-evaporation of the disk may induce rapid formation of planetesimals rather than inhibiting planet formation. In this scenario, the UV radiation of nearby stars may trigger planet formation because an increasing dust-to-gas ratio arises in the shielded interior, such that the dust becomes gravitationally unstable.

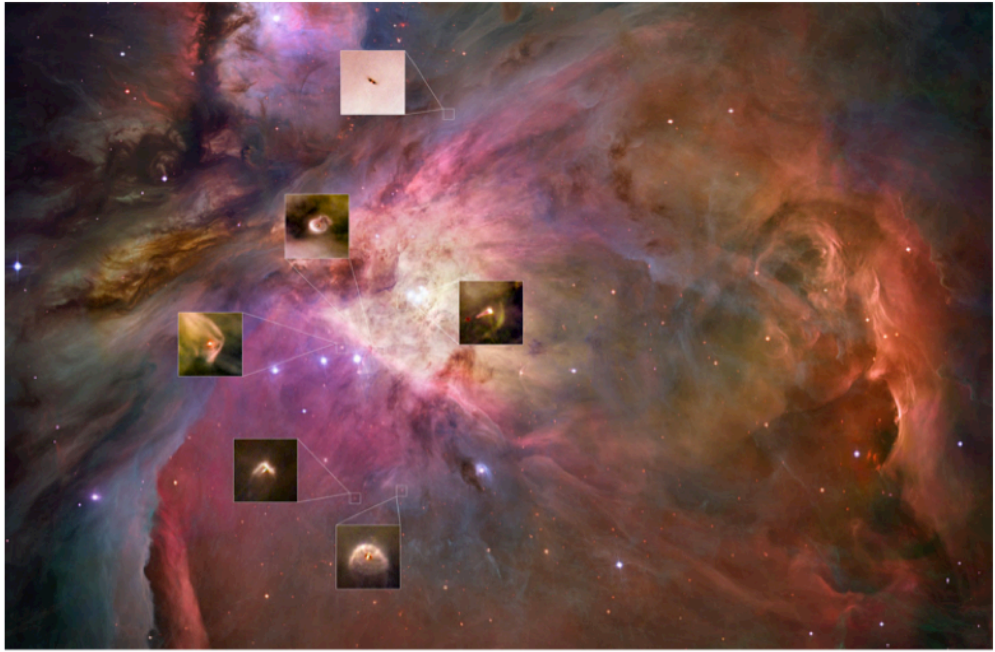

**Figure 26:** Evaporation of proto-planetary disks in Orion; most planets must have formed in such an environment.

## 8.2. Dynamical ejection of planets

The second important environmental effect is that of close encounters between stars and circumstellar disks. From detailed simulations of planet formation in clusters of different density, Bonnell et al. (2011) concluded that planet formation may be suppressed in globular clusters because planets with orbits wider than 0.1 AU would be ejected from the stellar system. In contrast, planet formation is relatively unaffected in open clusters, with only the wider-orbit planetary systems disrupted during the cluster's lifetime.

Simulations of systems like the Solar system indicate that Saturn, Uranus and Nepture would be affected by both encounters with stars and planet-planet scattering (Hao et al. 2013). On the other hand, Jupiter would be largely unaffected by direct encounters with neighbouring stars, as its mass is too large to be perturbed substantially by the other three. Hao et al. concluded that stellar encounters can account for the apparent scarcity of exoplanets in stellar clusters, both for those on wide orbits which are affected directly by stellar encounters, and also those close to the star which can disappear long after a stellar encounter has perturbed the planetary system (see Fig. 27).

Lastly, we note that Bate et al. (2009) investigated the formation of planets in the chaotic environment of star-forming regions, with variable accretion and dynamical interaction between stars. They found that misalignment primarily occurs due to truncation of the proto-planetary disk due to the dynamical interaction with stars. If the interaction is particularly strong, the disks can even be destroyed. Thus, the stellar density of a star-forming region appears to play an important role in both the formation and evolution of planets.

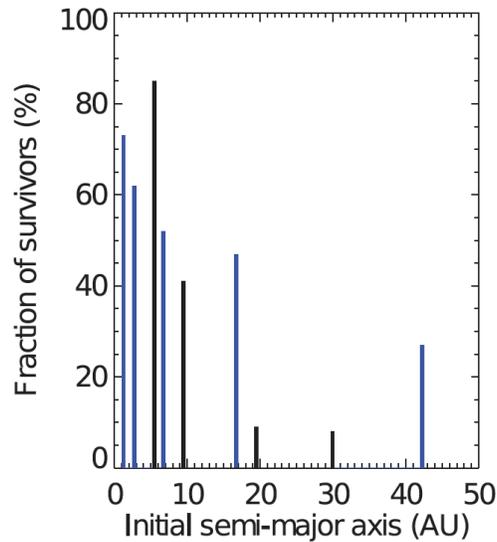

**Figure 27:** Fraction of planets surviving as a function of their initial semi-major axis, for two different models (blue/black) for a cluster like the Orion Nebula Cluster. Inner planets have a higher probability of surviving after the stellar encounters, while outer planets can more easily escape the system (Hao et al. 2013).

## 8.3. Studies of planets in clusters

It therefore appears clear that environment plays an important role in planet formation, but initial studies have had somewhat mixed results, as summarised:

- Hyades cluster: A large incidence of planets was expected in the Hyades because of its high metallicity, but an initial radial-velocity survey of 98 stars over 5 years did not yield any detections (Cochran et al. 2002; Guenther et al. 2005). Since then a planet of 7.6±0.2 $M_{Jup}$ (with a period of 594.9±5.3 days) was discovered around the intermediate-mass star ε Tau by Sato et al. (2007). Moreover, from detection of metal pollution, Farihi et al. (2013) concluded that rocky planetesimals are in orbit around two white dwarfs.

- M67: An initial survey of this open cluster by Pasquini et al. (2012) did not detect any planets, but new results over a longer baseline have recently discovered three planets around two G-type dwarfs and one K-type giant (Brucalassi et al. 2014).

- NGC 6252 & NGC 6791: Studies of these old, metal-rich, open clusters have also not detected any planets (Montalto et al. 2007; 2011).

- NGC 2423 & NGC 4349: Lovis & Mayor (2007) found two substellar objects with masses of 10.6 and 19.8 $M_{Jup}$ orbiting intermediate-mass stars in these intermediate-age, open clusters.

- Presaepe cluster (M44): An RV survey in M44 by Quinn et al. (2012) found two planets among 53 stars, giving a lower limit of $3.8_{-2.4}^{+5.0}$ % on the hot Jupiter frequency in this metal-rich open cluster. This is consistent with the frequency of hot Jupiters in the solar neighbourhood (~1%).

- NGC 6811: from their discovery of two Neptune-sized planets in this open cluster, Meibom et al. (2013) concluded that the frequency and properties of planets in open clusters are consistent with those of planets around field stars in the Galaxy.

A number of surveys of globular clusters have also been carried out. Gilliland et al. (2000) observed 47 Tuc for eight days with the *HST* to look for planetary transits. Ground-based telescopes have also been used to attempt similar detections, e.g., for 33 days in 47 Tuc (Weldrake et al. 2005) and for 25 days in ω Cen (Weldrake et al. 2008). No planets were detected by these studies, with Gilliland et al. concluding that planets in 47 Tuc must be a factor of ten rarer than for field stars. Indeed, there is only one known planetary system in a globular cluster – a hierarchical triple system with a Jupiter-mass planet in a wide orbit with a binary millisecond pulsar (PSR B1620-26; Backer 1993; Thorsett et al. 1999).

Thus, open clusters appear to have a comparable incidence of planets to that in the solar neighbourhood (which follows logically if the members of the solar neighbourhood also formed in clusters). In contrast, globulars seem to have a much lower frequency of planets. Interestingly, metal-rich open clusters do not appear to have a higher incidence of planets, as one might expect if the relation between metallicity and planet formation was the defining characteristic, perhaps a first hint that stellar density is more important than metallicity.

## 8.4. Example Observing Programmes for an ELT-MOS

An ELT-MOS will be able to undertake comprehensive studies of stars in a much broader range of clusters than possible with current facilities, e.g., in both open and globular clusters, spanning a range of densities and metallicities. There will be strong synergies with the clusters component of the Gaia-ESO Spectroscopic Survey which will provide quantitative physical parameters for cluster members in the coming years (vital for subsequent RV analysis), as well as identifying binaries to be excluded from further monitoring.

Such surveys will include both main-sequence and giant stars. The planetary companions of giant stars typically have masses of 2 to 10 $M_{Jup}$, and are typically orbiting at distances of 0.8 to 3 AU (Johnson et al. 2010). Their velocity semi-amplitudes, typically larger than 30 to 40 m/s, are shown in Fig. 28.

### 8.4.1. Globular clusters

As discussed above, search programmes in globular clusters suggest that the frequency of planets is very low. Is this because they have low metallictiies, or because of their high densities? Thus, example programmes could be RV-searches in 47 Tuc and NGC 6366, which have similar metallicities ([Fe/H] = –0.9 to –0.7) but different stellar densities. Equally, as noted earlier, we could investigate the population in relatively metal-rich globulars, e.g., NGC 6440, NGC 6441, and NGC 6388, with [Fe/H] = –0.5 to –0.4.

Requirements: G-type MS stars in 47 Tuc have *V*=18 mag, and K-giants have *V*=15 mag. A Jupiter-mass planet with an orbital period of P < 10 (100) days of a solar-like star would cause RV-variations with a semi-amplitude of 94 (44) m/s. Therefore, the requirement is to detect planets which have semi-amplitudes ≥ 40 m/s, for stars of *V*=18 (MS stars), or *V*=15 (giants).

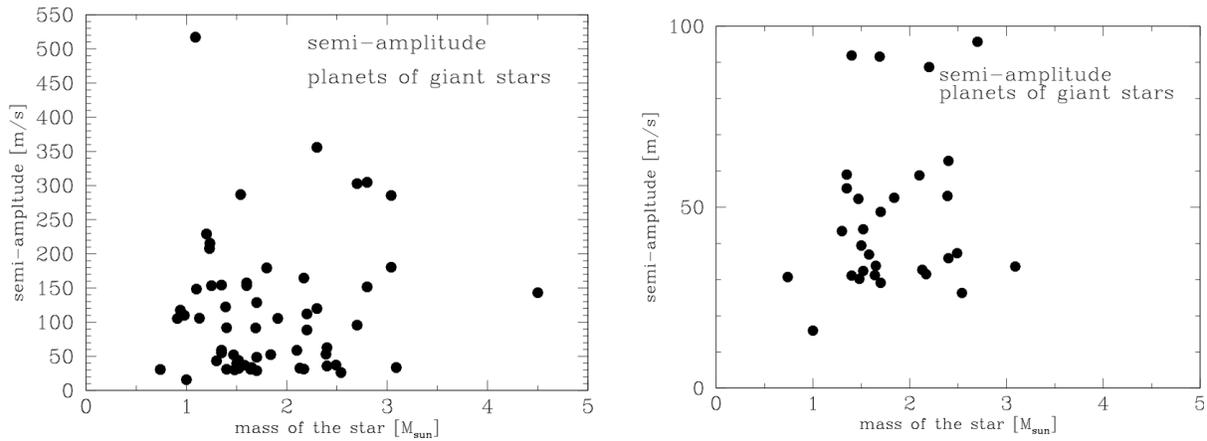

**Figure 28:** *Left:* Radial-velocity (RV) semi-amplitudes of known planet-hosting giant stars. These massive planets are relatively easy to detect compared to lower-mass planets around lower-mass stars. *Right:* Expanded view of results for lower RV semi-amplitudes – almost all the known planet-hosting giant stars have variations of ΔRV ≥ 30 m/s.

### 8.4.2. Frequency of planets vs. Galactocentric distance and in the bulge

Because of the metallicity gradient in our galaxy we would expect the number of planet-hosting stars to increase towards the inner galaxy. According to Reid (2006), the frequency of planets should be twice as common for stars at galactocentric distances of 7 kpc than at 9 kpc. Thus, we would like to be able to detect planets around stars 1 kpc from the Sun, which entails observations of MS stars at *V*=12 mag. Detecting planets in the Galactic bulge is more demanding because of the larger distance and the line-of-sight extinction; in this case giants have *V*=17 mag.

Requirements: As above, to detect a Jupiter-mass planet with orbital periods of P < 10 (100) days requires RV-detections of semi-amplitudes of 94 (44) m/s, down to *V*=17 mag in this case.

### 8.4.3. Exoplanets in Local Group galaxies

Thinking even further afield, all of the known exoplanets to date are confined to a few parsec from the Sun. In addition to the above questions relating to the role of environment on planet formation and evolution, we can also ask if exoplanets exist in other galaxies and, if so, whether they are similar to those found in the Milky Way. These questions can be addressed if we can measure precise RVs for 'hot Jupiters' around giant stars in Local Group galaxies. As an example case, giant stars in the Sagittarius Dwarf Elliptical galaxy (at ~20 kpc) have magnitudes in the range of *V* = 18-20, so are potentially within the grasp of RV monitoring with the E-ELT. This would extend our planet searches into a very different environment as the Sag. Dwarf has a much lower stellar density than the solar neighbourhood, and with a range of metallicities (−1.5 ≤ [Fe/H] ≤ 0.0).

Requirement: Detecting planets with a semi-amplitude of 40 m/s, for stars with *V* = 18-20 mag.

## 8.5. Feasibility

We now discuss some of the astrophysical processes which can impact on precision RV observations.

### 8.5.1. Star spots

RV periodicity can potentially be caused by spots on the stellar surface, thus if candidate planets are found we should check if the variations are comparable to the stellar rotation period. This can be done photometrically with a medium-sized telescope – while the signal caused by planets will always be the same, the properties of any spots will change. An additional check would be to determine the RVs for lines with different temperature sensitivities, to look for differences between the two sets of lines which might indicate that the RV-variations originate from a star spot. Other indications of stellar activity (e.g., the Ca H+K lines, changes in line widths) could also be looked for with, for instance, high-resolution spectroscopy from VLT-FLAMES.

### 8.5.2. Contaminating stars

A faint star contained within the PSF of a target star can also mimic a planet. Assuming that our main targets are K-type giants, early-type stars (with fewer spectral lines) will not significantly affect the RV measurements. From simulations, if a contaminating, later-type star is 2.5 mag fainter than the primary target, and with $\Delta RV \geq 20$ km/s, we expect a shift of $< 3$ m/s for the target. In this scenario, the two stars also produce two well-separated peaks from the cross-correlation RV analysis. In any case, high-resolution, AO-corrected images of all candidates should be taken (for which the angular resolution of the VLT is sufficient).

### 8.5.3. Stellar oscillations

The periods arising from stellar oscillations are typically of order a few minutes for solar-like stars and a few hours for giants. Such short orbital periods would be impossible for planets around such stars, so oscillations should not lead to any misidentifications. However, RV-variations from oscillations will be an important source of 'jitter' noise in planet-search observations. The oscillation amplitudes are much larger for giants than dwarfs, as they scale with both luminosity and mass; the maximum frequency is given by Kjeldsen & Bedding (1995).

As an example, $\beta$ Gem is a planet-hosting K0-type giant star (with M = 1.7±0.4 M$_\odot$), where the stellar oscillations have been studied in detail (Hatzes et al. 2006; Hatzes & Zechmeister 2007). The estimated planet mass is 2.3±0.5 M$_{Jup}$ with an orbital period of 590 days (k = 41±2 m/s). Seven oscillation modes have been identified, with the largest having an amplitude of 5 m/s. In total, oscillations cause an RV-uncertainty of 21 m/s (rms). Given the mass and luminosity of $\beta$ Gem, we would expect a maximum amplitude of ~6 m/s, and periods of a few cycles per day, in good agreement with the observations.

Thus, the RV-accuracy that can be achieved should not be limited by the instrument but by stellar oscillations. The semi-amplitude of a Jupiter-like planet orbiting a solar-like star at 1 AU is 28 m/s, thus stellar oscillations – at least for giants – should not prevent us detecting massive planets. While more RV measurements are required to characterise the stellar oscillations to detect a planet, they can then also be used to determine the masses of the host stars (Kjeldsen & Bedding 1995).

## 8.6. Performance Simulations

### 8.6.1. Signal-to-noise ratio & exposure time

To investigate the potential of detecting giant planets we undertook tests with a range of synthetic spectra (e.g., Fig. 29). Using the E-ELT exposure-time calculator we estimate that spectra with S/N = 100 for stars with $V$ = 18 can be obtained with an exposure of only 300s (assuming an instrument efficiency of 25% and using a K-type template). At $V$ = 19 and 20 the exposure times increase to 700s and 2200s, respectively; while these calculations are for the 42m telescope design, they highlight the feasibility of such observations.

### 8.6.2. Radial-velocity precision

Given a stable, well-calibrated spectrograph, the RV accuracy that can potentially be achieved for solar-like stars is given by the signal-to-noise (S/N), resolving power ($R$), and wavelength coverage (B, in Å) of the spectrograph (Hatzes and Cochran, 1992):

$$\delta RV \approx 1.45 \times 10^9 \times (S/N)^{-1} R^{-1.5} B^{-0.5} \text{ (m/s)}$$

Greater accuracy can be achieved for giant stars due to their generally stronger lines, while less accurate RVs are recovered for lower-metallicity stars due to their weaker lines. To quantify the required S/N we investigated the RV accuracy recovered for simulated spectra spanning a range of metallicities and gravities. The simulations employed 10 runs of synthetic spectra (with appropriate random noise characteristics) and assume a wavelength range of 500 to 600 nm (i.e., B = 100 nm in the above equation), $R$ = 20,000, and 3-pixel sampling. The simulated spectra were then analysed using the same cross-correlation methods used for real observations; the results are summarised in Table 6. Compared to dwarfs, greater accuracy can be achieved for giant stars due to their typically stronger absorption lines, while less accurate RVs will be recovered for metal-poor stars due to the weaker lines (see Fig. 29).

### 8.6.3. Number of spectra required per field

The detectability of planets has been studied by Cumming et al. (2004). To detect a planet with a semi-amplitude that is a factor of two larger than the noise-level, with a false-alarm probability of 1% (50%) requires 80 (30) measurements. If the semi-amplitude is four times the noise, only 28 measurements are required to reduce the false-alarm probability to 1%. With additional measurements, the false-alarm probability is quickly reduced, e.g., if we take 35 measurements it goes down to ~0.01%. With about 30 RV-measurements, planets with an RV semi-amplitude of 40 m/s can be detected. This corresponds to a planet with the mass of Jupiter with an orbital period of 130 days around a solar-like star (Fig. 30).

Table 6: Summary of potential RV accuracies (for S/N = 100 and 20) for a range of stellar metallicities and gravities.

| $T_{eff}$ | log(g) | [Fe/H] | δRV [m/s] S/N=100 | δRV [m/s] S/N=20 |
|---|---|---|---|---|
| **6000** | 4.00 | 0.0 | 11 | 57 |
| **6000** | 4.00 | -0.5 | 13 | 73 |
| **6000** | 4.00 | -1.5 | 29 | 170 |
| **5500** | 4.50 | 0.0 | 10 | 54 |
| **5500** | 4.50 | -0.5 | 15 | 41 |
| **5500** | 4.50 | -1.5 | 28 | 78 |
| **5000** | 2.50 | 0.0 | 8 | 40 |
| **5000** | 2.50 | -0.5 | 8 | 46 |
| **5000** | 2.50 | -1.5 | 15 | 69 |
| **4000** | 1.50 | 0.0 | 4 | 30 |
| **4000** | 1.50 | -0.5 | 9 | 64 |
| **4000** | 1.50 | -1.5 | 10 | 125 |

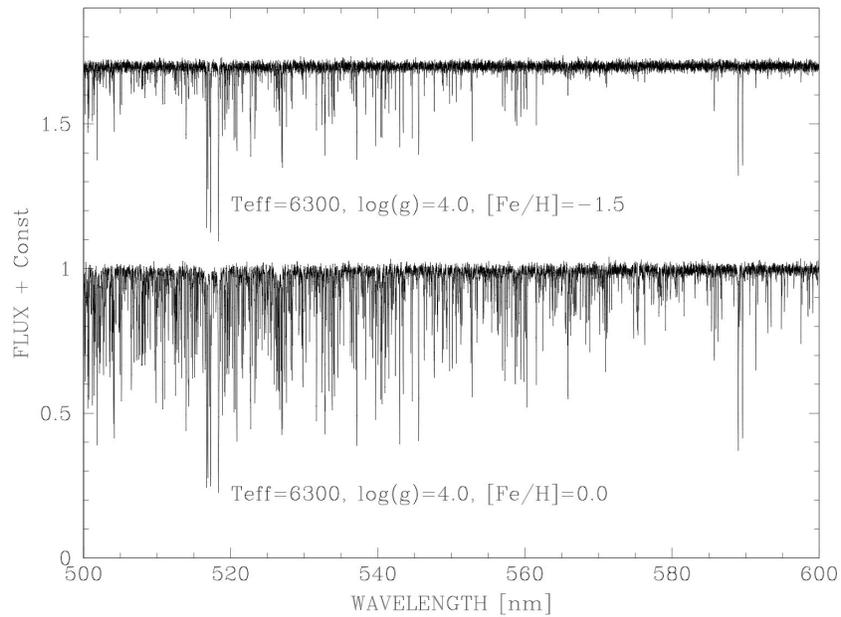

Figure 29: Simulated dwarf spectra with identical effective temperatures (Teff) and gravities log(*g*), but different metallicities. We can recover more precise radial-velocity (RV) estimates for the metal-rich spectrum given the greater intensity/density of absorption lines. Similarly, the 500-550nm region is richer in spectral features than 550-600nm.

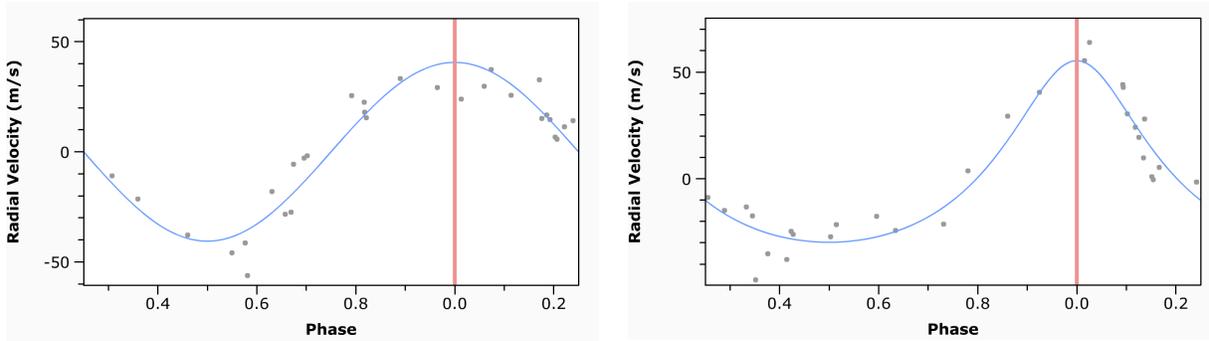

Figure 30: Simulated RV-data for a Jupiter-mass planet orbiting a solar-like star (period = 130 days) for which 30 RV-measurements have been obtained (with a precision of 10 m/s). *Left:* Circular orbit; *Right:* Eccentric orbit (ε = 0.3). (Calculated using the University of Nebraska-Lincoln Radial Velocity Simulator.)

### 8.6.4. Time-series Spectrophotometry

A further exoplanet case concerns the potential use of IFU spectroscopy to investigate the atmospheric properties of transiting planets and free-floating brown dwarfs. For examples of what is possible with current instrumentation:

- Crossfield et al. (2013) measured the radius of the warm ice giant GJ 3407b in six narrow bands from 2.09-2.36 µm using Keck-MOSFIRE, and could rule out both cloud-free or moderately metal-enriched atmospheric models from their results.

- Biller et al. (2013) used the MPG-ESO 2.2m GROND multi-band imager to obtain simultaneous six-band (*r'i'z'JHK*) photometric time series for the two closest known brown dwarfs, in the Luhman 16AB binary system (Luhman, 2013). The B component was known to display variability with a ~5 hr period (Gillon et al. 2013), and Biller et al. found 1-15% variability across multiple bands for both components, including phase offsets between bands for the B component. These phase offsets correlate with the model atmospheric pressure probed at each band, in other words, they reveal complex vertical cloud structures driving the variability.

Looking to the future, an AO-fed multi-IFU instrument would enable high-precision spectroscopic time-series work for directly-imaged exoplanet companions, as well as fainter brown dwarfs and transiting planets. To date, ground-based spectrophotometric time-series studies have been hindered by two factors: 1) instability due to changing telluric water absorption and 2) slit losses with changing parallactic angles, which prevent accurate photometric calibration across time series. The use of multi-IFU instruments directly solves both these problems: changing telluric water absorption can be tracked by observing multiple reference stars on different IFU arms, and a slitless IFU design would minimise slit losses. Pilot multi-IFU studies are currently underway: Biller and collaborators have recently acquired 7.5 hrs of time-series data with VLT-KMOS. Multi-IFU studies are only possible at present for comparatively bright objects, so a multi-IFU mode for MOSAIC would enable such studies for much fainter objects. In particular, the inclusion of AO should boost PSF-stability and also allow time-series work for close companions.

# ELT-MOS: TOP-LEVEL REQUIREMENTS

## 9.1. MOSAIC: A facility-class MOS for the E-ELT

As discussed in the cases above, in the context of the sources for MOS observations we can generally divide the cases into two types:

- 'High definition': Observations of tens of channels at fine spatial resolution, with MOAO providing high-performance AO for selected sub-fields in the focal plane (e.g. Rousset et al. 2010).
- 'High multiplex': Integrated-light (or coarsely resolved) observations of >100 objects at the spatial resolution delivered by GLAO.

These two types of observations entail two sets of requirements, which we are now studying within a common architecture toward the MOSAIC instrument. The requirements for the IGM case (SC2) and the Ly-α escape fraction case (in SC3) are slightly different as they require optical IFU observations, but without strong requirements on the spatial resolution (i.e. GLAO is sufficient). The best aperture for the single-object mode is that which maximises the signal-to-noise ratio for point-like sources in GLAO conditions; given the expected GLAO performance of the E-ELT this was argued to be ~0.6'' in the near-IR (see discussions in SC1 & SC3).

Table 7 summarises the top-level requirements that flow-down from each of the science cases. As noted earlier, we adopt the maximum field available in the current design of the E-ELT (equivalent to a 7' diameter) for the total on-sky patrol field from which science targets can be selected.

## 9.2. A 'survey' MOS for the E-ELT

The large-scale structure case in SC3 argues for a much larger multiplex than the other cases. Indeed, a much broader case can be made for a highly-multiplexed (≥1000 targets) MOS, for surveys of high-z galaxies over degree scales of the sky. In the longer-term plans for the E-ELT we propose that a MOS designed for such large-area surveys would lead to sufficiently distinct and exciting scientific breakthroughs to warrant the construction of a second MOS for the E-ELT.

Adopting a similar 'wedding cake' approach as used on current facilities, one can imagine 'wide', 'deep', and 'ultradeep' surveys, targeting progressively fainter targets over smaller regions of the sky. These would complement coordinated surveys at other wavelengths and provide large, comprehensive datasets to tackle fundamental questions with statistically-robust samples, while also identifying rare sources such as distant QSOs, very high-z galaxies, peculiar stars etc. One of the key attributes for such a 'survey' approach would be to minimise selection biases in spectroscopic samples of tens of thousands of galaxies, enabling:

- Estimates of physical parameters such as luminosities, stellar masses, metallicities, star-formation rates etc. and their relations with cosmic time (via precise redshifts);

- Detection of dynamically-bound, merger systems;

- Studies of the impact of environment (groups, clusters, large-scale structure etc);

- Investigation of the interconnection between galaxies and absorption-line systems of the IGM;

- Identification of targets at z>>1 for detailed follow-up with other facilities (e.g., by providing targets within a given redshift range so that their emission lines fall outside the OH sky lines);

- Calibration of larger studies based on photometric redshifts (BAO, weak-lensing surveys, etc).

## 9.3. A link to a high-resolution (HIRES) spectrograph

The two observation modes (high multiplex, high definition) imply different 'pick-offs' to select targets in the E-ELT focal plane, analogous to the single-object and IFU modes which feed the VLT-FLAMES Giraffe spectrograph. Within such architecture one could envisage a link to feed a high-resolution spectrograph (similar to that from FLAMES to UVES), which could provide observations in parallel to, e.g., high-multiplex observations, for high-resolution spectroscopy of QSO sight-lines, bright stars etc.

## 9.4. Simultaneous high-definition/high-multiplex observations

The high-definition and high-multiplex modes are complementary in the sense that the regions 'between' corrected MOAO sub-fields will have enhanced image quality (roughly equivalent to that from GLAO, see Evans et al. 2012). If the spectrographs for the two modes were independent, and the method of target selection were to enable parallel observations of different samples of targets, this could be an effective means to boost the operational efficiency of the E-ELT.

To expand on this idea, consider that multi-wavelength surveys have (necessarily) concentrated on a small number of deep fields for studies of distant galaxies, e.g., the HUDF, the CDFS, the COSMOS field, etc. Given the huge investment of resources in these deep fields, the E-ELT will almost certainly be used to observe high-z galaxies located within them. Moreover, even if new regions are observed to a comparable depth (and breadth of wavelengths) in the coming years, they will still be relatively limited in number. Thus, many of the high-z targets envisaged for the cases mentioned in SC1 and SC3 are in the same regions of the sky. The same is also true for the stellar populations cases, where there are common galaxies of interest.

Assuming the E-ELT operates on a similar operational model to the VLT (with provision for Large Programmes), one could imagine compelling cases for use of the two MOS modes in the same regions of the sky for several hundred hours. Although simultaneous observations are not strongly motivated on scientific grounds, *if* a robust technical and operational (e.g. sky-subtraction strategy) implementation for simultaneous use of the two modes can be realised, this could provide operational savings of order M€s (i.e. becoming comparable to meaningful shares of the budgets for individual instruments).

**Table 7: Summary of top-level requirements from each Science Case (with desirable reqs. in italics).**

| Case | Multiplex | FoV/target | Spatial sampling | λ-coverage (µm) | R |
|---|---|---|---|---|---|
| **SC1** First light | 20-40 | 2"x2" * | 40-90 mas | 1.0-1.8 *1.0-2.45* | 5,000 |
| | ≥150 | - | (GLAO: 0.6"Ø) | 1.0-1.8 *1.0-2.45* | >3,000 |
| **SC2** Large-scale structures | ≥10-15 | 2"x2" | (GLAO: IFU) | 0.4-0.6 *0.37-0.6* | >3,000 |
| | 50-100 | - | (GLAO) | 0.6-1.8 *0.6-2.45* | >3,000 |
| | >400 | - | (GLAO) | 0.4-1.4 *0.37-1.4* | >3,000 |
| **SC3** Gal. evolution | ≥10 | 2"x2" | 50-80 mas | 1.0-1.8 *1.0-2.45* | 5,000 |
| | ≥100 | - | (GLAO: 0.6"Ø) | 1.0-1.7 *0.8-2.45* | ≥5,000 *~10,000* |
| | ≥10 | 2"x2" | (GLAO: IFU) | 0.385-0.7 *0.37-0.7* | 5,000 |
| **SC4** AGN | ~10 | - | < 100 mas | 1.0-1.8 | >3,000 |
| **SC5** Extragal. stellar pops. | Dense | 1"x1" *1.5"x1.5"* | < 75 mas *20-40 mas* | 1.0-1.8 *0.8-1.8* | 5,000 |
| | 10s arcmin$^{-2}$ | - | (GLAO) | 0.4-1.0 | ≥5,000 *≥10,000* |
| **SC6** Gal. archaeol. | 10s arcmin$^{-2}$ | - | (GLAO) | 0.41-0.46 & 0.60-0.68 *0.38-0.46 & 0.60-0.68* | ≥15,000 *≥20,000* |
| **SC7** GC science | Dense | > 2"x2" | ~100 mas | 1.5-2.45 | ≥5,000 *≥10,000* |
| **SC 8** Planet form. | 10s | - | (GLAO) | 0.5-0.6 | ≥20,000 |

* Minimum size is 1"×1" if on/off sky subtraction is used.

# CONCLUSION

The Universe includes hundreds of billions of galaxies, each of which is populated by hundreds of billions of stars. Astrophysics aims to understand the complexity of an almost incommensurable number of stars, stellar clusters and galaxies, including their spatial distribution, their formation and their current interactions with the interstellar and intergalactic media. A considerable fraction of discoveries in these areas require statistics, which can only be provided by observations of large samples from a MOS.

This White Paper highlights a range of the key science cases, compiled with input from the astronomical community at workshops (e.g. the ELT-MOS meeting in Amsterdam in Oct. 2012), large conferences (e.g., the 'Shaping E-ELT Science & Instrumentation' meeting in Ismaning in Feb. 2013), as well as smaller meetings across Europe and in Brazil. The E-ELT will be the world's largest optical/IR telescope when complete, so we will aim for the largest possible discovery space when planning for a MOS, balanced by technical feasibility and cost. To enable the assembled cases we have defined two primary modes: a high multiplex mode (HMM) and a high (spatial) definition mode (HDM, with MOAO IFUs), and almost all of the science cases described here will benefit from both of these modes.

To illustrate the contributions of a MOS toward studies of galaxy formation and evolution, we summarise three overarching questions which are addressed by multiple cases:

- The epoch of galaxy formation: This is concomitant or driven by reionisation, the agents of which are still unclear. We will address these issues by age-dating (via spatially-resolved spectroscopy) the stellar populations of galaxies and their chemical enrichment (SC3) and pinning down the reionisation epoch as traced by Ly-$\alpha$ emitters (SC1). With SC1 we will also constrain the star-formation rates and metallicities of these very high-z galaxies, which will help us to connect the build-up of galaxies with the chemistry of the IGM and its redshift evolution (SC2).

- The process of galaxy formation: $\Lambda$-CDM argues that the process of galaxy formation should be hierarchical, and predicts a larger number of dwarf galaxies (much later than observed) with massive galaxies assembled late. Both predictions are in striking contradiction to observations and theoretical simulations. In particular, the chemistry of dwarf galaxies and spirals is different from those of large ellipticals, which challenges merger scenarios of formation. Thus, by observing the full mass-spectrum of galaxies (SC3), we can link the robust chemical analyses performed at z = 0 (SC6) to those in the more distant Universe.

- The evolutionary history of present-day galaxies: The Milky Way shows signs of recent accretion of satellites and streams - though these latecomers constitute a low fract of the total mass, it is not clear how local/representative the phenomenon is. What is the role of secular evolution vs. mergers/interactions in the Milky Way's history (SC6)? How do we best calibrate abundance diagnostics for more distant galaxies? Our only extra source of insight into these mechanisms is by studying the chemistry and dynamics of galaxies outside of the Local Group (SC5). We can also use these spatially-resolved studies as templates by which to study the sensitivity of different spectral tracers in integrated-light observations of more distant galaxies (SC3).

As discussed in SC7 and SC8, a MOS will also make exciting contributions to our understanding of the enigmatic Galactic Centre regions, and in the next steps of exoplanet research. In the last section we also summarised the case for a second MOS, with an even larger multiplex, to undertake large (degree scale) surveys of the high-z Universe.

Of course, the conceptual design of an ELT-MOS needs to be both feasible and affordable, thus it is unrealistic to attempt to simultaneously satisfy all of the 'goal' requirements presented here. We will prioritise the science requirements and take into account both technical and operational feasibilities as part of the proposed Phase A study of the MOSAIC concept, in which we aim to deliver a versatile and very capable MOS to the E-ELT as soon as possible in its operations, to enable the exciting and scientifically broad range of cases discussed here. We note that a visible/near-IR MOS with capabilities drawn from the cases in this White Paper is technically feasible, as demonstrated by recent studies of critical issues such as sky background subtraction and multi-object AO.

**Editors**
Chris Evans (UK ATC)
& Mathieu Puech (GEPI)